\documentclass[11pt]{article}
\usepackage[margin=1in]{geometry}
\usepackage[
backend=biber,
style=alphabetic,
sorting=nyt,
maxnames=7,
maxalphanames=7,
backref=true,
url=false
]{biblatex}
\usepackage{tikz}
\usepackage{scalerel}
\usepackage{pict2e}
\usepackage{setspace}
\usepackage{fancyhdr}
\usepackage{tkz-euclide}
\usetikzlibrary{calc}
\usetikzlibrary{patterns,arrows.meta}
\usetikzlibrary{shadows}
\usetikzlibrary{external}
 
\usepackage{pgfplots}
\pgfplotsset{compat=newest}
\usepgfplotslibrary{statistics}
\usepgfplotslibrary{fillbetween}

\usepackage{xcolor}

\usepackage{nicefrac}

\usepackage[colorlinks=true,citecolor=blue,linkcolor=blue]{hyperref}

\hypersetup{
    colorlinks,
    linkcolor={red!50!black},
    citecolor={blue!50!black},
    urlcolor={blue!80!black}
}

\usepackage[]{amsmath,amssymb,amsfonts,latexsym,amsthm,enumerate,fullpage,xcolor,bbm}
\usepackage{wrapfig}
\usepackage{float}
\usepackage[ruled]{algorithm2e}
\newcommand{\hrulealg}[0]{\vspace{1mm} \hrule \vspace{1mm}}
\usepackage[nameinlink, noabbrev, capitalize]{cleveref}
\usepackage{comment}
\usepackage{nicefrac}
\crefname{prop}{Proposition}{Propositions}
\crefname{ineq}{inequality}{inequalities}
\creflabelformat{ineq}{#2(#1)#3}
\usepackage{todonotes}
\usepackage{multicol}
\newtheorem{counter}{Counter}[section]
\newtheorem{theorem}[counter]{Theorem}
\newtheorem{thm}{Theorem}
\newtheorem{lemma}[counter]{Lemma}
\newtheorem{proposition}[counter]{Proposition}
\newtheorem{claim}[counter]{Claim}

\newtheorem{corollary}[counter]{Corollary}
\newtheorem{cor}{Corollary}
\newtheorem{definition}[counter]{Definition}
\newtheorem{observation}[counter]{Observation}
\newtheorem{remark}[counter]{Remark}

\usepackage{subcaption}
\usepackage{graphicx}
\usepackage{color, colortbl}
\definecolor{LightCyan}{rgb}{0.88,1,1}
\definecolor{Gray}{gray}{0.9}
\usetikzlibrary{positioning}
\usetikzlibrary{decorations.pathmorphing}
\usetikzlibrary{automata,positioning}
\usetikzlibrary{arrows}
\usetikzlibrary{arrows.meta}
\usetikzlibrary{calc}

\usepackage{enumitem}
\newlength\caselen
\newlist{casesenum}{enumerate}{2}
\setlist[casesenum,1]{label=\textbf{Case~\arabic*.}, 
  itemindent=*,leftmargin=0pt}
\setlist[casesenum,2]{label=\textbf{Case~\roman*.}, 
  itemindent=*,leftmargin=\parindent}

\newcommand{\N}{\mathbb{N}}

\newcommand{\F}{\mathbb{F}}

\newcommand{\R}{\mathbb{R}}
\newcommand{\poly}{\operatorname{poly}}

\newcommand{\pr}{{\prime}}

\newcommand{\U}{\mathbf{U}}
\newcommand{\X}{\mathbf{X}}
\newcommand{\Xpr}{\mathbf{X}^\pr}
\newcommand{\Y}{\mathbf{Y}}

\newcommand{\A}{\mathbf{A}}
\newcommand{\B}{\mathbf{B}}

\newcommand{\W}{\mathbf{W}}
\newcommand{\RR}{\mathbf{R}}
\newcommand{\ZZ}{\mathbf{Z}}

\newcommand{\IP}{\mathsf{IP}}

\renewcommand{\multicitedelim}{\addsemicolon\space}
\newcommand{\BPP}{\textsf{BPP}\xspace}

\newcommand{\zo}{\{0,1\}}

\newcommand{\abs}[1]{\left\lvert #1 \right\rvert}
\newcommand{\norm}[1]{\left\Vert #1 \right\Vert}

\usepackage{xspace}

\newcommand{\NOSF}[1][\relax]{%
\ifx\relax#1\textrm{NOSF source}\xspace%
\else\ensuremath{{\left(#1\right)}}\textrm{-NOSF source}%
\fi
}
\newcommand{\uniNOSF}[1][\relax]{%
\ifx\relax#1\textrm{uniform NOSF source}\xspace%
\else\textrm{uniform }\ensuremath{{\left(#1\right)}}\textrm{-NOSF source}%
\fi
}
\newcommand{\SHELA}[1][\relax]{%
\ifx\relax#1\textrm{oNOSF source}\xspace%
\else\ensuremath{{\left(#1\right)}}\textrm{-oNOSF source}%
\fi
}
\newcommand{\uniSHELA}[1][\relax]{%
\ifx\relax#1\textrm{uniform oNOSF source}\xspace%
\else\textrm{uniform }\ensuremath{{\left(#1\right)}}\textrm{-oNOSF source}%
\fi
}
\newcommand{\CG}[1][\relax]{%
\ifx\relax#1\textrm{aCG source}\xspace%
\else\ensuremath{{\left(#1\right)}}\textrm{-aCG source}%
\fi
}
\newcommand{\uniCG}[1][\relax]{%
\ifx\relax#1\textrm{uniform aCG source}\xspace%
\else\textrm{uniform }\ensuremath{{\left(#1\right)}}\textrm{-aCG source}%
\fi
}
\newcommand{\NOSFs}[1][\relax]{%
\ifx\relax#1\textrm{NOSF sources}\xspace%
\else\ensuremath{{\left(#1\right)}}\textrm{-NOSF sources}%
\fi
}
\newcommand{\uniNOSFs}[1][\relax]{%
\ifx\relax#1\textrm{uniform NOSF sources}\xspace%
\else\textrm{uniform }\ensuremath{{\left(#1\right)}}\textrm{-NOSF sources}%
\fi
}
\newcommand{\SHELAs}[1][\relax]{%
\ifx\relax#1\textrm{oNOSF sources}\xspace%
\else\ensuremath{{\left(#1\right)}}\textrm{-oNOSF sources}%
\fi
}
\newcommand{\uniSHELAs}[1][\relax]{%
\ifx\relax#1\textrm{uniform oNOSF sources}\xspace%
\else\textrm{uniform }\ensuremath{{\left(#1\right)}}\textrm{-oNOSF sources}%
\fi
}

\newcommand{\CGs}[1][\relax]{%
\ifx\relax#1\textrm{aCG sources}\xspace%
\else\ensuremath{{\left(#1\right)}}\textrm{-aCG sources}%
\fi
}
\newcommand{\uniCGs}[1][\relax]{%
\ifx\relax#1\textrm{uniform aCG sources}\xspace%
\else\textrm{uniform }\ensuremath{{\left(#1\right)}}\textrm{-aCG sources}%
\fi
}

\newcommand{\Ext}{\mathsf{Ext}}

\newcommand{\sExt}{\mathsf{sExt}}

\newcommand{\Cond}{\mathsf{Cond}}

\newcommand{\eps}{\varepsilon}
\newcommand{\epspr}{\varepsilon^\pr}

\newcommand{\Supp}{\mathsf{Supp}}
\newcommand{\edgecolor}{\chi}
\newcommand{\Nbr}{\text{Nbr}}
\newcommand{\colorcount}{\mathsf{count}}
\newcommand{\Adv}{\mathsf{Adv}}

\newcommand{\cX}{\mathcal{X}}

\renewcommand{\W}{\mathbf{W}}
\newcommand{\minH}{H_\infty}
\newcommand{\sminH}{\minH^\varepsilon}
\newcommand{\avgcondminH}{\widetilde{H}_\infty}
\newcommand{\floor}[1]{\left\lfloor#1\right\rfloor}
\newcommand{\ceil}[1]{\left\lceil#1\right\rceil}

\DeclareMathOperator{\supp}{Supp}

\DeclareMathOperator*{\E}{\mathbb{E}}

\usepackage{subfiles}
\newcommand{\dobib}{
    \printbibliography
}

\begin{document}
\renewcommand{\dobib}{}
\renewcommand*{\multicitedelim}{\addcomma\space}

\title{On the Existence of Seedless Condensers: Exploring the Terrain}
\author{}

\author{  Eshan Chattopadhyay\thanks{Supported by a Sloan Research Fellowship and NSF CAREER Award 2045576.}\\ Cornell University\\ \texttt{eshan@cs.cornell.edu}  \and Mohit Gurumukhani\footnotemark[1] \\ Cornell University\\ \texttt{mgurumuk@cs.cornell.edu} \and  Noam Ringach \thanks{Supported by NSF GRFP grant DGE – 2139899,  NSF CAREER Award 2045576 and a Sloan Research Fellowship.}   \\ Cornell University\\ \texttt{nomir@cs.cornell.edu} }
% \author{}
\date{}

% \begin{titlepage}
 \maketitle
 \pagenumbering{Roman}
 
 \begin{abstract}
While the existence of randomness extractors, both seeded and seedless, has been  studied for many sources of randomness, currently, not much is known regarding the existence of seedless condensers in many settings. Here, we prove several new results for seedless condensers in the context of three related classes of sources: Non-Oblivious Symbol Fixing (NOSF) sources, online NOSF (oNOSF) sources (originally defined as SHELA sources in  [AORSV, EUROCRYPT'20]), and almost Chor-Goldreich (CG) sources as defined  in [DMOZ, STOC'23]. We will think of these sources as a sequence of random variables $\mathbf{X}=\mathbf{X}_1,\dots,\mathbf{X}_\ell$ on $\ell$ symbols where at least $g$ out of these $\ell$ symbols are ``good'' (i.e., have some min-entropy requirement), denoted as a $(g,\ell)$-source, and the remaining ``bad'' $\ell-g$ symbols may adversarially depend on these $g$ good blocks. The difference between each of these sources is realized by restrictions on the power of the adversary, with the adversary in NOSF sources having no restrictions.

Prior to our work, the only known seedless condenser upper or lower bound in these settings is due to [DMOZ, STOC'23],  where they explicitly construct a seedless condenser for a restricted subset of $(g,\ell)$-adversarial CG sources. 

The following are our main results concerning seedless condensers for each of these  sources. 
\begin{enumerate}
    \item oNOSF sources
    \begin{enumerate}
        \item When $g\leq\ell/2$, we prove that condensing with error 0.99 above rate $\frac{1}{\lfloor \ell/g \rfloor}$ is impossible. In fact, we show that this is tight.
        \item Quite surprisingly, for $g> \ell/2$, we show the existence of  excellent condensers for uniform oNOSF sources. In addition, we show the existence of similar condensers for oNOSF sources with only logarithmic min-entropy. Our results are based on a new type of two-source extractors, called \emph{output-light two-source extractors}, that we introduce and prove the existence of. 
    \end{enumerate}
    \item Adversarial CG sources
    \begin{enumerate}
        \item We observe that uniform adversarial CG sources are equivalent to uniform oNOSF sources and consequently inherit the same results.
        \item We show that one cannot condense beyond the min-entropy gap of each block or condense low min-entropy CG sources above rate $1/2$.
    \end{enumerate}
    \item NOSF sources
    \begin{enumerate}
        \item We show that condensing with constant error above rate $\frac{g}{\ell}$ is impossible for uniform NOSF sources for any  $g$ and $\ell$, thus ruling out the possibility of any non-trivial condensing. This shows an interesting distinction between NOSF and oNOSF sources.
    \end{enumerate}
\end{enumerate}

% \begin{enumerate}
%     \item  When $g\leq\ell/2$, we prove for all three classes of sources that condensing with error 0.99 above rate $\frac{1}{\lfloor \ell/g \rfloor}$ is impossible. In fact, we show that this is tight for oNOSF sources and uniform almost CG sources.
%     \item  Quite surprisingly, for $g> \ell/2$, we show the existence of  excellent condensers for uniform oNOSF sources and uniform almost CG sources, thus proving a separation from NOSF sources. In addition, we show the existence of similar condensers for oNOSF sources with only logarithmic min-entropy. Our results are based on a new type of two-source extractors, called \emph{output-light two-source extractors}, that we introduce and prove the existence of. 
%     \item  We show that condensing with constant error above rate $\frac{g}{\ell}$ is impossible for uniform NOSF sources for any  $g$ and $\ell$. 
%     \item We show that condensing low min-entropy CG sources above rate $1/2$ is impossible.
% \end{enumerate}
These results make progress on several open question from [DMOZ, STOC'23], [AORSV, EUROCRYPT'20], and  [KN, RANDOM'23].
\end{abstract}

\maketitle
\newpage
%\begin{spacing}{0.91}
\setcounter{tocdepth}{2}
\tableofcontents
%\end{spacing}

\newpage
\clearpage
\pagenumbering{arabic}
%%% SECTIONS HERE %%%

\section{Introduction}
One of the most fruitful lines of research in computer science has been that of randomness. From the traditionally more applied areas of algorithm design (e.g., Monte Carlo simulations),  error-correcting codes and cryptography to the more theoretical areas of property testing, combinatorics, and circuit lower bounds, randomness has played a key role in seminal discoveries. In many of these works, the use of high-quality random bits, or alternatively, a way to convert low-quality randomness into high-quality randomness, is essential. In cryptography, the authors of \cite{dodis_impossibility_2004} showed that high-quality randomness is essential for tasks such as bit commitment schemes and secure two-party computation. On the other hand, being able to extract uniform bits from low-quality randomness allows us to simulate randomized algorithms \cite{zuckerman_general_1990}.

In most use-cases, randomness takes the form of uniformly random bits. These motivated the construction of randomness extractors,\footnote{In this paper, when we mention extractors/condensers, we usually mean seedless extractors/condensers.}
 functions that take low-quality randomness (which we often like to think of as natural processes) and convert it into uniformly random bits.
% There is a long line of works \cite{dvir_extensions_2013, dvirwigderson11merger, lu_extractors_2003} that construct seeded extractors, functions that take in low-quality randomness along with a small amount of uniform bits, with close to optimal parameters.
It is impossible to extract from the class of all sources and so extractors are constructed with respect to a restricted class of sources.

A number of works \cite{santha_generating_1986, chor_unbiased_1988, zuckerman_general_1990, reingold_note_2004} have shown that deterministic extraction is impossible for many natural classes of randomness sources.
The question that arises for such sources then is whether any improvement to their randomness can be made. That is, while it may not be possible to convert a source into uniform bits, maybe it is possible to condense a source into another source with a higher density of randomness. The central focus of our paper is in understanding the possibility of condensing for various natural models of weak sources where it is known that extraction is impossible. 

We first introduce the way that we measure randomness and the notions of extractors and condensers. The notion of randomness that is standard in this line of work is that of min-entropy. For a source $\X$ on $n$ bits,  we define its \emph{min-entropy} as $\minH(\X)=\min_{x\in\zo^n}\{-\log(\Pr[\X=x])\}$.
A source $\X$ over $n$ bits with min-entropy at least $k$ is called an \emph{$(n,k)$-source}. Given any two distributions $\X$ and $\Y$ on $\zo^n$, we define their statistical distance or total-variation (TV) distance as $\abs{\X-\Y}=\max_{Z\subseteq\zo^n}\abs{\Pr_{x\sim \X}[x\in Z]-\Pr_{y\sim\Y}[y\in Z]}$.  We also need the notion of   \emph{smooth min-entropy}: for a source $\X$ on $\zo^n$, it=s smooth min-entropy with smoothness parameter $\varepsilon$ is $\sminH(\X)=\max_{\Y:\abs{\X-\Y}\leq\varepsilon}\minH(\Y)$. Conceptually, smooth min-entropy asks that the source we are looking at be $\varepsilon$-close in TV-distance to some other source with the desired amount of min-entropy. We are now in a position to define randomness extraction and condensing.
\begin{definition}
Let $\cX$ be a family of distributions over $\zo^n$. A function $\Ext:\zo^n\to\zo^m$ is an \emph{extractor} for $\cX$ with error $\varepsilon>0$ if for all $\X\in\cX$ we have $\abs{\Ext(\X)-\U_m}\leq\varepsilon$. 
\end{definition}

For extractors to exist, we  require all sources in $\cX$ to have  entropy. When each source in $\cX$ is an $(n,k)$-source, we say that $\Ext$ is a $(k,\varepsilon)$-extractor for $\cX$.
For some classes, an extractor may not exist (such as for the class of all $(n, n-1)$-sources). Consequently, we turn to the looser requirements of condensing.
\begin{definition}
For a family of distributions $\cX$ over $\zo^n$,  a function $\Cond:\zo^n\to\zo^m$ is a \emph{condenser} with error $\varepsilon\geq0$ if for all $\X\in\cX$ we have that $\sminH(\Cond(\X))/m \ge \minH(\X)/n$. We say that $\Cond$ has \emph{entropy gap} $\Delta$ if $\sminH(\Cond(\X))\geq m-\Delta$. When $\cX$ is the class of $(n,k)$-sources and $k'=m-\Delta$, we say that $\Cond$ is a $(k, k', \varepsilon)$-condenser.
\end{definition}
%In words, the output of $\Cond$ is $\varepsilon$-close in statistical distance to some distribution with min-entropy rate higher than that of $\X$. This is a weaker requirement than extracting which requires the output distribution is $\eps$-close to the uniform distribution.  

Unfortunately, even this notion is too strong as we cannot condense with error $\varepsilon$ from the class of all $(n,k)$-sources so that the output entropy rate is larger than $k/n$. \footnote{Assuming $m\le n$, the output entropy can be shown to be most $k + m - n + \log(1 / (1-\eps))$. See \cref{lem:can't condense arbitrary n ; k source} for a proof of this fact.}
We thus study condensing from classes of sources which have some additional structure along with a min-entropy requirement.
In this paper, we explore the possibility of condensing from three related models of weak sources. These models, some of which have been studied since the 1980s, are very general and well-motivated by practical considerations.

The rest of the introduction is organized as follows: In \cref{intro:utility}, we present the case that condensers have many applications and are hence a natural direction of study, in particular when extraction is not feasible. In \cref{intro:models}, we discuss the models of weak sources that we study, present relevant prior work on these models, and discuss our results for each of them.

\subsection{The utility of condensing}\label{intro:utility}
We present two viewpoints in motivating our study of condensers. We compare what is possible via condensing in contrast to extracting and consider the utility of condensing for simulating \BPP algorithms.

\subsubsection{Condensing vs. extracting}\label{sec:intro-condensing vs extracting}

Condensers exist in many scenarios when it can be provably shown that deterministic extraction is not possible. Thus, they allow us to obtain randomness that is more useful than what we began with in cases where extracting uniform bits is impossible. One significant example is that of Santha-Vazirani (SV) sources \cite{santha_generating_1986} and their generalization, Chor-Goldreich (CG) sources \cite{chor_unbiased_1988}. 

Informally, an SV source is a string of random bits such that the conditional distribution of each bit on the bits that come before it is guaranteed to have some minimum amount of min-entropy; a CG source generalizes this to allow each bit to instead be a symbol in $\zo^n$. It is well known that deterministic extraction is impossible for both SV and CG sources \cite{santha_generating_1986, chor_unbiased_1988, reingold_note_2004}. The recent result of \cite{doron_almost_2023} with regards to condensing from CG sources  stands in contrast to these impossibility results for extraction. Other examples of sources for which deterministic extraction is not possible while deterministic condensing are the \emph{somewhat dependent} sources of \cite{ball_randomness_2022} and  block sources  \cite{ben-aroya_two-source_2019}.

We briefly mention that seeded condensers are known to achieve parameters unattainable by seeded extractors \cite{radhakrishnan_bounds_2000}. Further, seeded condensers have been extremely useful in excellent constructions of seeded extractors \cite{reingold_extracting_2006, zuckerman_linear_2007, ta-shma_lossless_2007, guruswami_unbalanced_2009}.
%Another important property of condensers is that they can attain parameters that are unachievable for extractors. In fact, there exist seeded condensers --- condensers that also take in a random seed as input --- that have the property that the min-entropy of their output is equal to that of their input, making them \emph{lossless} \cite{aviv_entropy_2019}. In contrast, the authors of \cite{radhakrishnan_bounds_2000} showed that no seeded extractor can be lossless. Furthermore, while condensers are important in their own right, it is worth mentioning that they are useful in the construction of extractors, such as in .

\subsubsection{Condensing for simulating \BPP algorithms}
Condensers with small entropy gap are useful in simulating randomized algorithms with low overhead \cite{doron_almost_2023}. There are two ways one can go about this. First, there exists an explicit seeded extractor $\Ext$ with seed length $d=O(\log(\Delta))$ that can extract from any $(n,k)$-source $\X$ with entropy gap $\Delta=n-k$ \cite{RVW}. Then, to simulate a randomized algorithm $\mathcal{A}$ in \BPP, we instead sample $x\sim \X$ and take the majority of the output of $\mathcal{A}$ on $\{\Ext(x,s)\}$ where we cycle over all seeds $s$ \cite{vadhan_pseudorandomness_2012}.

For some applications in randomized protocols, cryptography and interactive proofs, one cannot afford to compute $\Ext$ all $2^d$ times by cycling through every seed \cite{barak_leftover_2011, dodis_randomness_2012, dodis_overcoming_2013, dodis_key_2014}. 
Alternatively, we can simulate $\mathcal{A}$ using a ``one-shot'' method in which we do not iterate over all seeds. A result from \cite{dodis_key_2014}  allows us to simulate $\mathcal{A}$ on the condensed source $\X$ (with entropy gap $\Delta$) by reducing the error of $\mathcal{A}$ to $2^{-\Delta-1}\cdot \eps$ and then using $\X$ directly to simulate random bits in $\mathcal{A}$. Such a simulation will have error $\eps$.

\subsection{Models of weak sources and our results} \label{intro:models}
We consider three adversarial classes of sources motivated by weak sources that appear in practice as well as in various cryptographic settings. These   sources are natural generalizations of the well-studied independent sources wherein we allow for an adversarial dependence between sources. Changing the scope and power of the adversary in natural ways   gives rise to the three different classes of sources that we will consider.

The three randomness sources that we focus on in this work are all composed of blocks of bits, known as symbols, which vary in how they are permitted to relate to other symbols in the source. In these definitions, we will consider sources $\mathbf{X}=\mathbf{X}_1,\dots,\mathbf{X}_\ell$ of length $\ell$ where each $\mathbf{X}_i\in\{0,1\}^n$ is called a block. Generally, we will term blocks that have some minimum amount of randomness ``good'' and blocks that are chosen by an adversary as ``bad''. Next, we  discuss these three models of weak sources, presenting what was known from prior work and our new results for each of these models.

\subsubsection{Online non-oblivious symbol fixing sources}
The first class of adversarial sources that we will define is that of \emph{online non-oblivious symbol fixing (oNOSF) sources}. While these are a restriction of general NOSF sources, which we will define later, we introduce them first since they have the weakest adversary and, consequently, the strongest positive results. Formally, we define \SHELAs as follows.

\begin{definition}[\SHELAs, \cite{aggarwal_how_2020}]
    A \emph{\SHELA[g,\ell,n,k]} $\X=\X_1,\dots,\X_\ell$ on $(\zo^n)^\ell$ is such that $g$ out of the $\ell$ blocks are independently sampled $(n,k)$-sources (i.e., good), and the remaining $\ell-g$ bad blocks only depend on blocks with smaller indices (i.e., to their left).
    % Consequently, for all $i\in\mathcal{G}$ we require that $\X_i=\X_i\mid \X_1=x_1,\dots,\X_{i-1}=x_{i-1}$ for any prefix $x_1,\dots,x_{i-1}$. 
\end{definition}
If $k=n$, we call $\X$ a \emph{\uniSHELA[g,\ell]}.
\SHELAs form a natural class of sources to study when an adversary is working in real time and cannot predict the future. One such real-world example is that of blockchains. From \cite{oswald_bitcoin_2015, pass_analysis_2017}, we know that in a sequence of blocks, there will be some fraction of blocks that are chosen by honest players. Moreover, since these honest players are not working together, their chosen blocks may be considered as independent, fulfilling the requirement for good blocks for \SHELAs. The adversarial players, on the other hand, can only see blocks added to the blockchain thus far and do not know which values of blocks will be added in the future, fulfilling the requirements for bad blocks for \SHELAs. For more uses of \SHELAs, see \cite{aggarwal_how_2020}.

\paragraph{Previous work}
Prior to our work, the only results for condensing or extracting from \SHELAs are due to \cite{aggarwal_how_2020}. In \cite{aggarwal_how_2020}, the authors study Somewhere Honest Entropic Look Ahead (SHELA) sources, which are exactly convex combinations of \SHELAs (see \cref{prop:convex combination SHELA}). They (1) transform (not uniform) \SHELAs into \uniNOSFs and (2) show that for any $\gamma\in(0,1)$, there exists an $\ell$ such that extraction is not possible for \SHELAs[\floor{\gamma\ell},\ell]. 
% They conjectured that \SHELAs[g, \ell] cannot be transformed into \uniNOSFs[g^\pr, \ell^{\pr}] where $\frac{g^{\pr}}{\ell^\pr} > \frac{g}{\ell}$.
% (with parameters not meeting Ajtai-Linial, of course). {\color{red} Eshan: I don't think Ajtai-Linial has been discussed above}

\paragraph{Our results}
We prove the existence of condensers with excellent parameters when the majority of the blocks of a \uniSHELA are good.
\begin{theorem}[Informal version of \cref{thm:can-condense-from-uniform-shela-above-rate-half}]\label{thm:intro-can-condense-from-uniform-shela-above-rate-half}
For all constant $g, \ell$ and all $\eps$ such that $g>\ell/2$, there exists a condenser $\Cond:(\zo^n)^\ell\to\zo^m$ such that for any $\uniSHELA[g,\ell]$ $\X$, we have $\sminH(\Cond(\X))\geq m-O(\log(m/\eps))$ where $m=n$.
\end{theorem}
For our construction, we introduce a new type of two-source extractor\footnote{See \cref{def:two-source extractor} for a definition of two-source extractors} that we call a \emph{$R$-output-light} two-source extractor. Such a two-source extractor $2\Ext: \zo^{n_1} \times \zo^{n_2} \rightarrow \zo^m$ satisfies the additional guarantee that each output $z \in \zo^m$ can only be produced by $R$ inputs $x \in \zo^{n_1}$ in the first source (see \cref{def:output-light} for the formal definition). The existence of such extractors is not obvious, and we show that output-light two-source extractors exist with strong parameters in \cref{cor:great-output-light-two-source-extractors-exist}.  Our proof uses the observation that  $R$-output-lightness is implied by the notion of $R$-invertibility, which simply bounds $\norm{\Cond(\X)}_\infty$ by $R$ (see \cref{def:invertible-function} for a formal definition). Incidentally, this latter notion has been recently used in a different context, to construct explicit random access linear codes with constant rate and distance \cite{cook_explicit_2024}. While we are unable to explicitly construct such output-light two-source extractors, we do  construct an explicit output-light \emph{seeded} extractor, which we use to condense from $\uniSHELAs[2,3]$ and more (see \cref{sec:appendix-explicit condensers}).

In fact,  we can achieve a stronger result and show existence of condensers for \SHELAs with only logarithmic min-entropy guarantee in the good blocks.  
\begin{theorem}[Informal version of \cref{thm:can condense low min entropy shela with g > l/2+1}]\label{thm:intro-can condense low min entropy shela with g>l/2+1}
    For any constant $g, \ell$ and all $\eps$ such that $g>\ell/2+1$, there exists a condenser $\Cond:(\zo^n)^\ell\to\zo^m$ such that for any $\SHELA[g,\ell,n,k]$ $\X$ with $k\geq 2\log(n/\eps)$ we have $\sminH(\Cond(\X))\geq m-O(\log(m/\eps))$ where $m=\Omega(k)$.
\end{theorem}
To accomplish this, we transform logarithmic min-entropy \SHELAs to \uniSHELAs and then apply the condenser for \uniSHELAs.
We transform logarithmic min-entropy \SHELAs to \uniSHELAs by modifying the construction of a somewhere-extractor for high min-entropy SHELA sources by \cite{aggarwal_how_2020}.
These results imply that \SHELAs can be useful for low overhead simulation of \BPP algorithms. Furthermore, taken in tandem with the result that for all $\gamma>0$ there exists a large enough $\ell$ such that one cannot extract from \uniSHELAs[\floor{\gamma\ell},\ell] from \cite{aggarwal_how_2020}, we have shown that \SHELAs are one of the natural classes of sources that admit seedless condensing but not seedless extraction. This adds \SHELAs to the short list of such natural sources mentioned in \cref{sec:intro-condensing vs extracting}. 

In contrast, condensing in the regime of $g\leq\ell/2$ is more nuanced: some non-trivial condensing beyond rate $\frac{g}{\ell}$ is possible provided $g$ does not divide $\ell$, but condensing to a significantly higher rate is not possible.
\begin{thm}[\cref{thm:can't condense SHELA above 1/c}, restated]\label{thm:intro-can't condense when g < ell/2}
    For any function $f:(\zo^n)^\ell\to\zo^m$ and $\varepsilon>0$, there exists a constant $\delta$ and \uniSHELA[g,\ell] $\X$ with $g\leq \ell/2$ such that $\sminH(f(\X))\leq\frac{1}{\floor{\ell/g}}\cdot m+\delta$.
\end{thm}
This partially resolves\footnote{Our result on the existence  of condensers  falls short of completely resolving their conjecture as it does not  transform \uniSHELAs into \uniNOSFs.}
a conjecture of \cite{aggarwal_how_2020}: they conjectured that \SHELAs[g, \ell] cannot be transformed into \uniNOSFs[g^\pr, \ell^\pr] with $\frac{g^\pr}{\ell^\pr} > \frac{g}{\ell}$. Our condensing impossibility implies $\frac{g^\pr}{\ell^\pr} \le \frac{1}{\floor{\ell/g}}$ for any such transformation. 
% If we restrict ourselves to the case where $g$ divides $\ell$, then this theorem exactly reduces to:
% \begin{cor}
%     For any function $f:(\zo^n)^\ell\to\zo^m$ and for all $\varepsilon>0$ there exists a constant $\delta>0$ and \uniSHELA[g,\ell] $\X$ (where $g$ divides $\ell$) such that $\sminH(f(\X))\leq\frac{g}{\ell}\cdot m+\delta$.
% \end{cor}
This negative result is tight and we are able to condense \uniSHELAs[g, \ell] up to rate $\frac{1} {\floor{\ell/g}}$.
\begin{thm}(Informal version of \cref{thm:Condensing from uniform (g;l)-SHELA with g<=l/2})\label{thm:intro-Condensing from uniform (g;l)-SHELA with g<=l/2}
    For any constant $g, \ell$ and $\eps$ such that $\floor{\ell/g}=r$ and $\ell/g\neq r$, there exists a condenser $\Cond:(\zo^n)^\ell\to\zo^m$ such that for any \uniSHELA[g,\ell] $\X$ we have $\sminH(\Cond(\X))\geq\frac{1}{r}\cdot m-O(\log(m/\eps))$ where $m=\Omega(n)$.
\end{thm}

As before in \cref{thm:intro-Condensing from uniform (g;l)-SHELA with g<=l/2}, we can convert a logarithmic min-entropy \SHELA to a \uniSHELA and then apply \cref{thm:intro-Condensing from uniform (g;l)-SHELA with g<=l/2}. This yields:
\begin{thm}(Informal version of \cref{thm:Condensing from low entropy (g;l)-SHELA with g<=l/2})\label{thm:intro-Condensing from low entropy (g;l)-SHELA with g<=l/2}
    For all constant $g, \ell$ and $\eps$ such that $\floor{\frac{\ell-1}{g-1}}=r$ and $\frac{\ell-1}{g-1}\neq r$, there exists a condenser $\Cond:(\zo^n)^\ell\to\zo^m$ such that for any \SHELA[g,\ell,n,k] $\X$ with $k\geq 2\log(n/\eps)$, we have that $\sminH(\X)\geq \frac{1}{r}\cdot m-O(\log(m/\eps))$ with $m=\Omega\left(k\right)$.
\end{thm}
We note that \cref{thm:intro-can-condense-from-uniform-shela-above-rate-half} and \cref{thm:intro-can condense low min entropy shela with g>l/2+1} are special cases of \cref{thm:intro-Condensing from uniform (g;l)-SHELA with g<=l/2} and \cref{thm:intro-Condensing from low entropy (g;l)-SHELA with g<=l/2} in the case that $\floor{\ell/g}=r=1$, allowing us to state all of our condensing possibility results succinctly. 

Put together, our results demonstrate a sharp threshold at $g=\ell/2$ for condensing from \SHELAs with a small entropy gap. To our knowledge, there is no other set of sources that exhibits such behavior, making \SHELAs unique among both adversarial sources and general randomness sources.

\subsubsection{Adversarial Chor-Goldreich sources}
Next, we consider a generalization of \SHELAs, termed \emph{adversarial Chor-Goldreich (CG) sources}, that we obtain by strengthening the adversary's power. Adversarial CG sources share the motivation from \SHELAs that the adversary cannot predict the future. Rather than forcing the adversary to have its blocks only depend on blocks in the past (those with smaller indices), \CGs require that good blocks have some entropy conditioned on all blocks that came before them. In other words, bad blocks cannot expose all of the entropy of future good blocks.

\begin{definition}[Adversarial CG (aCG) sources, \cite{chor_unbiased_1988, doron_almost_2023}]
    We define a \emph{\CG[g,\ell,n,k]} $\X=\X_1,\dots,\X_\ell$ to be a distribution on $(\zo^n)^\ell$ such that there exists a set of good indices $\mathcal{G}\subseteq[\ell]$ of size at least $g$ for which $\minH(\X_i\mid \X_1=x_1,\dots,\X_{i-1}=x_{i-1})\geq k$ for all $i\in\mathcal{G}$ and all prefixes $x_1,\dots,x_{i-1}$.
\end{definition}

As before, if $k=n$, then we say that $\X$ is a \uniCG[g,\ell]. Observe that because the good blocks of a \SHELA are independent of all blocks before it, \SHELAs are trivially \CGs as well. As a consequence, our condensing impossibility results from \cref{thm:intro-can't condense when g < ell/2} immediately apply to \CGs as well. 
Moreover, a convenient fact that we later show in \cref{prop:uniSHELA equivalent to uniCG} and will rely on is that \uniCGs[g,\ell] and \uniSHELAs[g,\ell] are equivalent. 

CG sources are a well-studied class of sources introduced by \cite{chor_unbiased_1988} as a generalization of Santha-Vazirani sources \cite{santha_generating_1986}. Hence, the majority of the work done on CG sources has been in the non-adversarial setting in which $g=\ell$. Adversarial CG sources that contain bad blocks were only recently introduced in \cite{doron_almost_2023} (although they use the terminology ``almost'' CG sources), in which the authors show several condensing results for CG and adversarial CG sources. Our work can then be seen as a meaningful addition to this long line of research on CG sources and their generalizations.

\paragraph{Previous work}
The impossibility of extraction from both \SHELAs and \CGs due to \cite{aggarwal_how_2020, chor_unbiased_1988} naturally raises the question of whether there is a distinction between these two sources with regards to randomness condensing.

For CG sources, \cite{gavinsky_santha-vazirani_2020} showed that errorless condensing is impossible. In contrast, \cite{doron_almost_2023} proved several possibility results regarding condensing with error for CG sources.
Their results assume that the size of each block is very small (almost constant) compared to the number of blocks.
% For a lot of cases, this setting is arguably harder to condense from as compared to the setting where you are given small number of large sized blocks, since, given a large sized blocks, we can partition them and get many small sized blocks.

% The most significant of their results in the context of our work is that they constructed an explicit deterministic condenser with exponentially small error using the constant-degree lossless expanders given by \cite{capalbo_randomness_2002} for a subclass of \CGs they termed \emph{suffix-friendly almost CG sources}. These suffix-friendly almost CG sources are like our \CGs[g,\ell] except with the requirement that the $g$ good blocks be well-distributed among the $\ell$ total blocks. Their construction obtains a constant entropy gap for suffix-friendly almost CG sources when the fraction of bad blocks $b=\ell-g$ is quite small $b\leq10^{-8}\ell$, and otherwise they are not able to condense when $b\geq\ell/2$ or without the suffix-friendliness requirement. 

We also note that the authors of \cite{doron_almost_2023} considered various other relaxations of the definition of \CGs that we do not consider here. These include good blocks having only smooth min-entropy conditioned on previous blocks instead of the stronger condition of min-entropy, having smooth min-entropy conditioned on a constant fraction of prefixes of previous blocks instead of all prefixes, and having a Shannon entropy requirement instead of min-entropy requirement.

% In the regime of constant block sizes and an increasing number of blocks $\ell$, the authors of \cite{doron_almost_2023} constructed explicit condensers with a constant entropy gap for non-adversarial CG sources a subclass of \CGs, termed suffix-friendly almost-CG sources, \CGs in which the bad blocks are relatively evenly spaced, but in general achieved no non-trivial condensing when adversarial blocks can be arbitrarily placed. 

\paragraph{Our results}
In \cite{doron_almost_2023}, the authors pose the question of whether it is possible to condense from \CGs with a constant entropy gap.\footnote{In their paper, they phrase it as removing the requirement of suffix-friendliness.} We give a partially positive answer to this by showing that we can condense from \uniCGs[g,\ell] with $g>\ell/2$ with logarithmic entropy gap since \uniCGs[g,\ell] are equivalent to \uniSHELAs[g,\ell] and we can defer to \cref{thm:intro-Condensing from uniform (g;l)-SHELA with g<=l/2}. Of course, all of \cref{thm:intro-Condensing from uniform (g;l)-SHELA with g<=l/2} applies to \uniCGs, so we can condense any \uniCG[g,\ell] to rate $\frac{1}{\floor{\ell/g}}$. The generalization of these results in \cref{thm:intro-can condense low min entropy shela with g>l/2+1} do not hold for non-uniform \CGs since non-uniform \CGs need not be \SHELAs. Before our work, no non-trivial condensing was known for \uniCGs[g,\ell, n] even in the case of $g=\ell-1$. It is important to note that our results hold for comparatively large block sizes $n=2^{\omega(\ell)}$, in contrast to the results of \cite{doron_almost_2023} that hold for constant block sizes and increasing $\ell$.

As previously mentioned, since \SHELAs are a subclass of \CGs, our condensing impossibility results from \cref{thm:intro-can't condense when g < ell/2} transfer over. Thus, in the $g\leq\ell/2$ regime, we give a negative answer to the question of \cite{doron_almost_2023} by showing that good condensers do not exist for \uniCGs[g,\ell], let alone condensers with a constant entropy gap. Note that unlike our condensing possibility results that only apply to \uniCGs, our impossibility result applies to non-uniform \CGs as well.

In addition, we prove various condensing impossibility results that work even when there are no bad blocks (i.e., for non-adversarial, or just regular, CG sources): the first result of \cref{thm:intro-can't condense CG sources beyond entropy gap} is based on a reduction from general $(n, k)$-sources to CG sources and the second result uses a reduction from \uniSHELAs to low min-entropy CG sources. 

\begin{thm}[Informal version of \cref{thm:can't condense CG sources beyond entropy gap} and \cref{thm:can't condense CG sources}]\label{thm:intro-can't condense CG sources beyond entropy gap}
For all $\Delta > 0$ and for every function $f: (\zo^n)^{\ell}\to \zo^m$, there exists an \CG[\ell, \ell] $\X$ satisfying either of the following with $\varepsilon=0.99$:
\begin{itemize}
    \item The good blocks have min-entropy at least $n - \Delta - \log(\ell) - O(1)$ conditioned on all fixings of previous blocks and $\sminH(f(\X)) \le m - \Delta - \max(m - \ell n, 0) + O(1)$.
    \item  The good blocks have min-entropy at least $n/\ell - \log(\ell) - O(1)$ conditioned on all fixings of previous blocks and $\sminH(f(\X)) \le \frac{1}{2}\cdot m + O(1)$.
\end{itemize}
\end{thm}

It is important to note that the first bullet above does not subsume the second. 
% When the good blocks have low linear entropy, we can prove a stronger impossibility result that condensing beyond rate $1/2$ is impossible. 
In particular, the second bullet point from above gives a stronger result than the first in the setting when $m$ is much larger than $n$.

% \begin{thm}[Informal version of \cref{thm:can't condense CG sources}]\label{thm:intro-can't condense CG sources}
% For every function $f: (\zo^n)^{\ell}\to \zo^m$, there exists an \CG[\ell, \ell] $\X$ where the good blocks have min-entropy at least $n/\ell - \log(\ell) - O(1)$ conditioned on all fixings of previous blocks and $\sminH(f(\X)) \le \frac{1}{2}\cdot m + O(1)$ where $\eps = 0.99$.
% \end{thm}

We note that these results do not contradict the condensing result from \cite{doron_almost_2023} as in the parameter regimes for which \cref{thm:intro-can't condense CG sources beyond entropy gap} works, the condenser of \cite{doron_almost_2023} does not result in an entropy increase.
% Therefore, we have demonstrated the existence of a gap between the regime in which \cite{doron_almost_2023} shows CG sources can be condensed and in which \cref{thm:intro-can't condense CG sources} shows condensing above rate $1/2$ is impossible.
This also shows a separation between \CGs and \SHELAs since \cref{thm:intro-Condensing from low entropy (g;l)-SHELA with g<=l/2} can condense from \SHELAs in this parameter regime. 

\subsubsection{Non-oblivious symbol fixing sources}
Finally, we strengthen the adversary one last time by letting the bad blocks depend arbitrarily on all the good blocks. This gives rise to \NOSFs which themselves generalize the setting of non-oblivious bit-fixing (NOBF) sources \cite{chor_bit_1985} where each block is a bit (i.e., $n=1$).

\begin{definition}[\NOSFs]
    A \emph{\NOSF[g,\ell,n,k]} $\X=\X_1,\dots,\X_\ell$ on $(\zo^n)^\ell$ is such that $g$ out of the $\ell$ blocks are independently sampled $(n,k)$-sources (i.e., ``good'') while the other $\ell-g$ bad blocks may depend arbitrarily on the good blocks.
    % after they have been sampled.
    % We will denote the set of good indices by $\mathcal{G}\subseteq[\ell]$.
    % In addition, the adversary gets to select the indices of the good and bad blocks.
\end{definition}

When $k=n$ and $n$ is clear from context, we simply call $\X$ a \uniNOSF[g,\ell]. The adversary in \NOSFs clearly has a significant amount of power; every single good block is sampled before the adversary gets to decide what to place in the bad blocks. As \NOSFs are in the setting in which the adversary is the strongest, they are also the sources for which we are most motivated to be able to extract or condense as they are the most general. We note that much of the progress on explicit constructions of two-source extractors and condensers \cite{chattopadhyay_explicit_2019,ben-aroya_two-source_2019}, a major problem in the area of randomness extraction, is based on constructing extractors and condensers for NOSF sources (in a parameter regime where it was existentially known that extraction is possible). This further motivates our exploration of   condensing from NOSF sources in a more general parameter setting.

\paragraph{Previous work}
We can trace back study of extracting from NOBF sources to the seminal work of Ben-Or and Linial in \cite{ben-or_collective_1989}.\footnote{They used the terminology ``collective coin flipping protocol'' instead of ``NOBF extractor''.} They made the connection between NOBF extractors and the influence of sets of variables on Boolean functions. Together with the work of Kahn, Kalai, and Linial in \cite{kahn_influence_1988}, in which they demonstrated lower bounds on the influence of variables on Boolean functions, these works show that it is not possible to extract from uniform $(g,\ell)$-NOBF sources when the number of bad bits is $b=\ell-g=\Omega(\ell/\log \ell)$. While no analogous result is known for \NOSFs,\footnote{Although one is conjectured in \cite{friedgut_influences_2004} that attempts to recover what was initially proposed in \cite{bourgain_influence_1992}.} the extraction impossibility result from \cite{aggarwal_how_2020} for \SHELAs also applies for \NOSFs: for any $\gamma>0$ there exists a large constant $\ell$ such that it is impossible to extract even one bit from \uniNOSFs[\gamma\ell,\ell].

To attempt to match these lower bounds on extraction, resilient functions, introduced by \cite{ben-or_collective_1985}, have yielded the current best results. The resilient function of Ajtai and Linial in \cite{ajtai_influence_1993} and its explicit versions constructed by \cite{chattopadhyay_explicit_2019, meka_explicit_2017} achieve extractors for uniform $(g,\ell)$-NOBF sources when $b=O(\ell/\log^2\ell)$, leaving a $1/\log \ell$ gap between the lower and upper bounds. 

Noting that a \uniNOSF[g,\ell,n] is a uniform $(ng,n\ell)$-NOBF source, these results imply extractors when $g>\ell(1-1/C\log^2(n\ell))$, for some large enough constant $C$. This still leaves open whether condensing is possible for most settings of parameters. 

Related to this, the work of \cite{koppartyn23multimerger} explores what they call extracting multimergers, which we may consider as extractors for \uniNOSFs. For seedless extracting multimergers, their result implies that extracting from \uniNOSFs[2,3] is impossible.
%Their main result shows that seeded extracting mergers, multimergers in the case of \uniNOSFs[1,\ell], require $\log(n)$ seed length as is the case for seeded min-entropy extractors.

\paragraph{Our results}
As \SHELAs are also \NOSFs, our condensing impossibility result in \cref{thm:intro-can't condense when g < ell/2} also applies to \NOSFs[g,\ell] when $g\leq\ell/2$. However, we are able to show an even stronger result for any setting of $g$ and $\ell$ and thus extend existing lower bounds of extraction to condensing. 

\begin{thm}
[\cref{cor:constant rate (g:l) NOSF impossible} restated]\label{thm:intro-constant rate (g:l) NOSF impossible}
For all constant $g, \ell\in \N$, there exist constant $\eps, \delta > 0$ so the following holds:
for all $a, m, n\in \N$ and all functions $f: (\zo^n)^{a\ell}\to \zo^m$, there exists a \uniNOSF[ag, a\ell] $\X$ such that $\sminH(f(\X)) \le \frac{g}{\ell}\cdot m  + \delta$.
\end{thm}

% In the setting of $g>\ell/2$, we have:
% \begin{theorem}[\cref{thm:Can't condense (g;l)-NOSF} restated]\label{thm:intro-Can't condense (g;l)-NOSF}
% For all $g, m, n, \ell\ge 3\in \N$ with $\ell/2 < g < \ell$: the following holds:
% for any function $f:(\zo^n)^{\ell}\to \zo^m$, there exist $\eps = O(1), \delta = O(1)$ and \uniNOSF[g, \ell] $\X$ such that $\sminH(f(\X))\le \frac{g}{\ell}\cdot m + \delta$.
% \end{theorem}
% And in the setting of $g\leq\ell/2$ we get a stronger impossibility result than that achieved by \cref{thm:intro-can't condense when g < ell/2} for \NOSFs[g,\ell].

% \begin{theorem}[\cref{cor:(g;l) NOSF impossible} restated]\label{cor:intro-(g;l) NOSF impossible}
% For all $g, r\in \N$ with $r < g$, the following holds:
% for all $m, \ell, g\in \N$ where $\ell\mod g = r$, there exist $\eps = \left(\frac{1}{c(g+r)}\right)^r, \delta = c\cdot (r+g)^2\log(g+r)$ such that for all functions $f:(\zo^n)^\ell\to\zo^m$, there exists a \uniNOSF[g, \ell] $\X$ so that $\sminH(f(\X)) \le \frac{g}{\ell}\cdot m + \delta$.
% \end{theorem}

By varying $a$ above, we extend our result for any $g$ and $\ell$ to any rate $g/\ell$ \uniNOSF. These results together put \NOSFs in stark contrast with adversarial CG and \SHELAs since they can both be condensed in a useful manner for simulating \BPP algorithms, while we have shown that \NOSFs cannot be condensed in such a manner.

% ==================== OLD =========================

% We also explicitly construct an output-light seeded extractor to get such a condenser for \uniSHELAs[2,3] with asymptotically the same min entropy gap as the probabilistic construction:

% \begin{thm} [Informal version of \cref{thm:(2:3) SHELA explicit condense}]
% \label{thm:intro-(2;3)-SHELA explicit condense}
% For any $\eps > 0$, we can explicitly construct a condenser $\Cond:(\zo^n)^3\to\zo^m$ with $m = \frac{n}{16}$ such that for any \uniSHELA[2,3] $\X$, we have $\sminH(\Cond(\X))\geq m - O(\log( m/ \eps))$.
% \end{thm}

% Lastly, we show that \uniSHELAs[2,3] form another class of sources for which condensing is possible but extraction is not.
% \begin{thm}[Informal version of \cref{thm:can't extract (2;3)-SHELA}]\label{thm:intro-can't extract (2;3)-SHELA}
%     For any $f:(\zo^n)^3\to\zo$ there exists a \uniSHELA[2,3] $\X$ such that $\abs{f(\X)-\U_1}>0.08$.
% \end{thm}

\dobib

\section{Proof Overview}
  We   present the main ideas and techniques for proving our main condensing impossibility results in \cref{sec:overview-impossibility} and    possibility results in \cref{sec:overview-possibility}.
\subsection{Impossibility results}\label{sec:overview-impossibility}
% All of our condensing lower bounds ultimately stem from a similar bipartite graph covering argument that we then extend to a larger range of parameters. 
In this subsection, we will go over the main techniques used in proving the condensing impossibility result for the case that $g\leq\ell/2$ in \cref{subsubsec:overview-impossibility SHELA}, the condensing impossibility result for \uniNOSFs when $g>\ell/2$ in \cref{sec:overview-impossibility of condensing from uniform NOSFs}, and the condensing impossibility result for low min-entropy CG sources in \cref{sec:overview-impossibility condensing low min entropy CG}.

\subsubsection{Impossibility of condensing from \texorpdfstring{\uniSHELAs[g,\ell]}{uniform (g, l)-oNOSF sources} for \texorpdfstring{$g\leq\ell/2$}{g <= l/2}}\label{subsubsec:overview-impossibility SHELA}

We prove that when the number of good blocks $g$ is not more than half of the total number of blocks $\ell$, then condensing beyond rate $\frac{1}{\floor{\ell/g}}$ is impossible. Formally, we will prove the following statement.
\begin{theorem}[\cref{thm:can't condense SHELA above 1/c}, restated]\label{thm:overview-can't condense when g < ell/2}
For all $\eps$, there exists a $\delta$ such that for all $g, \ell\in \N$ with $g\le \ell/2$ and for all $f:(\zo^n)^\ell\to\zo^m$, there exists a \uniSHELA[g, \ell] $\X$ such that $\sminH(f(\X))\leq \frac{1}{\floor{\ell/g}}\cdot m + \delta$.
\end{theorem}

The steps we take to achieve the result of \cref{thm:overview-can't condense when g < ell/2} are, broadly, as follows:
\begin{enumerate}
    % \item\label{step:get lines}
    \item\label{step:shade in triangles}
    We first reduce proving the theorem to only proving it for the special case of $g = 1$.
    We show that if it is possible to condense $\uniSHELAs[g, \ell]$ to entropy-rate more than $\frac{1}{\floor{\ell/g}}$, then it is possible to condense $\uniSHELAs[1, \ell^\pr]$ to rate beyond $\frac{1}{\floor{\ell^\pr/1}} = \frac{1}{\ell^\pr}$ where $\ell^{\pr} = \floor{\ell/g}$.
    We do this by transforming any $\uniSHELA[1, \ell^{\pr}]$ to a $\uniSHELA[g, \ell]$.
    
    \item\label{step:get points}
    We prove the theorem for the special case of $g = 1$ and arbitrary $\ell$.
    We do this by using an ``induct or win" argument.
    We show either condensing from \uniSHELAs[1, \ell] is impossible (win) or we reduce to the case of condensing from \uniSHELAs[1, \ell-1] (induct).
    Either we will win at some point in our reduction or we will reach the base case of $g = \ell = 1$ where the claim trivially holds.
    Let $f$ be a candidate condenser and take cases on whether there exists a fixing of the first block in $f$ such that the partial function obtained by fixing $f$ to that values will have small support. If such a fixing exists, then we reduce the problem to condensing from \uniSHELAs[1, \ell-1]. If not, then we directly construct a \uniSHELA[1, \ell] where $f$ fails to condense from by reducing to a graph problem.
    
    \item\label{step:dominating set lemma}
    The graph problem we reduce to in the ``win'' case is the following:
    Let $G = (U, V)$ be a bipartite graph with $U = [N], V = [M]$ and such that $\deg(u) \ge c_0M^{\delta}$ for all $u\in U$ where $\delta > 0$ is some constant. Then, show there exists $D\subset V$ such that $\abs{\Nbr(D)} \ge c_1 N$ and $\abs{D} \le c_2\cdot M^{1 - \delta}$ where $c_0, c_1, c_2$ are some universal constants.
\end{enumerate}

We expand on these three steps and prove them.

\paragraph{Step 1}
In this step, we transform any \uniSHELA[1, \ell^{\pr}] $\X$ to a \uniSHELA[g, \ell] $\Y$ where $\ell^\pr = \floor{\ell/g}$.
Divide $\ell$ by $\ell^\pr$ so that $\ell = a\ell^\pr + r$ where $0\le r < \ell^\pr$. We compute that $a \ge g$.
We split the blocks of $\X$ as evenly as possible: split up the first $r$ blocks of $\X$ into $a + 1$ blocks and the remaining $\ell^\pr - r$ blocks into $a$ blocks. These $a\ell^\pr + r = \ell$ blocks that we obtained by splitting $\X$ will form $\Y$. If a block in $\X$ is uniform, then all the split up blocks will also be uniform. Similarly, if a block in $\X$ is bad and only depended on blocks appearing before it, so will all the blocks formed after splitting it. Also, as at least one block in $\X$ is good, $\Y$ must have at least $a \ge g$ good blocks in it. Hence, $\Y$ is indeed a \uniSHELA[g, \ell].

\paragraph{Step 2} 
In this step, we execute our induct or win argument.
Fix a candidate condenser function $f: (\zo^n)^{\ell}\to \zo^m$.
We proceed by contradiction and assume $f$ can condense from \uniSHELAs[1, \ell] beyond rate $1/\ell$.
We either directly construct a \uniSHELA[1, \ell] $\X$ where $f$ will fail to condense from or we show how using $f$, we can obtain a condenser for \uniSHELAs[1, \ell-1], which is a contradiction.

\begin{casesenum}
\item
There exists a fixing of the first block $x_1$ such that $\abs{f(x_1, y_{1}, \dots, y_{\ell-1}) | (y_1, \dots, y_{\ell-1})\in \zo^{n(\ell-1)}} \le 2^{m(1 - 1/\ell)}$. Then, by appropriately relabeling outputs, we can define $h:(\zo^n)^{\ell-1}\to \zo^{m(1 - 1/\ell)}$ as $h(y_1, \dots, y_{\ell-1}) = f(x_1, y_1, \dots, y_{\ell-1})$. 
We now show that $h$ will be a condenser for \uniSHELAs[1, \ell-1].
Let $\Y$ be arbitrary \uniSHELA[1, \ell-1]. We transform $\Y$ into a $\uniSHELA[1, \ell] \Y^\pr$ by letting the first block of $\Y^\pr$ be fixed to $x_1$ and the remaining $\ell-1$ blocks behave as $\Y$.
By assumption, $f$ can condense $\Y^\pr$ so that output entropy is more than $\frac{1}{\ell}\cdot m$.
However this implies $h$ can condense $\Y$ to have entropy more than $\frac{1}{\ell}\cdot m = \frac{1}{\ell-1}\cdot m(1 - 1/\ell)$. As $h$ outputs $m(1 - 1/\ell)$ bits, this is a contradiction.

\item
For every fixing of the first block $x_1: \abs{f(x_1, y_{1}, \dots, y_{\ell-1})} > 2^{m(1 - 1/\ell)}$.
To show $f$ fails to condense from $\X$, it suffices to show that with constant probability, $f(\X)$ will lie in a small set $D\subset \zo^m$ where $\abs{D} = O\left(2^{m(1/\ell)}\right)$ (see \cref{lem:TV-dist lower bound} for a formal version of this).
Consider the bipartite graph $H =(U=(\zo^n), V=\zo^m)$ where edge $(u,v)$ is included if there exist $y_1, \dots, y_{\ell-1}$ such that $f(x_1, y_1, \dots, y_{\ell-1}) = v$.
By assumption, for all $u\in U: \deg(u) > 2^{m(1 - 1/\ell)}$.
Our graph theoretic dominating set lemma from \hyperref[step:dominating set lemma]{Item 3.} guarantees that there exists $D\subset \zo^m$ such that $\abs{D} \le c_0 2^{m(1/\ell)}$ and $\abs{\Nbr(D)} \ge c_1 2^{n}$ where $c_0, c_1$ are universal constants.
Now, let $\X$ be \uniSHELA[1, \ell] where the first block is uniform and the remaining $\ell-1$ blocks are adversarial where the value of those $\ell-1$ blocks (depending on the value of the first block) is set so that $f$ outputs an element from $D$ if possible.
By the construction of the bipartite graph and the construction of $\X$, with probability $c_1$, $f(\X)$ will output an element in $D$. Hence, as $f$ outputs an element from a small set, $D$, with high probability, it fails to condense from $\X$.
\end{casesenum}

\paragraph{Step 3} 
We prove the dominating set lemma for bipartite graph in this step to conclude the proof of the ``win'' argument.
We construct $D$ by repeatedly adding the vertex from $V$ that has the highest degree, removing vertices incident to that vertex, and stopping until at least $c_1N$ many vertices from $U$ are incident to some vertex from $D$.
Whenever we attempt to add a vertex to $D$, the graph will have at least $(1-c_1)N$ many vertices and so at least $(1-c_1)N\cdot c_0\cdot M^{1-\delta}$ many edges.
This implies there will always be a vertex $v\in V$ such that $\deg(v) \ge c_0(1-c_1)\cdot \frac{N}{M^{\delta}}$.
This is true at each stage and we repeat this until at least $c_1 N$ many vertices are covered.
Hence, $\abs{D} \le c_2\cdot M^{1-\delta}$ for some universal constant $c_2$ as desired.

\subsubsection{Impossibility of condensing from \texorpdfstring{\uniNOSFs}{uniform NOSF sources}}\label{sec:overview-impossibility of condensing from uniform NOSFs}

We prove much stronger condensing impossibility result for \uniNOSFs: we prove that no non-trivial condensing is possible. We are able to do so since the bad blocks have no restrictions and can arbitrarily depend on any good block. Formally, we show the following:

\begin{theorem}
[\cref{cor:constant rate (g:l) NOSF impossible} restated]
\label{thm:overview-constant rate (g:l) NOSF impossible} 
For all fixed $g, \ell\in \N$, there exist fixed $\eps, \delta > 0$ so that the following holds:
for all $a, m, n\in \N$ and all functions $f: (\zo^n)^{a\ell}\to \zo^m$, there exists a \uniNOSF[ag, a\ell] $\X$ such that $\sminH(f(\X)) \le \frac{g}{\ell}\cdot m  + \delta$.
\end{theorem}

We prove \cref{thm:overview-constant rate (g:l) NOSF impossible} using the following strategy:
\begin{enumerate}
    \item\label{step:reduce to g more than half case}
    We reduce the general case to the special case of $a = 1$ and $g > \ell / 2$.
    
    \item
    Our high level strategy for this step is same as in \cref{step:get points} from \cref{subsubsec:overview-impossibility SHELA}.
    We perform an ``induct or win'' argument to show it is impossible to condense from \uniNOSFs[g, \ell] where $g > \ell/2$ beyond rate $g/\ell$.
    As earlier, we show either condensing from \uniNOSFs[g, \ell] is impossible (win) or we reduce to the case of condensing from \uniNOSFs[g, \ell-1] (induct).
    So, we recursively apply this argument and either win at some point or reach a base case of $g = \ell$ where the claim trivially holds.
    Let $f$ be a candidate condenser and take cases on whether there exists a block position $p$ such that for constant fraction of fixings of all other blocks, the partial function obtained by fixing $f$ to those values will have large support.
    If this holds, then we use the almost-dominating set argument from \cref{step:dominating set lemma} (from \cref{subsubsec:overview-impossibility SHELA}) to reduce to the case of condensing from \uniNOSFs[g, \ell-1].
    If such a position $p$ with these fixings do not exist, then we directly construct a \uniSHELA[g, \ell] where $f$ fails to condense from by reducing to a hypergraph problem.
    
    \item
    The hypergraph problem we reduce to in the ``win'' case is the following:
    Let $H = (V_1, \dots, V_t, E)$ be a $t$-uniform $t$-partite hypergraph with $V_1 = \dots = V_t = [N], \abs{E} = c_0N^t$.
    Let the edges of $H$ be colored in $M$ colors in a `locally light' way: such that for every position $p\in [T]$, and every $(t-1)$ tuples: $(v_1, \dots, v_{p-1}, v_{p+1}, \dots, v_t) \in [N]^{t-1}$, the number of distinct colored edges as entries in position $p$ vary is $\le c_1M^\delta$.
    Formally, $\abs{\edgecolor(v_1, \dots, v_{p-1}, y, v_{p+1}, \dots, v_t): y\in [N]} \le c_1M^{\delta}$.
    Then, there exists $D\subseteq [M]$ such that $\abs{D}\le c_2\cdot M^{t\delta}$ and at least $c_3 N^t$ edges in $H$ are colored in one of the colors from $D$.
    Here, $c_0, c_1, c_2, c_3$ are some constants.
\end{enumerate}

We expand on these three steps and prove them.

\paragraph{Step 1}
We show how to reduce to the case of $a = 1$. We do this using the same argument as in \cref{step:shade in triangles} from \cref{subsubsec:overview-impossibility SHELA}: we transform \uniNOSFs[g, \ell] into \uniNOSFs[ag, a\ell] by splitting up blocks; this way, a condenser for \uniNOSFs[ag, a\ell] will also condense from \uniNOSFs[g, \ell].

We next carefully examine the argument made in \cref{step:get points} and see that the induct or win argument made there can be generalized to show the following: either condensing from \uniNOSF[g, \ell] is impossible or we reduce to the case of condensing from \uniNOSF[g, \ell-g]. Applying this recursively to arbitrary $g, \ell$, we either win and show impossibility at some step or we end up reducing to showing impossibility for condensing from \uniNOSFs[g, \ell] where $g > \ell/2$.

Combining these two steps, we reduce the general case to the special case of $a = 1$ and $g > \ell/2$.

\paragraph{Step 2} 
In this step, we execute our induct or win argument.
We fix a candidate condenser function $f: (\zo^n)^{\ell}\to \zo^m$.
Proceed by contradiction and assume $f$ can condense from \uniNOSFs[g, \ell] beyond rate $g/\ell$.
We either directly construct a \uniNOSF[g, \ell] $\X$ where $f$ will fail to condense from or we show how using $f$, we can obtain a condenser for \uniNOSFs[g, \ell-1], which is a contradiction.
For $p\in [\ell]$, let $S_p$ be the set of $\ell-1$ tuples  $(x_1, \dots, x_{p-1}, x_{p+1}, \dots, x_{\ell})$ such that
\[
    \abs{\{f(x_1, \dots, x_{p-1}, y, x_{p+1}, \dots, x_{\ell}): y\in \zo^n \}} \ge c_0 2^{m/\ell}
\]

\begin{casesenum}
\item
There exists $p\in [\ell]$ such that $\abs{S_p} \ge c_1 2^{n(\ell-1)}$ where $c_1 > 0$ is a small constant.
Without loss of generality let $p = 1$.
Construct a bipartite graph $G = (U, V)$ where $U = S_1, V = \zo^m$ and edge $(u, v)$ if there exists a fixing $y$ of block $p$ such that $f(u, y) = v$.
Then, we see that $G$ satisfies the requirement for \cref{step:dominating set lemma} and hence, there exists $D\subset \zo^m$ such that $\abs{D}\le 2^{m(1 - 1/\ell)}$ which neighbors at least $c_3 2^{n(\ell-1)}$ vertices from $U$.
For the sake of presentation, assume $c_1 = c_3 = 1$. In the full proof, $c_1, c_3 > 0$ are small constants and we need to induct using a stronger inductive hypothesis. 
Now, define $h: \zo^{n(\ell-1)}\to \zo^{m(1 - 1/\ell)}$ as $h(y_1, \dots, y_{\ell-1}) = f(x_1, y_1, \dots, y_{\ell-1})$ where $x_1$ is such that $f(x_1, y_1, \dots, y_{\ell-1})\in D$ (as $c_1 = c_3 = 1$, such $x_1$ always exists). The output domain of $h$ can be made $\zo^{m(1 - 1/\ell)}$ instead of $D$ by appropriately relabeling the output.
We then show, similar to proof of case 1 of \cref{step:get points}, $h$ will be a condenser for \uniNOSFs[g, \ell-1] and get a contradiction.
% Let $\Y$ be arbitrary \uniNOSF[g, \ell-1].
% We transform $\Y$ into a \uniNOSF[g, \ell] $\Y^\pr$ by letting blocks $2, \dots, \ell$ of $\Y^\pr$ equal $\Y$ and the first block be such that, when other blocks have value $(y_1, \dots, y_{\ell-1})$, the first block outputs $x_1$ where $x_1$ is such that $f(x_1, y_1, \dots, y_{\ell-1})\in D$.
% By assumption, $f$ can condense $\Y^\pr$ so that output entropy is more than $\frac{g}{\ell}\cdot m$.
% However this implies $h$ can condense $\Y$ to have entropy more than $\frac{g}{\ell}\cdot m = \frac{g}{\ell-1}\cdot m(1 - 1/\ell)$. As $h$ outputs $m(1 - 1/\ell)$ bits, this is a contradiction.

\item
For all $p\in [\ell]$, $\abs{S_p} \le c_1 2^{n(\ell-1)}$.
We say $x = (x_1, \dots, x_{\ell})\in \zo^{n\ell}$ is bad if for some $p\in [\ell]$, removing position $p$ from $x$ makes it an element of $S_p$.
Let $B$ be set of such bad strings.
Then, $\abs{B}\le c_1\ell\cdot 2^{n\ell}$.
Let $H = (V_1, \dots, V_{\ell})$ where $V_i = \zo^n$ be $\ell$-uniform $\ell$-partite hypergraph where edge $v = (v_1, \dots, v_{\ell})$ is in $H$ if $v\not\in B$.
Then, $H$ has at least $(1 - c_1\ell)2^{n\ell}$ edges.
By an averaging argument, there exists $x = (x_1, \dots, x_{\ell-g})\in \zo^{n(\ell-g)}$ such that the number of edges in $H$ containing that $x$ is at least $(1 - c_1\ell)2^{ng}$.
Consider \uniSHELA[g, \ell] $\Y$ where the first $\ell-g$ blocks always output $x$ and the remaining $g$ blocks are uniform.
To show $f$ fails to condense from $\X$, it suffices to show: constant probability, $f(\X)$ will lie in a small set $D\subset \zo^m$ where $\abs{D} = O\left(2^{m(g/\ell)}\right)$ (see \cref{lem:TV-dist lower bound} for a formal version of this). 

Let $H^\pr = (U_1, \dots, U_g)$ where $U_i = \zo^n$ be $g$-uniform $g$-partite hypergraph where edge $u = (u_1, \dots, u_{g})$ is in $H^\pr$ if $(x_1, \dots, x_{\ell-g}, u_1, \dots, u_g)$ is in $H$.
Then, $H^{\pr}$ has at least $(1 - c_1\ell)2^{ng}$ edges.
Now, color $H^{\pr}$ into $2^m$ colors by coloring edge $(u_1, \dots, u_g)$ as $f(x_1, \dots, x_{\ell-g}, u_1, \dots, u_g)$.
By definition of $S_p$ and construction of $H^{\pr}$, we see that for every $\ell-1$ tuples $u$ in $\zo^{n(\ell-1)}$, the number of distinct colors in $H^\pr$ is at most $c_0\cdot 2^{m/\ell}$.
We apply the hypergraph lemma to $H^{\pr}$ and infer that there exists $D\subset \zo^m$ such that $\abs{D}\le c_2\cdot 2^{m(g/\ell)}$ at least $c_3\cdot 2^{ng}$ edges in in $H^\pr$ are colored in one of the colors from $D$.
Hence, we found a small set $D$ such that with constant probability, $f(\X)$ lies in $D$ as desired.
\end{casesenum}

\paragraph{Step 3} 
We finally solve the hypergraph problem to conclude the proof of the ``win'' argument.
We repeatedly pick the color which covers the most edges to $D$ until the number of edges covered is at least $c_3\cdot N^t$. At the last step of the process, $H$ must have at least $(c_0 - c_3)\cdot N^t$ edges. We show that at that stage, the chosen a color will cover at least $c_4 N^t / M^{t\delta}$ edges. This implies at each step before this, the chosen color must cover at least that many edges and hence, $\abs{D} \le \frac{1}{c_4}M^{t\delta}$ as desired.

So, our goal is to show that in a $t$-uniform $t$-partite hypergraph $H = (V_1, \dots, V_t)$ having at least $c_5 N^t$ edges and colored in $M$ colors in a `locally light' manner - on fixing any $t-1$ tuple, the number of colors adjacent to it as last entry varies is at most $c_1\cdot M^{\delta}$, there exists a color $\gamma$ covering at least $\Omega(N^t / M^{t\delta})$ edges. We induct on $t$ and show this. We sketch the idea below for bipartite graphs.

For every $v_2\in V_2$, let $C_{v_2}\subset [M]$ be the set of colors that have at most $c_6\cdot (N / M^{\delta})$ where $c_6$ is a very small constant.
We remove edge $(v_1, v_2)$ from $H$ if $(v_1, v_2)\in C_{v_2}$.
For each $v_2$, we remove at most $c_1c_6\cdot N$ edges incident to it.
Overall, we end up removing at most $c_1c_6\cdot N^2$ edges from $H$ and it still has $(c_5 - c_1c_6)N^2$ edges.
Doing this ensures that every color incident to every vertex $v_2$ in $V_2$ has at least $c_6\cdot (N / M^{\delta})$ edges incident to it.
We finally find such a popular color by doing the following:
By averaging argument, let $v_1^*\in V_1$ and $\gamma^*\in [M]$  be such that the number of edges incident to $v_1^*$ with color $\gamma$ is at least $\frac{c_5 - c_1c_6}{c_1}\cdot (N / M^{\delta})$.
Let $\Nbr_{\gamma}(v_1^*) = \{v_2\in V_2: (v_1^*, v_2) \textrm{ is colored with color } \gamma\}$.
Moreover, for every $v_2\in \Nbr_{\gamma}(v_1^*)$, the number of edges incident to them with color $\gamma$ is at least $c_6\cdot N / M^{\delta}$.
We are done as at least $c_6\cdot\frac{c_5 - c_1c_6}{c_1}\cdot N^2 / M^{2\delta} = \Omega(N^2 / M^{2\delta})$ edges in $H$ colored with color $\gamma$.

\subsubsection{Impossibility of condensing from low min-entropy \texorpdfstring{\CGs}{aCG sources}}\label{sec:overview-impossibility condensing low min entropy CG}
We provide two impossibility result for \CG[\ell, \ell], we only sketch proof for one of them as they both share many ideas. Our impossibility result \cref{thm:can't condense CG sources beyond entropy gap} is based on reduction from general $(n, k)$-sources and the fact that it is impossible to condense from such sources.

Here, we sketch a proof of the second impossibility result where we show that it is impossible to condense from non-adversarial CG sources when each block's min-entropy, conditioned on previous blocks, is roughly bounded by $n/(\ell+1)$.
\begin{theorem}[\cref{thm:can't condense CG sources} restated]\label{thm:overview-can't condense CG sources}
For all $0 < \eps < 1$ there exists a $\delta > 0$ such that the following holds:
for every function $f: (\zo^n)^{\ell}\to \zo^m$, there exists a \CG[\ell, \ell] $\X$ where the good blocks have min-entropy at least $\frac{n - \ell\log(2\ell / \eps)}{\ell+1}$ conditioned on all fixings of previous blocks and $\sminH(f(\X)) \le \frac{1}{2}\cdot m + \delta$.
\end{theorem}
The bulk of the proof is based on a transformation from a \uniSHELA[1,2] $\X=\X_1,\X_2$ to a source $\Y=\Y_1,\dots,\Y_\ell$ that is $\varepsilon/2$-close to an \CG[\ell,\ell]. With this transformation, applying \cref{thm:overview-can't condense when g < ell/2} with $\ell=2$ and $\varepsilon/2$ to $\X$ then allows us to infer that we also cannot condense from $\Y$ with error $\varepsilon$. Thus, we focus on how to construct $\Y$ next.

Briefly, to construct $\Y$, we will take substrings of $\X_1$ and $\X_2$ to place into each block of $\Y$. From $\X_2$, we will take constant sized chunks of size $t_2=\frac{n - \ell\log(2\ell / \gamma)}{\ell+1}$ where $\gamma=\frac{\varepsilon}{2\ell}$ to place into each $\Y_i$, and from $\X_1$ we will take blocks of increasing size $i\cdot t_1-1$ to place into each $\Y_i$ where $t_1=t_2+\log(1/\gamma)$. Our proof then finishes with an inductive argument to claim that $\Y$ is indeed $\varepsilon/2$-close to an \CG[\ell,\ell] source, as required.

\subsection{Possibility results}\label{sec:overview-possibility}

In this subsection, we will present our existential construction of condensers for \SHELAs and \uniCGs. We begin by describing the construction of our condenser for \uniSHELAs[g,\ell] and \uniCG[g,\ell] in the setting of $g>\ell/2$ in \cref{sec:overview-Condensing uniform online NOSF for g>l/2}. Then we generalize this result to any setting of $g$ and $\ell$ in \cref{sec:overview-condensing from SHELAs in g<=l/2}. Finally, we deal with logarithmic min-entropy \SHELAs in \cref{sec:overview-low min entropy SHELA condensing}.

\subsubsection{Condensing from \texorpdfstring{\uniSHELAs[g,\ell]}{uniform (g,l)-oNOSF sources} for \texorpdfstring{$g>\ell/2$}{g>l/2}}\label{sec:overview-Condensing uniform online NOSF for g>l/2}

Before we dive into the actual proof, it is instructive to see why a random function fails to be a condenser for \uniSHELAs[g,\ell]. In particular, let us consider \uniSHELAs[2,3]. For a random function $f:\zo^{3n}\to \zo^m$, with high probability over $x_1, x_2\in \zo^n$, we have $\abs{f(x_1, x_2, \cdot)} = 2^m$. Hence, if the adversary is in position 3, then it can depend on $x_1$ and $x_2$ to ensure the output of $f$ always lies in a small set. To overcome this, one can consider restricting the number of choices adversary has when it is in position $3$. This intuition indeed works out and we give further details.

\begin{theorem}[\cref{thm:can-condense-from-uniform-shela-above-rate-half} restated]\label{thm:overview-can-condense-from-uniform-shela-above-rate-half}
For all $g, \ell$ such that $g > \ell / 2$ and $\eps$, there exists a condenser $\Cond:(\zo^n)^{\ell}\to\zo^m$ such that for any \uniSHELA[g, \ell] $\X$, $\minH^{\eps}(\Cond(\X)) \ge m - (5^{\ell-g}-3)\log(gn/\eps)$ where $m = n - 2(5^{\ell-g}-1)\log(gn)$.
\end{theorem}
Our construction relies on a $(k_1,k_2,\varepsilon)$-two-source extractor $\Ext:\zo^{n_1}\times\zo^{n_2}\to\zo^m$ with a property that we term \emph{output-lightness}, the definition and importance of which we will see soon, and a clever choice of a partition and prefixes of our input source $\X$. We do not currently know of a construction of a two-source extractor with our desired min-entropy and error parameters that is also output-light, so our construction is currently based on an existential output-light two-source extractor that we show in \cref{cor:great-output-light-two-source-extractors-exist}. In particular, if we write $\X=\X_1,\dots,\X_\ell$ and we take $\Y_i$ to be the prefix of $\X_i$ containing the first $5^{\ell-i}\cdot 4\log(gn/\eps)$ bits, then we define our two inputs to $\Ext$ as $\ZZ_1=\X_1,\dots,\X_g$ and $\ZZ_2=\Y_{g+1},\dots,\Y_\ell$. Thus, our condenser becomes $\Cond(\X):=\Ext(\ZZ_1,\ZZ_2)$.

There are only two cases we must consider: when the adversary places at least one good block in $\X_{g+1},\dots,\X_\ell$ and when all of $\X_{g+1},\dots,\X_\ell$ are adversarial (so $\X_1,\dots,\X_g$ is uniform). In the latter case, we have that $\ZZ_1$ is just the uniform distribution on $gn$ bits and $\ZZ_2$ is fully controlled by the adversary. For $\Ext(\ZZ_1,\ZZ_2)$ to condense then, we would require that no element $h\in\zo^m$ have too much weight placed on it by the adversary. Recalling that $\ZZ_1$ is uniform in this case, this statement is equivalent to asking that the sum over all settings $z_1$ of $\ZZ_1$ of the number of $z_2$ such that $\Ext(z_1,z_2)=h$ is not larger than $R=2^{n_1+n_2-m+O(1)}$. This is precisely our definition of $R$-output-lightness (see \cref{def:output-light} for a formal definition). With this property, we use \cref{claim:small smooth entropy implies heavy set} to get that $\sminH(\Cond(\X))\geq n_1=\log(R/\varepsilon)$.

In the case that there is at least one good block among $\X_{g+1},\dots,\X_\ell$, then we notice that there must be one good block among $\X_1,\dots,\X_g$ because $g>\ell/2$, so $\minH(\ZZ_1)\geq n$. Without loss of generality, we also assume that we only have one good block $\X_j$ for $j\in\{g+1,\dots,\ell\}$. Consequently, we can define $\A=\Y_{g+1},\dots,\Y_{j-1}$ and $\B=\Y_{j+1},\dots\Y_\ell$ so that $\ZZ_2=\A\circ\Y_j\circ \B$ where the adversary controls both $\A$ and $\B$ but not $\Y_j$.  Since $\X$ is a \SHELA, $\Y_j$ remains uniform regardless of any fixing of $\A$, so $\minH(\Y_j\mid A)=\minH(\Y_j)=5^{\ell-j}\cdot 4\log(gn/\eps)$. In addition, since we chose $\A$ to be logarithmically small in $n$, the min-entropy chain rule (\cref{lem:min-entropy-chain-rule}) gives us that, with high probability over the fixings of $\A$, the min-entropy of $\ZZ_1$ is not decreased by too much more than the length of $\A$ which is at most $n_2$. In particular, for any of these good fixings $a\in\Supp(\A)$, we chose $k_1$ to be such that $\minH(\ZZ_1\mid \A=a)\geq k_1$. Then if we temporarily make the assumption that $\B$ is uniform, we have that $\minH(\ZZ_2\mid \A=a)=\minH(a,\Y_j,\B\mid \A=a)=\sum_{i=j}^\ell 5^{\ell-j}\cdot 4\log(gn/\eps)=(5^{\ell-i+1}-1)\log(gn/\eps)$. Since we can choose $k_2$ to be smaller than $\minH(\ZZ_2\mid \A=a)$, we get that $\Ext(\ZZ)$ is $\varepsilon$-close to $\U_m$. Of course, $\B$ may be adversarially chosen. To take this into account, we use \cref{lem:control-few-bits-can-still-condense}, which says that if only a few bits of a source are adversarially controlled then we can still condense, to reduce our output entropy by the length of $\B$ and multiplicatively increase our error by $2^{\operatorname{length}(\B)}$. Finally, because we constructed $\B$ to have $\sum_{i=j+1}^\ell 5^{\ell-j}\cdot 4\log(gn/\eps)=(5^{\ell-i}-1)\log(gn/\eps)$ bits, it is still short enough in comparison $\Y_j$ to allow us to condense with our desired error.

\subsubsection{Condensing from \texorpdfstring{\uniSHELAs[g,\ell] for any $g$ and $\ell$}{(g,l)-oNOSF sources for any g and l}}\label{sec:overview-condensing from SHELAs in g<=l/2}

While we can condense from \uniSHELAs[g,\ell] for $g> \ell/2$ as we saw above (\cref{thm:overview-can-condense-from-uniform-shela-above-rate-half}), we know from \cref{thm:overview-can't condense when g < ell/2} that when $g\leq\ell/2$ we cannot condense from \uniSHELAs[g,\ell] above rate $\frac{1}{\floor{\ell/g}}$. Here, we sketch the argument for a matching bound showing that this is indeed tight by generalizing \cref{thm:overview-can-condense-from-uniform-shela-above-rate-half}.

\begin{theorem}[\cref{thm:Condensing from uniform (g;l)-SHELA with g<=l/2} restated]\label{thm:overview-Condensing from uniform (g;l)-SHELA with g<=l/2}
For any $g, \ell, \eps$ such that $\floor{\ell/g}=r$ and $r < \ell/g$, there exists a condenser $\Cond:(\zo^n)^\ell\to\zo^m$ such that for any \uniSHELA[g,\ell] $\X$ we have $\sminH(\Cond(\X))\geq\frac{1}{r}\cdot m - 2(5^{\ell-g}-1)\log(gn/\eps)$ where $m = r(n-2(5^{\ell-g}-1)\log(gn))$.
\end{theorem}

Satisfyingly, we need no new tools to construct this condenser. Instead, we use $r$ instances of the condenser from \cref{thm:overview-can-condense-from-uniform-shela-above-rate-half}. We will prove this inductively on $r$, so let us consider the base case of $r=1$. Notice that $r=1$ implies that $g>\ell/2$, so we are exactly in a position to use the condenser $\Cond_1:(\zo^n)^\ell\to\zo^{m_1}$ from \cref{thm:overview-can-condense-from-uniform-shela-above-rate-half} without modification. Thus, we define our output block as $\mathbf{O}=\mathbf{O}_1=\Cond_1(\X)$.

To generalize to larger values of $r$, we perform induction on $r$ and take the inductive hypothesis of $r-1$ to be true. We consider two cases. Beginning with the case that all of $\X_1,\dots,\X_g$ are bad, we notice that $\X_{g+1},\dots,\X_{\ell}$ is a \uniSHELA[g,\ell-g] with $\floor{\frac{\ell-g}{g}}=r-1$ and $\frac{\ell-g}{g}\neq r-1$. Our inductive hypothesis then gives us $r-1$ output blocks $\mathbf{O}_2,\dots,\mathbf{O}_r$ on $(\zo^{m_r})^{r-1}$ where at least one is condensed. On the other hand, consider when at least one of $\X_1,\dots,\X_g$ is good and take $\Cond_1$ to be an instance of the condenser from \cref{thm:overview-can-condense-from-uniform-shela-above-rate-half} for $\X$, and define $\mathbf{O}_1$ to be $\Cond_1(\X)$ truncated to its first $m_r$ bits. Observe that if $\Cond_1(\X)$ succeeds and condenses $\X$ to some min-entropy $k$ source, then $\minH(\mathbf{O}_1)\geq k-(m_1-m_r)$, so we only lose as many bits of entropy in $\mathbf{O}_1$ as we truncate from $\Cond_1(\X)$, which we show in \cref{lem:removing d bits from source removes d bits of entropy}, and $m_1-m_r$ is still constant in $g$ and $\ell$. Then in this case we again get that $\mathbf{O}_1$ must be properly condensed by $2\Ext_1$ being output-light when all of $\X_1,\dots,\X_g$ are good or by $2\Ext_1$ being a two-source extractor when at least one of $\X_1,\dots,\X_g$ is good. Thus, if we let our output be $\mathbf{O}=\mathbf{O}_1,\dots,\mathbf{O}_r$, then at least one block is always condensed in any case.

\subsubsection{Condensing from logarithmic min-entropy \texorpdfstring{\SHELAs[g,\ell]}{(g,l)-oNOSF sources}}\label{sec:overview-low min entropy SHELA condensing}
We can extend \cref{thm:overview-can-condense-from-uniform-shela-above-rate-half} and \cref{thm:overview-Condensing from uniform (g;l)-SHELA with g<=l/2} to logarithmic min-entropy \SHELA by converting a logarithmic min-entropy \SHELA into a \uniSHELA via the following theorem.
\begin{theorem}[\cref{thm:existential low min entropy shela to uniform shela} restated]\label{thm:overview-existential low min entropy shela to uniform shela}
    For any $g, \ell, \eps$, there exists a function $f:(\zo^n)^\ell\to(\zo^m)^{\ell-1}$ with $m=\frac{k}{8\ell}$ such that for any \SHELA[g, \ell, k] $\X$ with $k\geq 2\log(n/\eps)$ there exists a \uniSHELA[g-1,\ell-1] $\Y$ such that $\abs{f(\X)-\Y}\leq \eps$.
\end{theorem}
Thus, if we take a \SHELA[g,\ell, n, k] $\X$ such that $g>\ell/2+1$ so $g-1>(\ell-1)/2$, we can simply apply $f$ from \cref{thm:overview-existential low min entropy shela to uniform shela} to $\X$ and then pass the result to our condenser from \cref{thm:overview-Condensing from uniform (g;l)-SHELA with g<=l/2} to condense from logarithmic min-entropy \SHELA. 
\begin{theorem}[\cref{thm:Condensing from low entropy (g;l)-SHELA with g<=l/2} restated]\label{thm:overview-Condensing from low entropy (g;l)-SHELA with g<=l/2}
    For all $g, \ell, r\in \N$ and $\eps$ such that $\floor{\frac{\ell-1}{g-1}} = r$ and $r < \frac{\ell-1}{g-1}$, there exists a condenser $\Cond:(\zo^n)^\ell\to\zo^m$ such that for any \SHELA[g,\ell,n,k] $\X$ with $k\geq 2\log(n/\eps)$, we have that $\sminH(\X)\geq \frac{1}{r}\cdot m - 2(5^{\ell-g}-1)\log\left(\frac{(g-1)k}{8\ell\eps}\right)$ with $m=r\left(\frac{k}{8\ell}-2(5^{\ell-g}-1)\log\left(\frac{(g-1)k}{8\ell}\right)\right)$.
\end{theorem}

All that is left then is to show how we convert a low min-entropy \SHELA to a \uniSHELA in \cref{thm:overview-existential low min entropy shela to uniform shela}. Our method here is based on the somewhere extractor for low-entropy \SHELA from \cite{aggarwal_how_2020} with two important modifications. First, we use a two-source extractor instead of a seeded extractor which enables us to handle logarithmic min-entropy in the good blocks of a \SHELA instead of just linear. Second, we require that the output of our function is not just somewhere random, but instead a \uniSHELA. To achieve this, we decrease the output length of our two-source extractor (which decreases the block length of our resulting \uniSHELA) to show that the good blocks in our resulting source are independent from all adversarial blocks before them. 

The construction of $f$ from \cref{thm:overview-existential low min entropy shela to uniform shela} is quite straightforward. For every $i\in\{2,\dots,\ell\}$, we use the same existential two-source extractor from \cref{cor:great-output-light-two-source-extractors-exist} that we used in the proof of \cref{thm:overview-can-condense-from-uniform-shela-above-rate-half} to define $2\Ext_i:(\zo^n)^{i-1}\times\zo^n\to\zo^m$ where $m=\frac{k}{8\ell}$ and $k\geq 2\log(n/\eps)$ is the min-entropy requirement of each good block in our \SHELA[g,\ell,n,k] $\X=\X_1,\dots,\X_\ell$. We then define our $\ell-1$ output blocks as $\mathbf{O}_i=2\Ext_i((\X_1,\dots,\X_{i-1}),\X_i)$, so $f(\X)=\mathbf{O}_2,\dots,\mathbf{O}_\ell$. Because there are $g$ good blocks in $\X$ at indices $G_1,\dots,G_g$, we are guaranteed that $\mathbf{O}_{G_2},\dots,\mathbf{O}_{G_g}$ are the outputs of a two-source extractor with a good block in each source. The crux of our argument then is to show that $\mathbf{O}_{G_2},\dots,\mathbf{O}_{G_g}$ are close to uniform and independent of the adversarial blocks before them. This part of our argument follows that of \cite{aggarwal_how_2020} closely, so we do not expand on it here except to note that shortening the length of our output blocks from $m=O(k)$, not depending on $\ell$, in \cite{aggarwal_how_2020} to $m=k/8\ell$ is what allows us to show that good output blocks are uniform and independent of output blocks before them.

\dobib

%\subfile{sections/new_outline}

\section{Preliminaries}\label{sec:preliminaries}
We will generally denote distributions or sources in a bold font, such as $\X$, and reserve $\U_m$ to be the uniform distribution on $m$ bits. When these sources are actually a sequence of sources, we use subscripts to denote blocks of that source as $\X=\X_1,\dots,\X_\ell$. In addition, since we often consider binary strings of length $n$ and $m$, we let $N=2^n$ and $M=2^m$. Often it is convenient to consider strings as labels, in which case we use the notation $[N]=\{1,2,\dots,N\}$.

\subsection{Basic probability lemmas}
Here, we first state a few basic probability facts that will be useful to us throughout. Our first one is a direct reverse Markov style inequality.
\begin{claim}[Reverse Markov]
\label{claim:reverse-markov}
Let $\X$ be a random variable taking values in $[0, 1]$. Then, for $0\le d < \E[X]$, it holds that
\[
\Pr[\X > d] \ge \frac{\E[\X] - d}{1 - d}
\]
\end{claim}

\begin{proof}
Let $\Y = 1 - \X$. Applying
 Markov's inequality to $\Y$ gives the required bound.
 \iffalse
\[
\Pr[\Y \ge 1 - d] \le \frac{\E[\Y]}{1 - d} = \frac{1 - \E[\X]}{1 - d}
\]
So, 
\[
\Pr[\X > d] = \Pr[\Y < 1 - d] = 1 - \frac{1 - \E[\X]}{1 - d} = \frac{\E[\X] - d}{1 - d}
\]
\fi
\end{proof}
We will use the following version of the Chernoff bound:
\begin{claim}[Chernoff Bound]
\label{claim:chernoff}
Let $\X_1, \dots, \X_n$ be independent random variables taking values in $\{0, 1\}$. 
Let $\X = \sum_i \X_i$.
Let $\mu = \E[\X]$.
Then, for all $\delta \ge 0$, the following holds:
\[
    \Pr[\X \ge (1 + \delta)\mu] \le \exp(-\delta^2 \mu / (2 + \delta))
\]
\end{claim}

Several of our impossibility results rely on a simple TV distance bound.
\begin{claim}[TV distance lower bound]\label{lem:TV-dist lower bound}
Let $\X\sim \zo^n$ and $S\subset \zo^n$ be such that $\Pr_{x\sim \X}[x\in S] \ge p$. Then, for $0 < \varepsilon < p$, it holds that $\sminH(\X)\leq \log\left(\frac{|S|}{p - \varepsilon}\right)$.
\end{claim}
\begin{proof}
Let $k = \log\left(\frac{|S|}{p - \varepsilon}\right)$.
Let $\Y\sim \zo^n$ be an arbitrary distribution with $\minH(\Y) \ge k$.
By the min entropy condition, for all $s\in S$, it holds that $\Pr[\Y = s] \le 2^{-k}$.
Hence, 
\begin{align*}
\abs{\X -\Y} \ge \Pr_{x\in \X}[x\in S] - \Pr_{y\in \Y}[y\in S] = p - 2^{-k}\cdot |S| = \varepsilon
\end{align*}
\end{proof}

We will utilize the very useful min entropy chain rule in our constructions.

\begin{lemma}[Min-entropy chain rule]\label{lem:min-entropy-chain-rule}
For any random variables $\X\sim X$ and $\Y\sim Y$ and $\eps>0$,
\[
    \Pr_{y\sim\Y}[H_\infty(\X\mid\Y=y)\geq H_\infty(\X)-\log|\Supp(\Y)|-\log(1/\eps)]\geq1-\eps.
\]
\end{lemma}

% We will also need the leftover hash lemma:
% \begin{theorem}\cite{impagliazzo89leftover}
 
% \end{theorem}

Lastly, we will later utilize a consequence of upper bounds on smooth min-entropy.
\begin{claim}[Lemma 8.8 from \cite{zuckerman_linear_2007}]
\label{claim:small smooth entropy implies heavy set}
Let $\X\sim \zo^n$ be such that $\sminH(\X) < k$. Then, there exists $D\subset \Supp(\X)$ such that $|D| < 2^k$ and $\Pr[\X\in D] \ge \eps$.
\end{claim}

\subsection{Extractors}
Let $\mathbf{A}\approx_\varepsilon\mathbf{B}$ mean that $\mathbf{A}$ and $\mathbf{B}$ are $\varepsilon$ close in statistical distance. Recall the definition of a seeded extractor.
\begin{definition}\label{def:seeded_ext}
A   $(k,\eps)$-seeded extractor $\Ext:\zo^n \times  \zo^d \to\zo^m$ satisfies the following:   for every $(n,k)$-source $\X$, and every $\Y=\U_d$,
$$\Ext(\X,\Y)\approx_{\eps}\U_m.$$
 $d$ is called the \emph{seed length} of $\Ext$.     $\Ext$ is called strong if 
$$\Ext(\X,\Y), \Y \approx_{\eps}\U_m, \Y.$$
\end{definition}

A useful fact about strong seeded extractors that they work even when the seed is not fully uniform. (See for example Lemma 6.4 from \cite{chattopadhyay_nonmalleable_2020} for a proof.)
\begin{lemma}\label{lem:CGL15 strong extractor bad seed}
Let $\Ext:\zo^n\times\zo^d\to\zo^m$ be a strong $(k,\varepsilon)$-seeded extractor. Let $\X$ be a $(n,k)$-source and let $\Y$ be a $(d,d-\lambda)$-source. Then,
\begin{align*}
    \abs{\Ext(\X,\Y), \Y- \U_m, \Y}\leq 2^\lambda\varepsilon.
\end{align*}
    
\end{lemma}

We will use the following construction of seeded extractors:

\begin{theorem}[Theorem 1.5 in \cite{guruswami_unbalanced_2009}]
\label{thm:GUV extractor}
For all constant $\alpha > 0$ and all $n, k, \eps$, there exists an explicit $(k, \eps)$-seeded extractor $\sExt: \zo^n \times \zo^d \rightarrow \zo^m$ with $d = O(\log (n / \eps))$ and $m \ge (1 - \alpha)k$.
\end{theorem} 
In addition, we will use a generalization of seeded extractors, two-source extractors, that only require the second source to be independent from the first and not necessarily be uniform.
\begin{definition}\label{def:two-source extractor}
    A function $2\Ext:\zo^{n_1}\times\zo^{n_2}\to\zo^m$ is a \emph{$(k_1,k_2,\varepsilon)$-two-source extractor} if for every $(n_1,k_1)$-source $\X_1$ and $(n_2,k_2)$-source $\X_2$ where $\X_1$ and $\X_2$ are independent of each other, we have
    \[
    2\Ext(\X_1,\X_2)\approx_\varepsilon \U_m.
    \]
    It is said to be \emph{strong in the first argument} if 
    \[
    2\Ext(\X_1,\X_2),\X_1\approx_\varepsilon \U_{m}, \X_1.
    \]
\end{definition}
Similarly, one can define $2\Ext$ that is strong in the second argument. If $2\Ext$ is strong in both arguments,  we simply say that it is \emph{strong}.
We  use the fact that inner product function  is a good two source extractor:
\begin{theorem}\cite{chor_unbiased_1988, Vazirani85, impagliazzo89leftover}
\label{inner product two source}
Let $\X, \Y \sim \zo^n$ with $\minH(\X) = k_1, \minH(\Y) = k_2$. Let $m = \frac{n}{r}$ for some $r\in \N$. Let $\IP(x, y): \zo^{2n}\rightarrow \zo^m$ be the function that interprets $x, y$ as elements of $\F_{2^m}^{r}$ and outputs the $m$ bit string corresponding to $x\cdot y$. Then, $\abs{\IP(\X, \Y) - \U_m} \le 2^{(n + m - k_1 - k_2) / 2}$.  
\end{theorem}

For a proof of the above theorem, see Theorem 2.5.3 in \cite{chattopadhyay_explicit_2016}.

\subsection{Randomness sources relevant to our work}
We now formally introduce the randomness sources that are relevant to our work. We begin with NOSF sources, which have no restrictions on the adversary producing the bad blocks. 

\begin{definition}[NOSF source]\label{def:NOSF}
    A \emph{\NOSF[g,\ell,n,k] (NOSF)} $\mathbf{X}$ with symbols in $\Sigma=\{0,1\}^n$ and length $\ell$ is over $\Sigma^\ell$, written as $\mathbf{X}=\mathbf{X_1,\dots, X_\ell}$, and has the following property: There exists a set of good blocks $\mathcal{G}\subseteq[\ell]$ such that $\abs{\mathcal{G}}\geq g$ and the random variables in $\{\mathbf{X}_i\}_{i\in\mathcal{G}}$ are each independently sampled $(n,k)$-sources.     We say that a block $\mathbf{X}_i$ is \emph{good} if $i\in\mathcal{G}$ and \emph{bad} otherwise. 
\end{definition}

Note that we have no restrictions on how bad blocks may depend on the good blocks. If $k=n$, we say that $\X$ is a \emph{uniform $(g,\ell,n)$-NOSF} source. When $n$ is implicit or not relevant, we simply call $\X$ a uniform $(g,\ell)$-NOSF source.
Next, we introduce \SHELAs by restricting the NOSF adversary.

\begin{definition}[Online NOSF source]\label{def:online NOSF}
    A \emph{\SHELA[g,\ell,n,k]} $\mathbf{X}$ with symbols in $\Sigma=\{0,1\}^n$ and length $\ell$ is over $\Sigma^\ell$, written as $\mathbf{X}=\mathbf{X_1,\dots, X_\ell}$, and has the following property: There exists a set of good blocks $\mathcal{G}\subseteq[\ell]$ such that $\abs{\mathcal{G}}\geq g$ and the random variables in $\{\mathbf{X}_i\}_{i\in\mathcal{G}}$ are each independently sampled $(n,k)$-sources such that $\X_i$ is independent of $\X_1,\dots,\X_{i-1}$.     We say that a block $\mathbf{X}_i$ is \emph{good} if $i\in\mathcal{G}$ and \emph{bad} otherwise. 
\end{definition}
 \begin{remark}\label{rmk:online NOSF are NOSF}
    Online NOSF sources are also NOSF sources because the adversary in \SHELAs is strictly weaker than that of \NOSFs.
\end{remark}

These \SHELAs are special cases of the SHELA sources from \cite{aggarwal_how_2020}. We now introduce SHELA sources in their full generality.
\begin{definition}[SHELA source \cite{aggarwal_how_2020}]\label{def:SHELA formal}
A distribution $\mathbf{X}$ over $(\zo^n)^\ell$ is a \emph{$(g,\ell,n, k)$-Somewhere Honest Entropic Look Ahead (SHELA) source} if there exists a (possibly randomized) adversary $\mathcal{A}$ such that $\mathbf{X}$ is produced by sampling $g$ out of $\ell$ indices to place independently sampled $(n,k)$-sources and then placing adversarial blocks in the other $\ell-g$ indices that may depend arbitrarily on any block that comes before it. 

Concretely, there must exist random variables $1\leq \mathbf{I}_1<\mathbf{I}_2<\cdots <\mathbf{I}_g\leq\ell$ with arbitrary joint distribution, denoting the indices of the independent $(n,k)$-sources, and $g$ independent $(n,k)$-sources $\mathbf{Z}_1,\mathbf{Z}_2,\dots,\mathbf{Z}_g$ such that $\mathbf{X}$ is generated in the following manner:
\begin{enumerate}
    \item Sample $(i_1,i_2,\dots,i_g)\sim(\mathbf{I}_1,\mathbf{I}_2,\dots,\mathbf{I}_g)$.
    \item For all $j\in[g]$ set $\mathbf{B}_{i_j}=\mathbf{Z}_j$.
    \item For all $i\in[\ell]\setminus\{i_1,i_2,\dots,i_g]$, the adversary sets $\mathbf{B}_i=\mathcal{A}(\mathbf{B}_1,\dots,\mathbf{B}_{i-1},i_1,\dots,i_g\}$.
    \item Finally, let $\mathbf{X}=(\mathbf{B}_1,\dots, \mathbf{B}_\ell)$.
\end{enumerate}
    We will generally call the blocks $\mathbf{Z}_1,\dots,\mathbf{Z}_g$ the ``good'' blocks and the remaining blocks ``bad'' blocks. 
\end{definition}
Similar to NOSF sources, when $k=n$ we will simply say $\mathbf{X}$ is a \emph{$(g,\ell,n)$-uniform SHELA} source, and when $n$ is implicit we will simplify further to a uniform $(g,\ell)$-SHELA source.

% In our construction of lower and upper bounds for SHELA sources, we will often think of \emph{fixed-index SHELA} sources instead since they are easier to reason about. We define them now.
% \begin{definition}[Fixed-index SHELA source]
%     A distribution $\mathbf{X}$ over $(\zo^n)^\ell$ is a \emph{$(g,\ell,n,k)$-fixed-index SHELA (fiSHELA)} source if there exists a (possibly randomized) adversary $\mathcal{A}$ such that $\mathbf{X}$ is produced by the adversary choosing $g$ out of $\ell$ indices to place independently sampled $(n,k)$-sources and then placing adversarial blocks in the other $\ell-g$ indices that may depend arbitrarily on any block that comes before it.

%     Concretely, the adversary $\mathcal{A}$ chooses $1\leq i_1<i_2<\cdots< i_g\leq\ell$, denoting the indices of the independent $(n,k)$-sources, and $g$ independent $(n,k)$-sources $\mathbf{Z}_1,\mathbf{Z}_2,\dots,\mathbf{Z}_g$ such that $\mathbf{X}$ is generated in the following manner:
%     \begin{enumerate}
%     \item For all $j\in[g]$ set $\mathbf{B}_{i_j}=\mathbf{Z}_j$.
%     \item For all $i\in[\ell]\setminus\{i_1,i_2,\dots,i_g]$, the adversary sets $\mathbf{B}_i=\mathcal{A}(\mathbf{B}_1,\dots,\mathbf{B}_{i-1},i_1,\dots,i_g\}$.
%     \item Finally, let $\mathbf{X}=(\mathbf{B}_1,\dots, \mathbf{B}_\ell)$.
%     \end{enumerate}
% \end{definition}

While working over \SHELAs is easier than working over general SHELA sources, all of our results still apply to general SHELA sources since SHELA sources are convex combinations of \SHELAs.
\begin{proposition}
    \label{prop:convex combination SHELA}
    Every $(g,\ell,n,k)$-SHELA source $\mathbf{X}$ is a convex combination of \SHELAs[g,\ell,n,k].
\end{proposition}
\begin{proof}
    Let $\mathbf{I}=\mathbf{I}_1,\mathbf{I}_2,\dots,\mathbf{I}_g$ be the distribution of indices used in the construction of $\mathbf{X}$. For a sample $\mathcal{I}\sim \mathbf{I}$, let $\mathbf{X}_{\mathcal{I}}$ be the \SHELA[g,\ell,n,k] in the construction of which the adversary chose the good blocks to be at indices $\mathcal{I}$ and the functions describing the bad blocks to be identical to those of $\mathbf{X}$ when the sample of indices from $I$ is $\mathcal{I}$. That is, when $\mathcal{I}$ is sampled in the construction of $\mathbf{X}$ we have for all $j\in[\ell]\setminus\mathcal{I}$ that $\mathbf{X}_j=(\mathbf{X}_{\mathcal{I}})_j$ as functions.

    With this setup, we directly have that $\mathbf{X}=\E_{\mathcal{I}\sim\mathbf{I}}[\mathbf{X}_{\mathcal{I}}]$, so $\mathbf{X}$ is a convex combination of $\mathbf{X}_{\mathcal{I}}$'s. 
\end{proof}

Lastly, we define adversarial Chor-Goldreich (CG) sources, which have an adversary like that of \SHELAs that can depend arbitrarily on past blocks, but the adversary of adversarial CG sources can have some effect on future blocks, unlike that of \SHELAs.

\begin{definition}[Adversarial CG source]\label{def:adversarial CG}
    A \emph{\CG[g,\ell,n,k]} $\mathbf{X}$ with symbols in $\Sigma=\{0,1\}^n$ and length $\ell$ is over $\Sigma^\ell$, written as $\mathbf{X}=\mathbf{X_1,\dots, X_\ell}$, and has the following property: There exists a set of good blocks $\mathcal{G}\subseteq[\ell]$ such that $\abs{\mathcal{G}}\geq g$ and the random variables in $\{\mathbf{X}_i\}_{i\in\mathcal{G}}$ have the property that for all prefixes $(a_1,\dots,a_{i-1})\in(\zo^n)^{i-1}$,
    \begin{align*}
        \minH(\mathbf{X}_i\mid \mathbf{X}_1,\dots,\mathbf{X}_{i-1}=a_1,\dots,a_{i-1})\geq k.
    \end{align*}
\end{definition}
As before, if $k=n$ then we simply call $\mathbf{X}$ a \uniCG[g,\ell,n], and we omit $n$ when it is implicit.

We have introduced all of these definitions since our results resolve open questions for each. The relationship between all these definitions is necessary to clearly see how our lower and upper bounds apply. In line with this, we show an equivalence between \uniSHELAs and \uniCGs.
\begin{proposition}\label{prop:uniSHELA equivalent to uniCG}
    A source $\mathbf{X}$ is a \uniSHELA if and only if it is a \uniCG.
\end{proposition}
\begin{proof}
    Say $\mathbf{X}$ is a \uniSHELA[g,\ell,n]. Then, because bad blocks may only depend on the good blocks that have a lower index than them and all the good blocks are sampled independently, the good blocks satisfy the prefix condition in \cref{def:adversarial CG} to give us that $\mathbf{X}$ is a \uniCG[g,\ell,n].

    On the other hand, say that $\mathbf{X}$ is a \uniCG[g,\ell,n]. Then the fact that for a good block $\mathbf{X}_i$ we have for all $(a_1,\dots,a_{i-1})\in(\zo^n)^{i-1}$ that $\minH(\mathbf{X}_i\mid\mathbf{X}_1,\dots,\mathbf{X}_{i-1}=a_1,\dots,a_{i-1})=n$, so $\mathbf{X}_i$ is uniform given any prefix, means that $\mathbf{X}_i$ is independent of all blocks that come before it. In particular, this means that bad blocks may only depend on the good blocks that come before them. In addition, the good blocks being uniform clearly means that they are independent from each other. Hence, $\mathbf{X}$ is a \uniSHELA[g,\ell,n] as well.
\end{proof}

% Putting both of these propositions together yields \cref{fig:sources} which depicts how our definitions interact.
% \begin{figure}[h]
%     \centering
%     \includegraphics[width=.5\textwidth]{Figures/sources.png}
%     \caption{Here we illustrate the containments between our sources along with the equivalence between uniform fiSHELA (U-fiSHELA) and uniform almost CG (U-almost CG) sources.}
%     \label{fig:sources}
% \end{figure}
Therefore, when we prove a condensing impossibility result by constructing a \SHELA, that same result applies to \NOSFs and \CGs sources as well. On the other hand, our condensing possibility results for \uniSHELAs also apply to \uniCGs, but our results for non-uniform \SHELAs may not apply to non-uniform \CGs.
\dobib

\section{Impossibility Results}\label{sec:impossibility}

In this section, we prove condensing impossibility results for \uniNOSFs and \uniSHELAs. First, in \cref{sec:Impossibility of Condensing When g <=ell/2} we demonstrate condensing impossibility results for all three classes of sources when $g\leq\ell/2$. Then, in \cref{sec:uniform NOSF condensing impossibility} we show a condensing impossibility result for \uniNOSFs[g,\ell] for arbitrary settings of $g$ and $\ell$. Finally, we use a result from \cref{sec:Impossibility of Condensing When g <=ell/2} to show the impossibility of condensing from low min-entropy CG sources in \cref{sec:Impossibility condensing low min-entropy CG}.

\subsection{Impossibility of condensing when \texorpdfstring{$g\leq\ell/2$}{g<=l/2}}\label{sec:Impossibility of Condensing When g <=ell/2}

% \begin{remark}
%     the results in this section accomplished by constructing a uniform fishela adversary. because uniform fishela sources are contained in all of the other sources we use in this work (\cref{fig:sources}), these results also apply to uniform nosf and uniform almost cg sources as well. thus, to be succinct, we will only state our results in terms of uniform shela sources.
% \end{remark}

We will prove that for $g\leq\ell/2$, it is impossible to condense from \uniSHELAs[g, \ell] to rate more than $\frac{1}{\floor{\ell/g}}$. As we noted in \cref{rmk:online NOSF are NOSF} and \cref{prop:uniSHELA equivalent to uniCG}, these results apply to \uniNOSFs and \uniCGs as well.

\begin{theorem}\label{thm:can't condense SHELA above 1/c}
For all $\eps > 0$, there exists a $\delta > 0$ such that for all $g, \ell\in \N$ with $g\le \ell/2$ and for all $f:(\zo^n)^\ell\to\zo^m$, there exists a \uniSHELA[g, \ell] $\X$ so that $\sminH(f(\X))\leq \frac{1}{\floor{\ell/g}}\cdot m + \delta$.
\end{theorem}

This implies that for the special case when $g$ divides $\ell$, any non-trivial condensing is impossible.
\begin{corollary}\label{cor:can't condense from SHELA where g | ell}
For all $\varepsilon>0, g, \ell\in \N$ with $g\mid\ell$, there exists a $\delta > 0$ such that: for all functions $f:(\zo^n)^\ell\to\zo^m$, there exists a \uniSHELA[g, \ell] $\X$ such that $\sminH(f(\mathbf{X})) < \frac{g}{\ell}\cdot m + \delta$.
\end{corollary}

\begin{proof}
Follows immediately from \cref{thm:can't condense SHELA above 1/c}.
\end{proof}

The proof of \cref{thm:can't condense SHELA above 1/c} involves two ingredients. First, we show that for the special case of $g = 1$, condensing above rate $\frac{1}{\ell}$ is impossible for \uniSHELAs[1, \ell]. Second, we extend these results to \uniSHELAs[g, \ell] with $g\leq\ell/2$ by showing that if it is impossible to condense from \uniSHELAs[1, \ell'], then it is impossible to condense above rate $\frac{1}{\ell'}$ from \uniSHELAs[g, \ell] when $\frac{g}{\ell}\leq\frac{1}{\ell'}$.

Formally, these two lemmas are as follows:

\begin{lemma}
\label{lem:(1:l) impossible}
For all $\eps > 0$, there exists a $\delta > 0$ such that for all functions $f:(\zo^n)^\ell\to\zo^m$, there exists a \uniSHELA[1, \ell] $\X$ so that $\sminH(f(\X)) < \frac{1}{\ell}\cdot m + \delta$.
\end{lemma}

\begin{lemma}\label{lem:can't condense SHELA above any upper limit 1/l'}
Let $g, \ell, \ell^\pr, n^\pr, n, m\in \N$ be such that $\ell'\leq\ell, \frac{g}{\ell}\leq\frac{1}{\ell'}, \ceil{\ell/\ell'}n < n^\pr$. 
Let $0 < \eps < 1, \delta > 0$ be such that:
for any function $f:(\zo^{n^\pr})^{\ell'}\to\zo^{m}$, there exists a \uniSHELA[1, \ell'] $\Y$ so that $\sminH(f(\Y)) < \frac{1}{\ell'}\cdot m+\delta$.
Then, for any function $h:(\zo^n)^{\ell}\to\zo^{m}$, there exists a \uniSHELA[g, \ell] $\X$ such that $\sminH(h(\X))\leq\frac{1}{\ell'}\cdot m + \delta$.
\end{lemma}

Our main theorem follows by combining these two lemmas.

\begin{proof}[\textbf{Proof of \cref{thm:can't condense SHELA above 1/c}}]
    Divide $\ell$ by $g$ so that $\ell=c\cdot g+r$ where $c \ge 1,r < g \in\N$.
    We can derive our desired impossibility result by applying \cref{lem:can't condense SHELA above any upper limit 1/l'} to the result of \cref{lem:(1:l) impossible} \uniSHELAs[1, c].
\end{proof}

We defer the proof of \cref{lem:can't condense SHELA above any upper limit 1/l'} until \cref{subsec:impossibility-deferred-proofs}. In the next subsubsection, we will focus on proving \cref{lem:(1:l) impossible}.

\subsubsection{Proving main theorem for the case of \texorpdfstring{$g = 1$}{g=1}}

We prove this lemma by showing that if one cannot condense from \uniSHELAs[g, \ell], then one cannot condense from \uniSHELAs[g, \ell+g].

\begin{lemma}\label{lem:(g:l) implies (g:l+g)}
Let $c_0, c_1, \eps, \delta\in \R$ and $g, n, \ell\in \N$ be such that $g\le \ell, 0 < c_0 < 1, \eps < c_1 < 1$. 
Assume that for all $A\in \N$ and function $f: (\zo^{n})^{\ell} \rightarrow [A]$, there exists a \uniSHELA[g, \ell] (\uniNOSF[g, \ell], respectively) $\X$ such that $\sminH(f(\X))\le \frac{g}{\ell}\cdot \log(A) + \delta$.
Then, for all $M\in \N$ and every function $h: (\zo^n)^{\ell+g} \rightarrow [M]$, there exists a \uniSHELA[g, \ell+g] (\uniNOSF[g, \ell], respectively) $\Y$ such that $\sminH(h(\Y)) \le \frac{g}{\ell+g}\cdot m + \delta^\pr$ where $\delta^\pr = \max\left(\log\left(\frac{c_1}{(1 - c_1)c_0(c_1 - \varepsilon)}\right), \delta + \frac{\log(c_0)g}{\ell}\right)$ and $m = \log(M)$.
\end{lemma}

We remark that \cref{lem:(g:l) implies (g:l+g)} paves the way for an inductive argument and we instantiate it to prove \cref{lem:(1:l) impossible} as follows:

\begin{proof}[Proof of \cref{lem:(1:l) impossible}]
We inductively apply \cref{lem:(g:l) implies (g:l+g)} with $g = 1$ and arbitrary $\ell$ to prove the claim.
Notice that for all distributions $\X$ on $\zo^m$, $\sminH(\X)\le m$.
For the base case of $g = 1, \ell=1$: for any function $f$ and \uniSHELA[1, 1] $\W$, it must be that $\sminH(f(\W))\leq m$.
Now, inductively apply \cref{lem:(g:l) implies (g:l+g)} by setting $c_0 = 1, c_1 = \frac{1 + \eps}{2}, \delta = \log\left(\frac{2(1+\eps)}{(1-\eps)^2}\right)$ to infer the claim for \uniNOSFs[1, \ell].
\end{proof}

\subsubsection{Recursive impossibility lemma}

To prove \cref{lem:(g:l) implies (g:l+g)}, we find a dominating set in dense bipartite graphs with left degree lower bound. We will use it to construct a \uniSHELA that will serve as a counterexample for a candidate condenser.

\begin{lemma}[Small Dominating Set in Bipartite Graph]
\label{lem:greedy-covering}
Let $c_0 > 0, 0 < c_1 < 1, \delta > 0\in \R, N, M\in \N$ be arbitrary.
Let $G = (U, V, E)$ be a bipartite graph with $\abs{U} = N$, $\abs{V} = M$, such that for all $u\in U:\deg(u)\geq c_0\cdot M^\delta$. Then, there exists $D\subseteq V$ with $\abs{D}\le \frac{c_1}{(1-c_1)c_0}\cdot M^{1-\delta}$ such that $\abs{\Nbr(D)}\geq c_1N$.
\end{lemma}

Using this dominating set lemma, we prove \cref{lem:(g:l) implies (g:l+g)}.

\begin{proof}[Proof of \cref{lem:(g:l) implies (g:l+g)}]
Fix a function $h:(\zo^n)^{\ell+g} \to [M]$. We will construct a \uniSHELA[g, \ell+g] (\uniNOSF[g, \ell+g] respectively) $\Y$ such that $\sminH(f(Y)) < \frac{g}{\ell+g}\cdot m + \delta^\pr$.
Let $N = 2^n$.
We consider two cases:
\begin{casesenum}
\item For all $(x_1, \dots, x_g) \in (\zo^n)^g: |\supp(h(x_1, \dots, x_g, \U_{\ell}))| \ge c_0 M^{\ell / (\ell + g)}$.\\
Consider an undirected bipartite graph $G = (U, V, E)$ where $U = (\zo^n)^g$ and $V = [M]$ with edge $e = (u, v)\in E$ where $u = (x_1, \dots, x_g)\in U$ and $v\in V$ iff there exist $x_{g+1}, \dots, x_{\ell + g}$ such that $h(x_1, \dots, x_{\ell+g}) = v$.
Applying \cref{lem:greedy-covering} to $G$, there exists $D\subset [M]$ such that $\abs{D} \le \frac{c_1}{(1-c_1)c_0}M^{g / (\ell + g)}$ and for $c_1 N^g$ many tuples $(x_1, \dots, x_g)\in(\zo^n)^g$, there exist $y_1, \dots, y_{\ell}\in (\zo^n)^{\ell}$ such that $h(x_1, \dots, x_g, y_1, \dots, y_{\ell}) \in D$.
Let $\Adv: (\zo^n)^g\rightarrow (\zo^n)^{\ell}$ be defined as:
\[
\Adv(x_1, \dots, x_g) =
\begin{cases}
(y_1, \dots, y_{\ell}) & \textrm{if there exist $y_1, \dots, y_{\ell}$ such that $h(x_1, \dots, x_g, y_1, \dots, y_{\ell})\in D$}\\
(0^n)^\ell & \textrm{otherwise}
\end{cases}
\]
Consider the \uniSHELA[g, \ell+g] (\uniNOSF[g, \ell+g] respectively) $\X = (\X_1, \dots, \X_{\ell+g})$ such that $\X_1, \dots, \X_g$ are independent uniform distributions and $(\X_{g+1}, \dots, \X_{g+\ell}) = \Adv(\X_1, \dots, \X_g)$.
Then, with probability $c_1$, $h(\X)\in D$. Applying \cref{lem:TV-dist lower bound}, we infer that $\sminH(\X)\le \log\left(\frac{c_1 M^{g / (\ell + g)}}{(1 - c_1)c_0(c_1 - \varepsilon)}\right) = \frac{g}{\ell+g}\cdot m + \log\left(\frac{c_1}{(1 - c_1)c_0(c_1 - \varepsilon)}\right)\le \frac{g}{\ell+g}\cdot m + \delta^\pr$.

\item There exist $x_1, \dots, x_g \in (\zo^n)^g$ such that $|\supp(h(x_1, \dots, x_g, \U_{\ell}))| \le c_0 M^{\ell / (\ell + g)}$.\\
Let $S = \supp(h(x_1, \dots, x_g, \U_{\ell}))$.
Define $f: \zo^{\ell}\rightarrow S$ by $f(y_1, \dots, y_{\ell}) = h(x_1, \dots, x_g, y_1, \dots, y_{\ell})$.
Then, by assumption, there exists \uniSHELA[g, \ell] (\uniNOSF[g,\ell], respectively) $\Y$ such that $\sminH(f(\Y)) \le \frac{g}{\ell}\cdot \log(\abs{S}) + \delta$.
Consider \uniSHELA[g, \ell+g] $\X = (\X_1, \dots, \X_{\ell+g})$ where distributions $\X_1, \dots, \X_g$ always output $x_1, \dots, x_g$ and $\X_{g+1}, \dots, \X_{\ell+g}$ are distributed as $\Y$. 
Then, 
\[
\sminH(f(\X))\le \frac{g}{\ell}\cdot \log(\abs{S}) + \delta \le \frac{g}{\ell+g}\cdot m + \delta + \frac{\log(c_0)g}{\ell} \le \frac{g}{\ell+g}\cdot m + \delta^\pr
\]
\end{casesenum}
\end{proof}

\subsubsection{Finding small dominating set in bipartite graphs}

We now directly prove \cref{lem:greedy-covering}.

\begin{proof}[Proof of \cref{lem:greedy-covering}]
    We construct $D$ via a greedy algorithm specified in \cref{alg:greedy covering}. This algorithm greedily chooses right vertices in $V$ with highest degree, adds them to $D$, and stops once the neighborhood of $D$, gets large enough.
    \begin{algorithm}[ht]
        \caption{}\label{alg:greedy covering}
        $i\gets 0$\\
        $D\gets \varnothing$\\
        $G_0=(U_0,V_0,E_0)\gets G=(U,V,E)$\\
        \While{$\abs{\Nbr(D)} < c_1N$}{
        Let $v_i\in V_i$ be the vertex of maximum degree in $G_i$\\
        $D\gets D\cup\{v_i\}$\\
        $V_{i+1}\gets V_i\setminus\{v_i\}$\\
        $U_{i+1}\gets U_i\setminus\Nbr(v_i)$\\
        $E_{i+1}\gets E_i\setminus\{(u,v)\in E: v=v_i\text{ or }u\in\Nbr(v_i)\}$\\
        $G_{i+1}\gets (U_{i+1},V_{i+1},E_{i+1})$
        }
    \end{algorithm}
    To analyze this algorithm, we can use loose bounds on the number of edges and vertices at any one step. As the algorithm stops once at least $c_1N$ vertices are removed from $U$, for all iterations $i$, $\abs{U_i}\geq(1-c_1)N$. In addition, because left vertices are only removed when one of their neighbors in $V$ is added to $D$, the remaining vertices in $U$ always have their original degrees intact. So, for all iterations $i$ and for all $u\in U_i$, it holds that $\deg(u)\geq c_0\cdot M^\delta$. So, 
    \[
        \abs{E_i}\geq\abs{U_i}c_0\cdot M^{\delta}\geq (1 - c_1)Nc_0\cdot M^\delta
    \]
    Observe that for all $i$, $\abs{V_i}\leq \abs{V} = M$. So,
    \[
        \deg(v_i)\ge 
        \frac{\abs{E_i}}{\abs{V_i}}\geq\frac{(1-c_1)Nc_0\cdot M^\delta}{M}=\frac{(1-c_1)c_0N}{M^{1-\delta}}.
    \]
    The algorithm terminates when at least $c_1N$ vertices are added to $D$ and at each step $\deg(v_i)$ vertices are added to $D$. Hence, the number of iterations for which the algorithm runs is at most
    \[
    \frac{c_1N}{\frac{(1-c_1)c_0N}{M^{1-\delta}}}=\frac{c_1}{(1-c_1)c_0}\cdot M^{1-\delta}
    \]
    The claim follows since exactly $1$ vertex is added to $D$ in each iteration.
\end{proof}

\subsection{Impossibility of condensing from \texorpdfstring{\uniNOSFs[g,\ell]}{uniform (g,l)-NOSF sources}}\label{sec:uniform NOSF condensing impossibility}

Our main theorem in this subsection is that it is impossible to condense from \uniNOSFs[g, \ell] where $g\ge \frac{\ell}{2}+1$. Using it and previous results, we obtain impossibility results for all $g, \ell$.

\begin{theorem}\label{thm:Can't condense (g;l)-NOSF}
There exists a universal constant $c > 0$ such that for all $g, \ell, m, n\in \N$ with $\ell/2 < g < \ell$, there exist $\eps = \left(\frac{1}{c\ell}\right)^{\ell-g}, \delta = c\cdot \ell^2\log(\ell)$ so that the following holds:
for any function $f:(\zo^n)^{\ell}\to \zo^m$, there exists a \uniNOSF[g, \ell] $\X$ such that $\sminH(f(\X))\le \frac{g}{\ell}\cdot m + \delta$.
\end{theorem}

We also infer the following useful corollary that shows that \uniNOSFs[g, \ell] cannot be condensed beyond rate $1 - 1/\ell^\pr$ with error $O(1/\ell^\pr)$ where $\ell^\pr$ is the smallest integer such that $g/\ell \le 1 - 1/\ell^\pr$.

\begin{corollary}\label{cor:cannot condense NOSF beyond l-1/l}
There exists a universal constant $c$ such that the following holds:
For all $g, \ell, \ell^{\pr}, m, n\in \N$ where $\ell^\pr$ is the smallest integer such that $\frac{g}{\ell}\le \frac{\ell^\pr-1}{\ell^\pr}$, there exist $\eps = \frac{1}{c\ell^\pr}, \delta = c\cdot \left(\ell^\pr\right)^{2}\log(\ell^\pr)$ so that the following holds:
for all functions $f: (\zo^n)^{\ell}\to \zo^m$, there exists a \uniNOSF[g, \ell] $\X$ such that $\sminH(f(\X)) \le \frac{1}{\ell^\pr}\cdot m + \delta$.
\end{corollary}

We also get a stronger impossibility result for \uniNOSFs[g, \ell] (compared to condensing impossibility for \uniSHELAs[g, \ell] proved in \cref{thm:can't condense SHELA above 1/c}) for the regime $g\le \ell/2$.

\begin{corollary}\label{cor:(g;l) NOSF impossible}
There exists a universal constant $c$ such that for all $\ell, g, r, m, n\in \N$ with $\ell\mod g = r$, there exist $\eps = \left(\frac{1}{c(g+r)}\right)^r, \delta = c\cdot (r+g)^2\log(g+r)$ so that the following holds:
for all functions $f:(\zo^n)^\ell\to\zo^m$, there exists a \uniNOSF[g, \ell] $\X$ such that $\sminH(f(\X)) \le \frac{g}{\ell}\cdot m + \delta$.
\end{corollary}

\begin{proof}
For $\ell = g + r$, apply \cref{lem:Can't condense (g;l)-NOSF} (see below) to infer there exists a universal constant $c$ such that the claim holds for $\eps = \left(\frac{1}{c(g+r)}\right)^r$, $\delta = c\cdot (r+g)^2\log(g+r)$.
Now, recursively apply \cref{lem:(g:l) implies (g:l+g)} with these $\eps, \delta$, setting $c_0 = 1, c_1 = \frac{1+\eps}{2}$ to infer the claim. When applying \cref{lem:(g:l) implies (g:l+g)}, we take advantage of the fact that $\eps < 3/4$ and that $c$ is a large enough constant to get that $\delta \ge \log\left(\frac{2(1+\eps)}{(1-\eps)^2}\right)$.
\end{proof}

We obtain impossibility result for all \uniNOSFs[ag, a\ell] where $g$ and $\ell$ are constants and $a\in \N$ is arbitrarily large.

\begin{corollary}
\label{cor:constant rate (g:l) NOSF impossible}
For all fixed $g, \ell\in \N$, there exist constants $\eps, \delta > 0$ so that the following holds:
for all $a, m, n\in \N$ and for all functions $f: (\zo^n)^{a\ell}\to \zo^m$, there exists a \uniNOSF[ag, a\ell] $\X$ such that $\sminH(f(\X)) \le \frac{g}{\ell}\cdot m  + \delta$.
\end{corollary}

We also record the special case of when the total number of blocks $\ell$ is a constant.

\begin{corollary}\label{cor:constant blocks (g:l) NOSF impossible}
For all fixed $g, \ell\in \N$, there exist constants $\eps, \delta > 0$ so that the following holds:
For all $m, n\in \N$ and for all functions $f: (\zo^n)^{\ell}\to \zo^m$, there exists a \uniNOSF[g, \ell] $\X$ such that $\sminH(f(\X)) \le \frac{g}{\ell}\cdot m  + \delta$.
\end{corollary}

\begin{proof}
Directly follows by setting $a = 1$ in \cref{cor:constant rate (g:l) NOSF impossible}.
\end{proof}

We prove our main theorem using the following general version of the theorem which we denote as our main lemma:

\begin{lemma}\label{lem:Can't condense (g;l)-NOSF}
There exists universal constants $c$ such that for all $c_0 > 0, g, \ell, M, n\in \N$ with $\ell/2 < g < \ell$, and for all $A\subset (\zo^n)^{\ell}$ with $\abs{A} = c_0 (2^n)^{\ell}$, the following holds:
for any function $f:(\zo^n)^{\ell}\to [M]$, there exists a \uniNOSF[g, \ell] $\X$, $A^\pr\subset A\cap \supp(\X)$ and $D\subset [M]$ such that $f(A^\pr) \subset D$ where $\abs{A^\pr} \ge c_0\cdot \left(\frac{1}{c\ell}\right)^{\ell-g}\cdot N^{g}$, and $\abs{D} \le \left(c\ell\right)^{\ell^2}\cdot \left(\frac{2}{c_0}\right)^{g}\cdot M^{g/\ell}$.
\end{lemma}

Using this main lemma, the theorem follows:

\begin{proof}[Proof of \cref{thm:Can't condense (g;l)-NOSF} assuming \cref{lem:Can't condense (g;l)-NOSF}]
Applying \cref{lem:Can't condense (g;l)-NOSF} with $A = M = \zo^m$, we infer that there exists \uniNOSF[g, \ell] $\X$, universal constant $c_0$ and $D\subset \zo^m$ such that $\abs{D} \le (c_0\ell)^{\ell^2}\cdot 2^g\cdot M^{g/\ell}$ and $\Pr[f(\X)\in D] \ge \left(\frac{1}{c_0\ell}\right)^{\ell-g}$.
Applying \cref{lem:TV-dist lower bound} with $\eps = \frac{1}{2}\cdot \left(\frac{1}{c_0\ell}\right)^{\ell-g}$, we infer that 
\begin{align*}
\sminH(f(\X)) 
& \le \log\left(\frac{|D|}{\varepsilon/2}\right)\\
& \le \frac{g}{\ell}\cdot m + \ell^2\log(c_0\ell) + g + (\ell-g)\log(c_0\ell) + 1\\
& \le \frac{g}{\ell}\cdot m + c\cdot \ell^2\log(\ell)\\
\end{align*}
where $c$ is a large enough universal constant.
As $\eps = \frac{1}{2}\cdot \left(\frac{1}{c_0\ell}\right)^{\ell-g} \ge \left(\frac{1}{c\ell}\right)^{\ell-g}$, we infer the claim.
\end{proof}

To prove our corollary regarding condensing \uniNOSFs[g, \ell] where $\frac{g}{\ell}$ is a large constant, we will use the following lemma:

\begin{lemma}\label{lem:can't condense NOSF above any upper limit l'-1/l'}
Let $g, \ell, \ell^\pr, n^\pr, n, m\in \N$ be such that $\frac{g}{\ell}\leq\frac{\ell^\pr-1}{\ell^\pr}, \ceil{\ell/\ell'}n < n^\pr$. 
Let $0 < \eps < 1, \delta > 0$ be such that:
for any function $f:(\zo^{n^\pr})^{\ell'}\to\zo^{m}$, there exists a \uniNOSF[\ell^\pr-1, \ell'] $\Y$ so that $\sminH(f(\Y)) \le \frac{\ell^\pr-1}{\ell^\pr}\cdot m + \delta$.
Then, for any function $h:(\zo^n)^{\ell}\to\zo^{m}$, there exists a \uniNOSF[g, \ell] $\X$ such that $\sminH(h(\X))\leq\frac{\ell^\pr-1}{\ell^\pr}\cdot m + \delta$.
\end{lemma}

We will prove this lemma in a later in \cref{subsec:impossibility-deferred-proofs}. Using it, the corollary immediately follows:

\begin{proof}[Proof of \cref{cor:cannot condense NOSF beyond l-1/l}]
We apply \cref{thm:Can't condense (g;l)-NOSF} to \uniNOSFs[\ell^\pr-1, \ell^\pr] and use it in \cref{lem:can't condense NOSF above any upper limit l'-1/l'} to infer the claim.
\end{proof}

To prove our corollary regarding condensing \uniNOSFs[ag, a\ell] where $g$ and $\ell$ are constants and $a$ is arbitrary, we will use the following lemma that allows us to generalize the impossibility result:

\begin{lemma}\label{lem:impossibility scales by arbitrary constant}
Let $g, \ell\in \N$ and $0 < \eps < 1, \delta > 0$ be such that for all $n, m\in \N$ and  all functions $f: (\zo^n)^{\ell}\to \zo^m$, there exists an \uniNOSF[g, \ell] $\X$ such that $\sminH(f(\X)) \le \frac{g}{\ell}\cdot m + \delta$.
Then, for all $a, n, m\in \N$ and all functions $f: (\zo^n)^{a\ell}\to \zo^m$, there exists an \uniNOSF[ag, a\ell] $\X$ such that $\sminH(f(\X)) \le \frac{g}{\ell}\cdot m + \delta$.
\end{lemma}

We will also prove this lemma in \cref{subsec:impossibility-deferred-proofs}. Using it, the corollary immediately follows:

\begin{proof}[Proof of \cref{cor:constant rate (g:l) NOSF impossible}]
We apply \cref{cor:(g;l) NOSF impossible} with $g, \ell$ to infer that there exist $0 < \eps < 1, \delta > 1$ such that for all $n, m\in \N$ and all functions $f: (\zo^n)^{\ell} \to \zo^m$, there exists an \uniNOSF[g, \ell] $\X$  such that $\sminH(f(\X))\le \frac{g}{\ell}\cdot m + \delta$. Finally, we apply \cref{lem:impossibility scales by arbitrary constant} to infer the claim.
\end{proof}

\subsubsection{Proving the main lemma}

Here, we will prove \cref{lem:Can't condense (g;l)-NOSF}. 
We first introduce some helpful notation for this part.
For an edge $e\in E$, let $\edgecolor(e)$ denote the color of $e$ in $H$.
For a vertex $x\in H$, let
\[
    \Nbr_H(x) = \{y\in H: (x, y)\in E\}.
\]
Similarly, for a vertex $x\in H$, and color $\gamma\in [M]$, let
\[
    \Nbr_H(x, \gamma) = \{y\in H: (x, y)\in E \text{ and } \edgecolor(x, y) = \gamma\}.
\]

To prove the main lemma, we will utilize the following special case of the main lemma, corresponding to the case of $g = \ell-1$, that we prove later:

\begin{lemma}\label{lem:Can't condense (l-1;l)-NOSF}
There exists a universal constant $c > 0$ such that for all $M, n, \ell\ge 3\in \N$, and $A\subset (\zo^n)^{\ell}$ with $\abs{A} = c_0 (2^n)^{\ell}$, the following holds:
for any function $f:(\zo^n)^{\ell}\to\zo^m$, there exists a \uniNOSF[\ell-1, \ell] $\X$, $A^\pr\subset A\cap \supp(\X)$ with $\abs{A^\pr} \ge \frac{1}{c}\cdot \frac{c_0}{\ell}\cdot N^{\ell-1}$, and $D\subset [M]$ with $\abs{D} \le c\cdot \frac{1}{\ell^2}\cdot \left(\frac{2}{c_0}\right)^{\ell-2}\cdot M^{(\ell-1)/\ell}$ such that $f(A^\pr) \subset D$.
\end{lemma}

The main lemma follows by an inductive argument where the special case above is the base case.

\begin{remark}
In proof of \cref{lem:Can't condense (g;l)-NOSF}, one can use $g = \ell$ as the base case as well. However, for clarity's sake we use $g = \ell-1$ as the base case. For first time readers, it will be helpful to first read the direct non-inductive proof of \cref{lem:Can't condense (l-1;l)-NOSF} presented in \cref{subsubsec:proving main lemma for g = l-1} before reading the proof of \cref{lem:Can't condense (g;l)-NOSF} as both these proofs share a lot of ideas.
\end{remark}

\begin{proof}[Proof of \cref{lem:Can't condense (g;l)-NOSF}]
Let $N = 2^n$. We will often identify $\zo^n$ with $[N]$ wherever convenient.
We let $c$ be a very large universal constant.

We proceed by induction on $b = \ell-g$. Formally, for $b\ge 1\in \N$ we will prove the claim for $\ell, g\in \N$ with $\ell/2 < g < \ell$ such that $b = \ell-g$.

For the base case, we let $b = 1$ and apply \cref{lem:Can't condense (l-1;l)-NOSF} to infer the claim.

For the inductive step, say we want to prove the hypothesis for $b \ge 2$ assuming the hypothesis holds for $b-1$. Fix some $g, \ell$ such that $\ell - g = b$.
Let $c_1 = \frac{1}{4\ell}, c_2 = \left(\frac{3}{2}\right)^{1/\ell^2}-1, c_3 = \frac{2}{3}, c_4 = \frac{c_0}{4}$.
By binomial approximation, there exists a constant $\alpha \ge 1$ such that $\frac{1}{\alpha}\cdot \frac{0.4}{\ell^2} \le c_2 \le \alpha\cdot \frac{0.6}{\ell^2}$ for all $\ell$.
For all positions $p\in [\ell]$, let $S_p\subset [N]^{\ell-1}$ be defined as follows: $(x_1, \dots, x_{p-1}, x_{p+1}, \dots, x_{\ell})\in S_p$ if and only if
\[
    \abs{\{f(x_1, \dots, x_{p-1}, y, x_{p+1}, \dots, x_{\ell}): y\in \zo^n \land (x_1, \dots, x_{p-1}, y, x_{p+1}, \dots, x_{\ell}) \in A\}} \ge c_2 M^{1/\ell}
\]
We consider various cases:
\begin{casesenum}
\item
There exists $p\in [\ell]$ such that $\abs{S_p} \ge c_0c_1  N^{\ell-1}$.\\
Consider the bipartite graph $G = (U, V, E)$ where $U = S_p, V = [M]$ and edge $e = (u, v) = ((x_1, \dots, x_{p-1}, x_{p+1}, \dots, x_{\ell}), z)\in E$ if and only if there exists $y\in \zo^n$ such that $f(x_1, \dots, x_{p-1}, y, x_{p+1}, \dots, x_{\ell}) = z$ and $(x_1, \dots, x_{p-1}, y, x_{p+1}, \dots, x_{\ell})\in A$. Then by assumption, for all $u\in U$, it holds that $\deg(u)\ge c_2 M^{1/\ell}$. 
We apply \cref{lem:greedy-covering} to $G$ and infer that there exists $D_{end}\subset V$ such that $\abs{D_{end}} \le \frac{c_3}{c_2(1-c_3)}M^{(\ell-1)/\ell}$ and $\Nbr_G(D_{end}) \ge c_0c_1c_3\abs{U} \ge c_0c_1c_3 N^{\ell-1}$.

Let $A^{\pr}_{end}\subset A$ be defined as follows: for a vertex $u = (x_1, \dots, x_{p-1}, x_{p+1}, \dots, x_{\ell})\in \Nbr_G(D_{end})$, we add $(x_1, \dots, x_{p-1}, y, x_{p+1}, \dots, x_{\ell})$ to $A^{\pr}_{end}$ where $y$ is such that $(x_1, \dots, x_{p-1}, y, x_{p+1}, \dots, x_{\ell})\in A$ and  $f(x_1, \dots, x_{p-1}, y, x_{p+1}, \dots, x_{\ell})\in D_{end}$ (we only pick one such $y$ per $u$ and  if multiple such $y$ exist, we break ties arbitrarily).
% By construction, we have that $A^{\pr}_{end}\subset A$ and
% \[
% \abs{A^{\pr}_{end}} \ge c_0c_1c_3 N^{\ell-1}
% \]
% Moreover, $f(A^{\pr}_{end}) \in D_{end}$ and
% \[
%     \abs{D_{end}} \le \frac{c_3}{c_2(1-c_3)}M^{(\ell-1)/\ell} 
% \]
% Let $\Adv_{end}: (\zo^n)^{\ell-1}\rightarrow \zo^{n}$ be defined as follows:
% \[
% \Adv_{end}(x_1, \dots, x_{p-1}, x_{p+1}, \dots, x_{\ell}) =
% \begin{cases}
% y & \exists y : (x_1, \dots, x_{p-1}, y, x_{p+1}, \dots, x_{\ell})\in A^{\pr}_{end}\\
% 0^n & \textrm{otherwise}
% \end{cases}
% \]
Let $z\in D_{end}$ be an arbitrary element. Let $f_{end}: (\zo^n)^{\ell-1}\to D_{end}$ be defined as follows:
\begin{align*}
& f_{end}(x_1, \dots, x_{p-1}, x_{p+1}, \dots, x_{\ell}) =\\
& \begin{cases}
f(x_1, \dots, x_{p-1}, y, x_{p+1}, \dots, x_{\ell}) & \exists y: (x_1, \dots, x_{p-1}, y, x_{p+1}, \dots, x_{\ell}) \in A^{\pr}_{end}\\
z & \textrm{otherwise}
\end{cases}
\end{align*}

We now use inductive hypothesis on the candidate function $f_{end}$ (having range $D_{end}$), and restriction set $\Nbr_G(D_{end})$. Notice that $\abs{\Nbr_G(D_{end})}\ge c_0^{end} = c_0c_1c_3$. We infer there exists \uniNOSF[g, \ell-1] $\X^{ind}$, $A^\pr_{ind}\subset \Nbr_G(D_{end})\cap \supp(\X^{ind})$, and $D_{ind}\subset D_{end}$ such that $f(A^\pr_{ind})\subset D_{ind}$. 
Let $\Adv: (\zo^n)^{\ell-1}\rightarrow \zo^{n}$ be defined as follows:
\[
\Adv_{end}(x_1, \dots, x_{p-1}, x_{p+1}, \dots, x_{\ell}) =
\begin{cases}
y & \exists y : (x_1, \dots, x_{p-1}, y, x_{p+1}, \dots, x_{\ell})\in A^{\pr}_{end}\\
0^n & \textrm{otherwise}
\end{cases}
\]
Let $\X^{ind} = (\X^{ind}_1, \dots, \X^{ind}_{p-1}, \X^{ind}_{p+1}, \dots, \X^{ind}_{\ell})$.
Now, define
\[
\X = (\X^{ind}_1, \dots, \X^{ind}_{p-1}, \Adv_{end}(\X^{ind}_1, \dots, \X^{ind}_{p-1}, \X^{ind}_{p+1}, \dots, \X^{ind}_{\ell}), \X^{ind}_{p+1}, \dots, \X^{ind}_{\ell})
\]
Similarly, define
\[
A^\pr = \{(x_1, \dots, x_{p-1}, y, x_{p+1}, \dots, x_{\ell})\mid (x_1, \dots, x_{p-1}, y, x_{p+1}, \dots, x_{\ell})\in A^\pr_{end}\land (x_1, \dots, x_{p-1}, x_{p+1}, \dots, x_{\ell})\in A^\pr_{ind}\}
\]
Let $D = D_{ind}$.
By construction, $\X$ is a \uniNOSF[g, \ell] where $A^\pr\subset A\cap\sup(\X)$ and $f(A^\pr)\in D$.
Moreover,
\begin{align*}
\abs{A^\pr}
&\ge c_0^{prev}\cdot \left(\frac{1}{c(\ell-1)}\right)^{\ell-g-1}\cdot N^g\\
&\ge c_0\cdot\frac{1}{4\ell}\cdot\frac{2}{3}\cdot\left(\frac{1}{c\ell}\right)^{\ell-g-1}\cdot N^g\\
&\ge c_0\cdot \left(\frac{1}{c\ell}\right)^{\ell-g}\cdot N^g\\
\end{align*}
Also,
\begin{align*}
\abs{D}
&\le \left(c(\ell-1)\right)^{(\ell-1)^2}\cdot \left(\frac{1}{c^{end}_0}\right)^{g}\cdot\left(\abs{D_{end}}\right)^{g / (\ell-1)}\\
&\le \left(c(\ell-1)\right)^{(\ell-1)^2}\cdot \left(\frac{1}{c_0c_1c_3}\right)^{g}\cdot\left(\frac{c_3}{1-c_3}\cdot\frac{1}{c_2}\cdot M^{(\ell-1)/\ell}\right)^{g / (\ell-1)}\\
&\le \left(c\ell\right)^{(\ell-1)^2}\cdot \left(\frac{1}{c_0}\cdot (4\ell)\cdot \frac{3}{2} \right)^{g}\cdot\left(2\cdot \frac{\alpha\cdot\ell^2}{0.4}\cdot M^{(\ell-1)/\ell}\right)^{g / (\ell-1)}\\
&\le \left(c\ell\right)^{(\ell-1)^2}\cdot \left(\frac{1}{c_0}\right)^{g}\cdot \left(6\ell\right)^{\ell}\cdot (5\alpha\cdot \ell^2)\cdot M^{g/\ell}\\
&\le \left(c\ell\right)^{\ell^2}\cdot \left(\frac{1}{c_0}\right)^{g}\cdot M^{g/\ell}\\
\end{align*}
Hence, the inductive step is proven for this case. 

\item
The above case does not happen, i.e., for all $p\in [\ell]$, $\abs{S_p} < c_0c_1 N^{\ell-1}$.\\
Let $S \subset [N]^{\ell}$ be defined as follows: $(x_1, \dots, x_{\ell})\in S$ if and only if there exists $p\in [\ell]$ such that $x_1, \dots, x_{p-1}, \dots, x_{p+1}, \dots, x_{\ell}\in S_p$.
Then,
\[
    \abs{S}\le \sum_{p=1}^{\ell} \abs{S_p}\cdot N \le c_0c_1\ell\cdot N^{\ell}
\]
Consider the $\ell$-uniform $\ell$-partite hypergraph $H = (V_1, \dots, V_{\ell}, E)$ where $e = (v_1, \dots, v_{\ell})\in E$ if and only if $e\in A\setminus S$.
As $\abs{S}\le c_0c_1\ell\cdot N^{\ell}$, it must be that $\abs{E} \ge c_0(1-c_1\ell)\cdot N^{\ell}$.
This implies there exists $(v_1^*, \dots, v_{\ell-g}^*)\in (V_1, \dots, V_{\ell-g})$ such that $\deg(v_1^*, \dots, v_{\ell-g}^*) \ge c_0(1-c_1\ell)\cdot N^g$.
Consider the $g$-uniform $g$-partite hypergraph $H^\pr = (V_{\ell-g+1}, \dots, V_{\ell}, E^\pr)$ where $e = (v_{\ell-g+1}, \dots, v_{\ell})\in E^\pr$ if and only if $(v_1^*, \dots v_{\ell-g}^*, v_{\ell-g+1}, \dots, v_{\ell})\in E$.
Then, $\abs{E^\pr} \ge c_0(1-c_1\ell)\cdot N^{g}$
Now, color the edges of $H^\pr$ with colors from $[M]$ such that $\edgecolor(v_{\ell-g+1}, \dots, v_{\ell}) = f(v_1^*, \dots, v_{\ell-g}^*, v_{\ell-g+1}^*, \dots, v_{\ell})$.
By assumption, for every position $p\in [\ell-g+1, \ell]$, and every $(\ell-1)$ tuple $(v_{\ell-g+1}, \dots, v_{p_1}, v_{p+1}, \dots, v_{\ell}) \in [N]^{\ell-1}$: the number of distinct colored edges as entries in position $p$ vary in $H^\pr$ is $\le c_2M^{1/\ell}$.
Formally, $\abs{\edgecolor_{H^\pr}(v_{\ell-g+1}, \dots, v_{p-1}, y, v_{p+1}, \dots, v_{\ell}): y\in [N]} \le c_2M^{1 / \ell}$.
Applying \cref{lem:greedy-edge-covering} to $H$, we infer that there exists $D\subset [M]$ such that $\abs{D}\le \frac{c_4c_2 (c_2 + 1)^{g(g+1) / 2 - 1}}{(c_0(1-c_1\ell)-c_4)^{g}} M^{g/\ell}$ and $c_4 N^g$ edges in $H$ are colored in one of the colors from $D$. 

Let $A^\pr = \{(v_1^*, \dots, v_{\ell-g}^*, x_{\ell-g+1}, \dots, x_{\ell}) : \edgecolor_{H^\pr}(x_{\ell-g+1}, \dots, x_{\ell})\in D\}$.
Then, $A^\pr\subset A$ and
\[
\abs{A^\pr} \ge c_4 N^{g} \ge \frac{1}{c}\cdot c_0\cdot N^{g} \ge c_0\cdot \left(\frac{1}{c\ell}\right)^{\ell-g}\cdot N^{g}
\]
Moreover, $f(A^\pr)\subset D$ and
\[
\abs{D}\le \frac{c_4c_2 (c_2 + 1)^{g(g+1)/ 2 - 1}}{(c_0(1-c_1\ell)-c_4)^{g}} M^{g/\ell} \le (c\ell)^{\ell^2}\cdot \left(\frac{1}{c_0}\right)^g\cdot M^{g/\ell}
\]

We now create \uniNOSF[\ell-1, \ell] $\X = (\X_1, \dots, \X_{\ell})$ which will have the desired properties.
Let $\Adv: \zo^{(\ell-1)n}\rightarrow \zo^{n}$ be defined as follows:
\[
\Adv(x_{\ell-g+1}, \dots, x_{\ell}) =
\begin{cases}
v_1^*, \dots, v_{\ell-g}^* & \textrm{if }(x_{\ell-g+1}, \dots, x_{\ell})\in E^\pr\\
0^n & \textrm{otherwise}
\end{cases}
\]
Let $\X_{\ell-g+1} = \dots = \X_{\ell} = \U_n$ and let $\X_1, \dots, \X_{\ell-g} = \Adv(\X_{\ell-g+1}, \dots, \X_{\ell})$.
Then, $A^\pr\subset \supp(\X)$ as desired, completing the inductive step for this case as well.
\end{casesenum}
\end{proof}

\subsubsection{Proving the main lemma for \texorpdfstring{$g=\ell-1$}{g = l-1}}\label{subsubsec:proving main lemma for g = l-1}

We will prove our main lemma for the case of $g = \ell-1$ using a color covering lemma for dense $t$-partite $t$-uniform hypergraphs colored in some special way:

\begin{lemma}[Small Color Covering for Hypergraphs]
\label{lem:greedy-edge-covering}
    Let $0 < c_0 \le 1, 0 < c_1, 0 < \eps < c_0$ be arbitrary.
    Let $H = (V_1, \dots, V_t, E)$ be a $t$-uniform $t$-partite hypergraph with $V_1 = \dots = V_t = [N], \abs{E} = c_0N^t$.
    Let the edges of $H$ be colored in one of $M$ colors so that for every position $p\in [T]$, and every $(t-1)$ tuples: $(v_1, \dots, v_{p_1}, v_{p+1}, \dots, v_t) \in [N]^{t-1}$, the number of distinct colored edges as entries position $p$ vary is $\le c_1M^\delta$.
    Formally, $\abs{\edgecolor(v_1, \dots, v_{p-1}, y, v_{p+1}, \dots, v_t): y\in [N]} \le c_1M^{\delta}$.
    Then, there exists $D\subseteq [M]$ such that $\abs{D}\le \frac{\eps c_1(c_1 + 1)^{t(t+1)/2 - 1}}{(c_0 - \eps)^t}\cdot M^{t\delta}$ and at least $\eps N^t$ edges in $H$ are colored in one of the colors from $D$.
\end{lemma}

We prove this color covering lemma later. Using it, we prove our main lemma for the case of $g = \ell-1$:

\begin{proof}[Proof of \cref{lem:Can't condense (l-1;l)-NOSF}]
Let $N = 2^n$. We will often identify $\zo^n$ with $[N]$ wherever convenient.
Let $c_1 = \frac{1}{4\ell}, c_2 = \left(\frac{3}{2}\right)^{1/\ell^2}-1, c_3 = \frac{2}{3}, c_4 = \frac{c_0}{4}$.
We let $c$ be a very large universal constant.
By binomial approximation, there exists a constant $\alpha \ge 1$ such that $\frac{1}{\alpha}\cdot \frac{0.4}{\ell^2} \le c_2 \le \alpha\cdot \frac{0.6}{\ell^2}$ for all $\ell$.
For all positions $p\in [\ell]$, let $S_p\subset [N]^{\ell-1}$ be defined as follows: $(x_1, \dots, x_{p-1}, x_{p+1}, \dots, x_{\ell})\in S_p$ if and only if
\[
    \abs{\{f(x_1, \dots, x_{p-1}, y, x_{p+1}, \dots, x_{\ell}): y\in \zo^n \land (x_1, \dots, x_{p-1}, y, x_{p+1}, \dots, x_{\ell}) \in A\}} \ge c_2 M^{1/\ell}
\]
We consider various cases:
\begin{casesenum}
\item
There exists $p\in [\ell]$ such that $\abs{S_p} \ge c_0c_1  N^{\ell-1}$.\\
Consider the bipartite graph $G = (U, V, E)$ where $U = S_p, V = [M]$ and edge $e = (u, v) = ((x_1, \dots, x_{p-1}, x_{p+1}, \dots, x_{\ell}), z)\in E$ if and only if there exists $y\in \zo^n$ such that $f(x_1, \dots, x_{p-1}, y, x_{p+1}, \dots, x_{\ell}) = z$ and $(x_1, \dots, x_{p-1}, y, x_{p+1}, \dots, x_{\ell})\in A$. Then by assumption, for all $u\in U$, it holds that $\deg(u)\ge c_2 M^{1/\ell}$. 
We apply \cref{lem:greedy-covering} to $G$ and infer that there exists $D\subset V$ such that $\abs{D} \le \frac{c_3}{c_2(1-c_3)}M^{(\ell-1)/\ell}$ and $\Nbr_G(D) \ge c_0c_1c_3\abs{U} \ge c_0c_1c_3 N^{\ell-1}$.

We now construct set $A^\pr\subset A$. For each vertex $u = (x_1, \dots, x_{p-1}, x_{p+1}, \dots, x_{\ell})\in \Nbr_G(D)$, we add $(x_1, \dots, x_{p-1}, y, x_{p+1}, \dots, x_{\ell})$ to $A^\pr$ where $y$ is such that $(x_1, \dots, x_{p-1}, y, x_{p+1}, \dots, x_{\ell})\in A$ and  $f(x_1, \dots, x_{p-1}, y, x_{p+1}, \dots, x_{\ell})\in D$ (we only pick one such $y$ per $u$ and  if multiple such $y$ exist, we break ties arbitrarily).
By construction, we have that $A^\pr\subset A$ and
\[
\abs{A^\pr} \ge c_0c_1c_3 N^{\ell-1} \ge \frac{1}{c}\cdot \frac{c_0}{\ell}\cdot N^{\ell-1}
\]
Moreover, $f(\A^\pr) \in D$ and
\[
    \abs{D} \le \frac{c_3}{c_2(1-c_3)}M^{(\ell-1)/\ell} \le c\cdot \frac{1}{\ell^2}\cdot \left(\frac{2}{c_0}\right)^{\ell-2}\cdot M^{(\ell-1)/\ell}
    % c\cdot \ell^2\cdot M^{(\ell-1)/\ell}
\]

We now construct \uniNOSF[\ell-1, \ell] $\X = (\X_1, \dots, \X_{\ell})$ with the desired properties.
Let $\Adv: (\zo^n)^{(\ell-1)}\rightarrow \zo^{n}$ be defined as follows:
\[
\Adv(x_1, \dots, x_{p-1}, x_{p+1}, \dots, x_{\ell}) =
\begin{cases}
y & \exists y: (x_1, \dots, x_{p-1}, y, x_{p+1}, \dots, x_{\ell})\in A^\pr\\
0^n & \textrm{otherwise}
\end{cases}
\]
Let $\X_1 = \dots = \X_{p-1} = \X_{p+1} = \dots = \X_{\ell} = \U_n$.
Let $\X_p = \Adv(\X_1, \dots, \X_{p-1}, \dots, \X_{p+1}, \dots, \X_{\ell})$.
Then, $\A^\pr\subset \supp(\X)$ as desired. 

\item
The above case does not happen, i.e., for all $p\in [\ell]$, $\abs{S_p} < c_0c_1 N^{\ell-1}$.\\
Let $S \subset [N]^{\ell}$ be defined as follows: $(x_1, \dots, x_{\ell})\in S$ if and only if there exists $p\in [\ell]$ such that $x_1, \dots, x_{p-1}, \dots, x_{p+1}, \dots, x_{\ell}\in S_p$.
Then,
\[
    \abs{S}\le \sum_{p=1}^{\ell} \abs{S_p}\cdot N \le c_0c_1\ell\cdot N^{\ell}
\]
Consider the $\ell$-uniform $\ell$-partite hypergraph $H = (V_1, \dots, V_{\ell}, E)$ where $e = (v_1, \dots, v_{\ell})\in E$ if and only if $e\in A\setminus S$.
As $\abs{S}\le c_0c_1\ell\cdot N^{\ell}$, it must be that $\abs{E} \ge c_0(1-c_1\ell)\cdot N^{\ell}$.
This implies there exists $v_1^*\in V_1$ such that $\deg(v_1^*) \ge c_0(1-c_1\ell)\cdot N^{\ell-1}$.
Consider the $(\ell-1)$-uniform $(\ell-1)$-partite hypergraph $H^\pr = (V_2, \dots, V_{\ell}, E^\pr)$ where $e = (v_2, \dots, v_{\ell})\in E^\pr$ if and only if $(v_1^*, \dots, v_{\ell})\in E$.
Then, $\abs{E^\pr} = \deg(v_1^*) \ge c_0(1-c_1\ell)\cdot N^{\ell-1}$
Now, color the edges of $H^\pr$ with colors from $[M]$ such that $\edgecolor(v_2, \dots, v_{\ell}) = f(v_1^*, v_2, \dots, v_{\ell})$.
By assumption, for every position $p\in [2, \ell]$, and every $(\ell-1)$ tuple $(v_2, \dots, v_{p_1}, v_{p+1}, \dots, v_{\ell}) \in [N]^{\ell-1}$: the number of distinct colored edges as entries in position $p$ vary in $H^\pr$ is $\le c_2M^{1/\ell}$.
Formally, $\abs{\edgecolor_{H^\pr}(v_2, \dots, v_{p-1}, y, v_{p+1}, \dots, v_{\ell}): y\in [N]} \le c_2M^{1 / \ell}$.
Applying \cref{lem:greedy-edge-covering} to $H$, we infer that there exists $D\subset [M]$ such that $\abs{D}\le \frac{c_4c_2 (c_2 + 1)^{(\ell-1)\ell / 2 - 1}}{(c_0(1-c_1\ell)-c_4)^{\ell-1}} M^{(\ell-1)/\ell}$ and $c_4 N^{\ell-1}$ edges in $H$ are colored in one of the colors from $D$. 

Let $A^\pr = \{(v_1^*, x_2, \dots, x_{\ell}) : \edgecolor_{H^\pr}(x_2, \dots, x_{\ell})\in D\}$.
Then, $A^\pr\subset A$ and
\[
\abs{A^\pr} \ge c_4 N^{\ell-1} \ge \frac{1}{c}\cdot c_0\cdot N^{\ell-1} \ge \frac{1}{c}\cdot \frac{c_0}{\ell}\cdot N^{\ell-1}
\]
Moreover, $f(\A^\pr)\in D$ and
\[
\abs{D}\le \frac{c_4c_2 (c_2 + 1)^{(\ell-1)\ell / 2 - 1}}{(c_0(1-c_1\ell)-c_4)^{\ell-1}} M^{(\ell-1)/\ell} \le c\cdot \frac{1}{\ell^2}\cdot \left(\frac{2}{c_0}\right)^{\ell-2}\cdot M^{(\ell-1)/\ell}
\]

We now create \uniNOSF[\ell-1, \ell] $\X = (\X_1, \dots, \X_{\ell})$ which will have the desired properties.
Let $\Adv: \zo^{(\ell-1)n}\rightarrow \zo^{n}$ be defined as follows:
\[
\Adv(x_2, \dots, x_{\ell}) =
\begin{cases}
v_1^* & \textrm{if }(x_2, \dots, x_{\ell})\in E^\pr\\
0^n & \textrm{otherwise}
\end{cases}
\]
Let $\X_2 = \dots = \X_{\ell} = \U_n$ and let $\X_1 = \Adv(\X_2, \dots, \X_{\ell})$.
Then, $A^\pr\subset \supp(\X)$ as desired. 
\end{casesenum}
\end{proof}

\subsubsection{Finding a small color covering in locally-light hypergraphs}

We consider dense $t$-uniform $t$-partite hypergraphs where all edges are colored and  the hypergraph satisfies a ``locally-light" condition: all $t-1$-tuples are adjacent to a small number of colors. The covering lemma finds small set of colors that covers constant fraction of edges in the hypergraph. We do this by finding a popular color in such a hypergraph.

\begin{lemma}[Popular Color in Locally-Light Hypergraphs]
\label{lem:locally-light-popular-element}
    Let $0 < c_0 \le 1, 0 < c_1$ be arbitrary.
    Let $t\ge 2\in \N$.
    Let $H=(V_1, \dots, V_{t},E)$ be a $t$-uniform $t$-partite hypergraph with $\abs{V_1} = \dots = \abs{V_t} = N, \abs{E} = c_0N^t$. Let the edges of $H$ be colored in one of $M$ colors so that for every position $p\in [T]$, and every $(t-1)$ tuples: $(v_1, \dots, v_{p_1}, v_{p+1}, \dots, v_t) \in [N]^{t-1}$, the number of distinct colored edges as entries position $p$ vary is $\le c_1M^\delta$.
    Formally, $\abs{\edgecolor(v_1, \dots, v_{p-1}, y, v_{p+1}, \dots, v_t): y\in [N]} \le c_1M^{\delta}$.
    Then, there exists a color $\gamma\in [M]$ such that at least $\frac{c_0^{t}}{c_1(c_1+1)^{t(t+1)/2 - 1}}\cdot N^t / M^{t\delta}$ edges in $H$ are colored with color $\gamma$.
\end{lemma}

Using this lemma, our color covering lemma for hypergraph follows by repeatedly finding such popular colors.

\begin{proof}[Proof of \cref{lem:greedy-edge-covering}]
    We introduce some additional notation: for a color $\gamma\in [M]$, let
    \[
        \colorcount_H(\gamma) = \abs{\{e\in H: \edgecolor(e) = \gamma\}}.
    \]
    We will construct $D$ by a greedy algorithm where we add the most popular color in $H$ to $D$, remove all edges of that color, and repeat. Further details are specified in \cref{alg:greedy edge covering}.
    \begin{algorithm}[ht]
        \caption{}\label{alg:greedy edge covering}
        $i\gets 0$, $D\gets \emptyset$\\
        $H^{(0)} = (V_1^{(0)}, \dots, V_t^{(0)},E^{(0)})\gets H = (V_1, \dots, V_t, E)$\\
        \While{$\colorcount_H(D) \le \eps N^t$}{
        Let $\gamma_i\in [M]$ be the color that maximizes $\colorcount_{H_i}(\gamma_i)$.\\
        $D\gets D\cup\{\gamma_i\}$\\
        $E^{(i+1)}\gets E^{(i)}\setminus\{e\in E\mid \edgecolor(e) = \gamma_i\}$\\
        $H^{(i+1)}\gets (V_1, \dots, V_t, E^{(i+1)})$\\
        $i\gets i+1$
        }
    \end{algorithm}
    
    We observe that $\colorcount_H(\gamma_1) \ge \dots \ge \colorcount_H(\gamma_{\abs{D}})$.
    At the iteration number $\abs{D}$ of the loop, the number of uncovered edges in $H$ is at least $(c_0 - \eps)N^2$.
    Applying \cref{lem:locally-light-popular-element} on $H^{(|D|-1)}$, we infer that
    \[
    \colorcount_H(\gamma_{\abs{D}}) \ge \frac{(c_0 - \eps)^{t}}{c_1(c_1+1)^{t(t+1)/2 - 1}}\cdot N^t / M^{t\delta}
    \]
    
    Hence, the number of edges covered in each iteration of the loop is at least 
    $\frac{(c_0 - \eps)^{t}}{c_1(c_1+1)^{t(t+1)/2 - 1}}\cdot N^t / M^{t\delta}$
    As the loop stops when the number of edges covered is at least $\eps N^t$, the number of iterations to terminate is at most
    \[
        \frac{\eps N^t}{\frac{(c_0 - \eps)^{t}}{c_1(c_1+1)^{t(t+1)/2 - 1}}\cdot N^t / M^{t\delta}}
        = \frac{\eps c_1(c_1 + 1)^{t(t+1)/2 - 1}}{(c_0 - \eps)^t}\cdot M^{t\delta}
    \]
    As the number of iterations of the loop equals $\abs{D}$, we indeed infer the claim.
\end{proof}

\subsubsection{Finding a popular color in locally-light hypergraphs}

For the base case, we find such a popular color in graphs:

\begin{lemma}[Popular Color in Locally-Light Graphs]
\label{lem:locally-light-popular-element-base-case}
    Let $0 < c_0 \le 1, 0 < c_1$ be arbitrary.
    Let $H=(U,V,E)$ be a bipartite graph with $\abs{U} = \abs{V} = N, \abs{E} = c_0N^2$. Let the edges of $H$ be colored in one of $M$ colors so that for every vertex $x\in H$, the number of distinct colored edges incident on $x$ is $\le c_1M^\delta$. Then, there exists a color $\gamma\in [M]$ such that at least $\frac{c_0^2}{(c_1+1)^2c_1}\cdot N^2 / M^{2\delta}$ edges in $H$ are colored with color $\gamma$.
\end{lemma}

Using this, we inductively find a popular color in locally-light hypergraphs. 

\begin{proof}[Proof of  \cref{lem:locally-light-popular-element}]
    We prove this result by induction on $t$ with the inductive hypothesis for $t$ being that such a popular color exists for graphs with this property.
    
    For the base case, $t = 2$. We apply \cref{lem:locally-light-popular-element-base-case} directly on $H$ and infer the claim.
    
    For the inductive step, assume that we have proven the hypothesis for $t-1$ and using it, we prove the hypothesis for $t$ where $t\ge 3$.
    Let $c_2 = \frac{c_0}{c_1+1}$. Let
    \[
        E^\pr = \{e = (v_1, \dots, v_t)\in E: \abs{\Nbr_H((v_2, \dots, v_t), \edgecolor(e))}\ge c_2 N / M^{\delta}\}
    \]
    Let $H^\pr = (V_1, \dots, V_t, E^\pr)$.
    We now lower bound the number of edges in $E^\pr$.
    Fix arbitrary $(t-1)$-tuple $v = (v_2, \dots, v_{t})\in [N]^{t-1}$. By assumption, $\abs{\{\gamma\in [M]: \abs{\Nbr_H(v, \gamma)} > 0\}} \le c_1 M^{\delta}$. In $H^\pr$, we excluded all edges in $H$ incident to the tuple $v$ with color $\gamma$ such that $\abs{\Nbr_H(v, \gamma)} < c_2 N / M^{\delta}$. Hence, we exclude at most $c_1c_2 N$ such edges incident to $v$ in $H$.
    As there are at most $N^{t-1}$ such tuples, the total number of edges we excluded in $E^\pr$ is at most $c_1c_2N^t$.
    Hence, $\abs{E^\pr} \ge (c_0 - c_1c_2) N^t$.

    As $\abs{V_1}\le N$, there exists $v_1^*\in V_1$ such that $\Nbr_{H^\pr}(v_1^*) \ge \abs{E^\pr} / N \ge (c_0 - c_1c_2)N^{t-1}$.
    Let $G = (V_2, \dots, V_t, E_G)$ be a $(t-1)$-uniform $(t-1)$-partite hypergraph where $(v_2, \dots, v_t)\in E_G$ if and only if $(v_1^*, v_2, \dots, v_t)\in E^\pr$. We see that $G$ satisfies the conditions of the inductive hypothesis for $t-1$. Hence, there exists a color $\gamma^*\in [M]$ such that at least $\frac{(c_0 - c_1c_2)^{t-1}}{c_1(c_1+1)^{t(t-1)/2 - 1}}\cdot N^{t-1} / M^{(t-1)\delta}$ edges in $G$ are colored by $\gamma^*$.
    For each edge $e = (v_2, \dots, v_t)$ in $G$ with $\edgecolor(e) = \gamma^*$, by property of $E^\pr$ and the fact that edge $(v_1^*, v_2, \dots, v_t)\in E^\pr$, $\abs{\Nbr_{H^\pr}((v_2, \dots, v_t), \gamma^*)} \ge c_2N/M^{\delta}$.
    Thus, the number of edges in $H^\pr$, and hence $H$ colored with $\gamma^*$ is at least
    \[
        \frac{(c_0 - c_1c_2)^{t-1}}{c_1(c_1+1)^{t(t-1)/2 - 1}}\cdot N^{t-1} / M^{(t-1)\delta}\cdot c_2N/M^{\delta} \ge \frac{c_0^{t}}{c_1(c_1+1)^{t(t+1)/2 - 1}}\cdot N^T / M^{t\delta}
    \]
    Hence, the inductive hypothesis holds for $t$, completing the inductive step, proving the claim.
\end{proof}

Finally, we directly argue a popular color exists in dense locally-light bipartite graphs.

\begin{proof}[Proof of \cref{lem:locally-light-popular-element-base-case}]
    Let $c_2 = \frac{c_0}{c_1+1}$. Let
    \[
        E^\pr = \{e = (u, v)\in E: \abs{\Nbr_H(v, \edgecolor(e))}\ge c_2 N / M^{\delta}\}
    \]
    Let $H^\pr = (U, V, E^\pr)$.
    We now lower bound the number of edges in $E^\pr$.
    Fix arbitrary vertex $v\in V$. By assumption, $\abs{\{\gamma\in [M]: \abs{\Nbr_H(v, \gamma)} > 0\}} \le c_1 M^{\delta}$. In $H^\pr$, we excluded all edges in $H$ incident to $v$ with color $\gamma$ such that $\abs{\Nbr_H(v, \gamma)} < c_2 N / M^{\delta}$. Hence, we exclude at most $c_1c_2 N$ such edges incident to $v$ in $H$.
    As $|V|\le N$, the total number of edges we excluded in $E^\pr$ is at most $c_1c_2N^2$.
    Hence, $\abs{E^\pr} \ge (c_0 - c_1c_2) N^2$.

    As $\abs{U}\le N$, there exists $u^*\in U$ such that $\Nbr_{H^\pr}(u^*) \ge \abs{E^\pr} / N \ge (c_0 - c_1c_2)N$.
    As the number of distinct colored edges incident on $u^*$ is at most $c_1M^{\delta}$, there exists a color $\gamma^*\in [M]$ such that $\abs{\Nbr_{H^\pr}(u^*, \gamma^*)} \ge \frac{(c_0 - c_1c_2)N }{c_1 M^{\delta}}$. For each $v\in \Nbr_{H^\pr}(u^*, \gamma^*)$, by definition of $E^\pr$, $\abs{\Nbr_{H^\pr}(v, \gamma^*)} \ge \frac{c_2N}{M^{\delta}}$. Hence, the number of edges $e\in H^\pr$ colored with $\gamma^*$ is at least
    \[
        \sum_{v\in \Nbr_{H^\pr}(u^*, \gamma^*)} \abs{\Nbr_{H^\pr}(v, \gamma^*)} \ge \abs{\Nbr_{H^\pr(u^*, \gamma^*)}}\cdot \frac{c_2N}{M^{\delta}} \ge \frac{(c_0 - c_1c_2)c_2N^2 }{c_1 M^{2\delta}} = \frac{c_0^2 N^2}{(c_1+1)^2c_1 M^{2\delta}}
    \]
\end{proof}

\subsection{Impossibility of condensing from CG sources}\label{sec:Impossibility condensing low min-entropy CG}

We prove two impossibility results regarding impossibility of condensing from \CGs[\ell, \ell]. Our first result \cref{thm:can't condense CG sources beyond entropy gap} states that any candidate condenser cannot decrease the entropy gap present in the blocks of CG sources. Our second result in contrast, states that when blocks have linear entropy, then condenser cannot condense beyond rate $1/2$. The latter result is much stronger than the former in regimes where $m$ is comparatively larger than $n$ (say $m = O(n\ell)$ and $\ell = \omega(1)$).

\subsubsection{Impossibility of non-trivial condensing beyond min-entropy gap}

We will use the fact that it is impossible to condense from general $(n, k)$-sources.

\begin{lemma}\label{lem:can't condense arbitrary n ; k source}
For all $n, k, m\in \N$ and $\eps > 0$ the following holds:
For all functions $f: \zo^n\to \zo^m$, there exists an $(n, k)$ source $\X$ such that $\sminH(f(\X)) \le m - (n - k) + \log(1 / (1-\eps)) - \max(m - n, 0)$.
\end{lemma}

We believe a result of this form is well-known but we were unable to find a good reference. Thus, for the sake of completeness, we prove this lemma at the end of this subsection. Using this, we prove our impossibility result for \CGs[\ell, \ell].

\begin{theorem}\label{thm:can't condense CG sources beyond entropy gap}
For all $0 < \eps < 1, \Delta$ and $\ell, m, n\in \N$, the following holds:
for every function $f: (\zo^n)^{\ell}\to \zo^m$, there exists a \CG[\ell, \ell] $\X$ where the good blocks have min-entropy at least $n - \Delta - \log(\ell/\eps) - O(1)$ conditioned on all fixings of previous blocks and $\sminH(f(\X)) \le m - \Delta + \log(2 / (2 - \eps)) - \max(m - \ell n, 0)$.
\end{theorem}

\begin{proof}
Let $\X$ be an arbitrary $(t, k)$-source where $t = \ell n$ and $k = n - \Delta$. We transform $\X$ into a source $\Y = (\Y_1, \dots, \Y_{\ell})$ with block lengths $n$ that is $\eps/2$-close to an \CG[\ell, \ell] where the good blocks have min-entropy at least $n - \Delta - \log(2(\ell-1)/\eps)$ conditioned on all fixings of the previous blocks. We then apply \cref{lem:can't condense arbitrary n ; k source} to infer that it is impossible to condense \CGs[\ell, \ell] so that the output distribution is $\eps$-close to having min-entropy more than $m - \Delta + \log(2 / (2 - \eps) - \max(m - \ell n, 0)$ as desired.

Let $\gamma = \frac{\eps}{2(\ell-1)}$.
For $1\le i\le \ell$, we define each block $\Y_i$ as $\Y_i = \X[(i-1)\cdot n, i\cdot n]$.
We prove that $\Y$ is $\eps/2$-close to a block source with $\ell$ blocks of length $n$ each and each of them has entropy at least $n - \Delta - \log(1/\gamma)$ conditioned on all previous blocks.
We will inductively prove that for all $j$ down from $\ell + 1$ to $1$, there exist blocks $\ZZ_j, \dots, \ZZ_{\ell}$ such that:
\begin{enumerate}
    \item 
    \[
        (\Y_1, \dots, \Y_{\ell}) \approx_{(\ell-j+1)\gamma} (\Y_1, \dots, \Y_{j-1}, \ZZ_j, \dots, \ZZ_{\ell})
    \]
    
    \item 
    For all $w\in \N$ such that $j\le w\le \ell$, conditioned on every fixing of $\Y_1, \dots, \Y_{j-1}, \ZZ_j, \dots, \ZZ_{w-1}$: the min-entropy of $\ZZ_w$ is at least $n - \Delta - \log(1/\gamma)$.

    \item 
    \[
        \minH((\Y_1, \dots, \Y_{j-1})) \ge n(j-1) - \Delta.
    \]
\end{enumerate}

The base case of $j = \ell+1$ is trivially true.

For the inductive step, assume the claim holds for $j+1 \le \ell+1$. We prove the claim for $j$.
We begin by proving the third requirement is satisfied.
By assumption, we know that $\minH((\Y_1, \dots, \Y_{j})) \ge jn - \Delta$.
Looking ahead, we apply \cref{lem:removing d bits from source removes d bits of entropy} to $(\Y_1, \dots, \Y_{j})$ and to it's projection onto first $n(j-1)$ bits ; we infer that $\minH((\Y_1, \dots, \Y_{j-1})) \ge n(j-1) - \Delta$ as desired.

Applying \cref{lem:min-entropy-chain-rule}, we get that with probability at least $1 - \gamma$ over fixings of $\Y_1, \dots, \Y_{j-1}$, the source $(\Y_1, \dots, \Y_{j})$ will have conditional min-entropy at least $n - \Delta - \log(1/\gamma)$.
So, with probability at least $1 - \gamma$ over fixings of $(\Y_1, \dots, \Y_{j-1})$, $\Y_j$ will have conditional min-entropy at least $n - \Delta - \gamma$.
Let these good fixings of $\Y_1, \dots, \Y_{j-1}$ be $S$.
By the inductive hypothesis, there exist blocks $\ZZ_{j+1}, \dots, \ZZ_{\ell}$ that satisfy the conditions of the inductive hypothesis.
Define the distribution $(\ZZ^\pr_{j}, \dots, \ZZ^\pr_{\ell})$ as being same as the conditional distribution of $(\Y_{j}, \ZZ_{j+1}, \dots, \ZZ_{\ell})$ when $(\Y_1, \dots, \Y_{j-1})\in S$ and equal to $(\U_n)^{\ell-j+1}$ otherwise.
Then,
\[
(\Y_1, \dots, \Y_j, \ZZ_{j+1}, \dots, \ZZ_{\ell}) \approx_{\gamma} (\Y_1, \dots, \Y_{j-1}, \ZZ^\pr_j, \dots, \ZZ_{\ell})
\]
Hence,
\begin{align*}
\abs{(\Y_1, \dots, \Y_{\ell}) - (\Y_1, \dots, \Y_{j-1}, \ZZ^\pr_j, \dots, \ZZ^\pr_{\ell})}
& \le  \abs{(\Y_1, \dots, \Y_{\ell}) - (\Y_1, \dots, \Y_j, \ZZ_{j+1}, \dots, \ZZ_{\ell})}\\
& +  \abs{(\Y_1, \dots, \Y_j, \ZZ_{j+1}, \dots, \ZZ_{\ell}) - (\Y_1, \dots, \Y_{j-1}, \ZZ^\pr_j, \dots, \ZZ^\pr_{\ell})}\\
& \le (\ell-j)\gamma + \gamma\\
& \le (\ell-j+1)\gamma\\
\end{align*}

We now prove that the second condition of the inductive hypothesis for $j$ is satisfied.

When $(\Y_1, \dots, \Y_{j-1})\not\in S$, for all $w\in \N$ with $j\le w\le \ell$, $\ZZ^\pr_w$ will be independent and uniform and so will have entropy at least $n - \Delta - \log(1/\gamma)$ conditioned on all fixings of blocks before it.

When $(\Y_1, \dots, \Y_{j-1})\in S$, then the conditional distribution $(\ZZ^\pr_j, \dots, \ZZ^\pr_w)$ equals the conditional distribution $(\Y_j, \ZZ_{j+1}, \dots, \ZZ_{\ell})$.
By the definition of $S$, $\ZZ^\pr_j$ will have min-entropy at least $n - \Delta - \log(1/\gamma)$. Moreover, for all $w\in \N$ with $j+1\le w\le \ell$, by the inductive hypothesis, on every fixing of $(\Y_j, \ZZ_{j+1}, \ZZ_{w-1})$, we have that $\ZZ_w$ will have min-entropy at least $n - \Delta - \log(1/\gamma)$. 
Thus, on every fixing of $(\ZZ^\pr_j, \dots, \ZZ^\pr_{w-1})$, we have that $\ZZ^\pr_w$ will have min-entropy at least $n - \Delta - \log(1/\gamma)$. 

Hence, all $3$ conditions are satisfied and the inductive step is proven.
\end{proof}

Lastly, we provide the proof that no non-trivial condensers exist for arbitrary $(n, k)$-sources.

\begin{proof}[Proof of \cref{lem:can't condense arbitrary n ; k source}]
Let $N = 2^n, M = 2^m, K = 2^k$.
We identify $\zo^n, \zo^m$ with $[N], [M]$ respectively.
We prove that for $m\le n$, $\sminH(f(\X)) \le k + m - n + \log(1 / (1-\eps))$.
If $m > n$ then we observe that $\abs{\supp(f)} \le N$. So, we relabel $f$ so that it's co-domain is $\zo^n$, apply the mentioned claim, and  infer that $\sminH(f(\X)) \le k + \log(1 / (1-\eps))$.
Hence, it suffices to prove the said claim for $m\le n$.

For $z\in [M]$, let $\chi(z) = \{x\in [N]: f(x) = z\}$ and $w(z) = \abs{\chi(z)}$.
Without loss of generality, let $w(1)\ge w(2)\ge \dots \ge w(M)$.
For $i\in [M]$, let $S_i = \sum_{j=1}^i w(j)$.
Let $i^*\in [M]$ be the smallest integer such that $S_{i^*} \ge K$.
Let $r = K - S_{i^* - 1}$.
Let $A = \cup_{j=1}^{i^* - 1} \chi(j)$.
Moreover, add arbitrary $r$ elements from $\chi(i^*)$ into $A$.
Let $\X$ be the $(n, k)$ source that is uniform over the set $A$.
Let $k_{out} = k + m - n + \log(1 / (1-\eps))$.
Then, we claim that $\sminH(f(\X)) \le k_{out}$.
As $w(1) \ge \dots \ge w(M)$, it must be that for all $1\le j\le M: \frac{S_j}{j} \ge \frac{S_M}{M} = \frac{N}{M}$.
Hence, $S_j \ge \frac{jN}{M}$.
In particular, if $j \ge \frac{KM}{N}$, then $S_j \ge K$.
As $i^*$ is the smallest integer such that $S_{i^*}\ge K$, it must be that $i^*\le \frac{KM}{N}$.
Hence, with probability $1$, $f(\X)\in [KM / N]$.
Applying \cref{lem:TV-dist lower bound}, we infer that $\sminH(f(\X)) \le k + m - n + \log(1 / \eps)$ as desired.
\end{proof}

\subsubsection{Impossibility of condensing beyond rate $1/2$}

Using condensing impossibility result for \uniSHELAs[1, 2], we prove a condensing impossibility result for \CG[\ell, \ell] (which are just CG sources, with no adversarial blocks) where the good blocks have min-entropy at least $O(n/\ell)$ conditioned on every fixing of previous blocks.

\begin{theorem}\label{thm:can't condense CG sources}
For all $0 < \eps < 1$ there exists a $\delta > 0$ such that the following holds:
for every function $f: (\zo^n)^{\ell}\to \zo^m$, there exists a \CG[\ell, \ell] $\X$ where the good blocks have min-entropy at least $\frac{n - \ell\log(2\ell / \eps)}{\ell+1}$ conditioned on all fixings of previous blocks and $\sminH(f(\X)) \le \frac{1}{2}\cdot m + \delta$.
\end{theorem}

\begin{proof}
Let $\X = (\X_1, \X_2)$ be an arbitrary \uniSHELA[1, 2] where the length of the blocks is at least $\frac{\ell(\ell+1)}{2}\cdot \left(\frac{n - \ell\log(2\ell / \eps)}{\ell+1} + \log(2\ell / \eps)\right)$. We transform $\X$ into a source $\Y = (\Y_1, \dots, \Y_{\ell})$ with block lengths $n$ that is $\eps/2$-close to an \CG[\ell, \ell].
We will then apply \cref{lem:(1:l) impossible} with $\ell=2$ and error $\eps/2$ to infer that it is impossible to condense such \CGs[\ell, \ell] so that the output distribution is $\eps$-close to having min-entropy more than $\frac{1}{2}\cdot m + \delta$.  

Let $\gamma = \frac{\eps}{2\ell}, t_2 = \frac{n - \ell\log(1 / \gamma)}{\ell+1}, t_1 = t_2 + \log(1 / \gamma)$. Define each block $\Y_i$ as follows:
\[
\Y_i = \X_2[(i-1)\cdot t_2 + 1, \dots, (i)\cdot t_2]\circ \X_1[(i)(i-1)/2\cdot t_1 + 1, \dots, (i)(i+1)/2\cdot t_1]\circ 0^{n - (t_2 + i\cdot t_1)}
\]
We claim that $\Y$ is $\eps/2$-close to a CG source where each good block has min-entropy at least $t_2$ for every fixing of the previous blocks.
We consider cases on the location of the good block in $\X$.
\begin{casesenum}
\item Block $\X_2$ is good.\\
As $\X$ is a \uniSHELA, $\X_1$ and $\X_2$ are independent. As $\X_2$ is uniform, this implies each sub-source from $\X_2$ is also uniform conditioned on every fixing of all other bits in $\X$.
Hence, for all $i$, and $(y_1, \dots, y_{i-1})\in (\zo^{n})^{i-1}:\minH\left(\Y_i \mid (\Y_1, \dots, \Y_{i-1}) = (y_1, \dots, y_{i-1})\right) \ge t_2$.

\item Block $\X_1$ is good.\\
We will use the following claim that most fixings of previous blocks preserve min-entropy:
\begin{claim}\label{claim: CG-impossibility-helpful-claim-fixing-previous-blocks-leaves-min-entropy}
For all $1\le i\le \ell$: with probability at least $1 - \gamma$ over fixings of $\Y_1, \dots, \Y_{i-1}: \minH(\Y_i) \ge t_2$  
\end{claim}

We will prove this claim later.
% Using it, we will inductively prove that for all $1\le i\le \ell$, $(\Y_1, \dots, \Y_i)$ is $(i-1)\cdot \gamma$ close to a block source where each block has min-entropy $t_2$ conditioned on every fixing of the previous blocks.
% For the base case, $j = 1$ and the claim is trivially true.
% For the inductive step, assume the claim is true for $j-1 \ge 1$; we will prove the claim for $j$. By inductive hypothesis, there exists a block source $(\ZZ_1, \dots, \ZZ_{j-1})$ such that $\abs{(\ZZ_1, \dots, \ZZ_{j-1}) - (\Y_1, \dots, \Y_{j-1})} \le (j-2)\gamma$. By \cref{claim: CG-impossibility-helpful-claim-fixing-previous-blocks-leaves-min-entropy}, with probability $1 - \gamma$ over fixings of $(\Y_1, \dots, \Y_{j-1})$, $\Y_j$ will have min-entropy at least $t_2$. By triangle inequality, with probability at least $1 - (j-1)\gamma$ over fixings of $\ZZ_1, \dots, \ZZ_{j-1}$, $\minH(\Y_j) \ge t_2$. Let $\ZZ_j$ be the source which equals $\Y_j$ conditioned on these good fixings of $\ZZ_1, \dots, \ZZ_{j-1}$ and equal uniform distribution otherwise. Then, $\abs{\ZZ_1, \dots, \ZZ_{j-1}, \ZZ_j}$

Using it, we will inductively to prove that for all $j$ starting from $\ell$ down to $1$, there exist blocks $\ZZ_j, \dots, \ZZ_{\ell}$ such that:
\begin{enumerate}
    \item 
    $(\Y_1, \dots, \Y_{\ell}) \approx_{(\ell-j+1)\gamma} (\Y_1, \dots, \Y_{j-1}, \ZZ_j, \dots, \ZZ_{\ell})$.
    \item 
    For all $t\in \N$ such that $j\le t\le \ell$, conditioned on every fixing of $\Y_1, \dots, \Y_{j-1}, \ZZ_j, \dots, \ZZ_{t-1}$: the min-entropy of $\ZZ_t$ is at least $t_2$.
\end{enumerate}
The overall claim exactly corresponds to the inductive hypothesis for $j = 1$ and hence, it suffices to prove this.

For the base case of $j = \ell$, proceed as follows: 
Applying \cref{claim: CG-impossibility-helpful-claim-fixing-previous-blocks-leaves-min-entropy}, with probability at least $1 - \gamma$ over fixings of $\Y_1, \dots, \Y_{\ell-1}$, $\Y_{\ell}$ will have conditional min-entropy at least $t_2$. Let these good fixings of $\Y_1, \dots, \Y_{\ell-1}$ be $S$.
Define the distribution $\ZZ_{\ell}$ as being same as the conditional distribution $\Y_{\ell}$ when $(\Y_1, \dots, \Y_{\ell-1})\in S$ and equals $\U_n$ otherwise.
Then,
\[
(\Y_1, \dots, \Y_{\ell}) \approx_{\gamma} (\Y_1, \dots, \Y_{\ell-1}, \ZZ_{\ell})
\]
Moreover, conditioned on every fixing of $(\Y_1, \dots, \Y_{\ell-1})$: $\ZZ_{\ell}$ will have entropy at least $t_2$.
Hence, both conditions are satisfied and the base case is proven.\\

For the inductive step, assume the claim holds for $j+1 \le \ell$. We prove the claim for $j$.
Applying \cref{claim: CG-impossibility-helpful-claim-fixing-previous-blocks-leaves-min-entropy}, with probability at least $\gamma$ over fixings of $\Y_1, \dots, \Y_{j-1}$, $\Y_{j}$ will have conditional min-entropy at least $t_2$. Let these good fixings of $\Y_1, \dots, \Y_{j-1}$ be $S$.
By the inductive hypothesis, there exists blocks $\ZZ_{j+1}, \dots, \ZZ_{\ell}$ that satisfy both conditions laid out in the inductive hypothesis.
Define the distribution $(\ZZ^\pr_{j}, \dots, \ZZ^\pr_{\ell})$ as being same as the conditional distribution of $(\Y_{j}, \ZZ_{j+1}, \dots, \ZZ_{\ell})$ when $(\Y_1, \dots, \Y_{j-1})\in S$ and equal to $(\U_n)^{\ell-j+1}$ otherwise.
Then,
\[
(\Y_1, \dots, \Y_j, \ZZ_{j+1}, \dots, \ZZ_{\ell}) \approx_{\gamma} (\Y_1, \dots, \Y_{j-1}, \ZZ^\pr_j, \dots, \ZZ_{\ell})
\]
Hence,
\begin{align*}
\abs{(\Y_1, \dots, \Y_{\ell}) - (\Y_1, \dots, \Y_{j-1}, \ZZ^\pr_j, \dots, \ZZ^\pr_{\ell})}
& \le  \abs{(\Y_1, \dots, \Y_{\ell}) - (\Y_1, \dots, \Y_j, \ZZ_{j+1}, \dots, \ZZ_{\ell})}\\
& +  \abs{(\Y_1, \dots, \Y_j, \ZZ_{j+1}, \dots, \ZZ_{\ell}) - (\Y_1, \dots, \Y_{j-1}, \ZZ^\pr_j, \dots, \ZZ^\pr_{\ell})}\\
& \le (\ell-j)\gamma + \gamma\\
& \le (\ell-j+1)\gamma\\
\end{align*}
We now prove that the second condition of the inductive hypothesis for $j$ is satisfied.

When $(\Y_1, \dots, \Y_{j-1})\not\in S$, for all $t\in \N$ with $j\le t\le \ell$, $\ZZ^\pr_t$ will be independent and uniform and hence will have entropy at least $t_2$ conditioned on all fixings of blocks before it.

When $(\Y_1, \dots, \Y_{j-1})\in S$, then the conditional distribution $(\ZZ^\pr_j, \dots, \ZZ^\pr_t)$ equals the conditional distribution $(\Y_j, \ZZ_{j+1}, \dots, \ZZ_{\ell})$.
By the definition of $S$, $\Y_j$ and hence $\ZZ^\pr_j$ will have min-entropy at least $t_2$. Moreover, for all $t\in \N$ with $j+1\le t\le \ell$, by the inductive hypothesis, on every fixing of $(\Y_j, \ZZ_{j+1}, \ZZ_{t-1})$, we have that $\ZZ_t$ will have min-entropy at least $t_2$. 
Equivalently, on every fixing of $(\ZZ^\pr_j, \dots, \ZZ^\pr_{t-1})$, we have that $\ZZ^\pr_t$ will have min-entropy at least $t_2$. 

Hence, both conditions are satisfied and the inductive step is proven.

\end{casesenum}
\end{proof}

Lastly we prove our claim that most fixings of previous blocks preserve min-entropy in the later block.
\begin{proof}[Proof of \cref{claim: CG-impossibility-helpful-claim-fixing-previous-blocks-leaves-min-entropy}]
Let $s = \frac{i(i-1)}{2}\cdot t_1$.
As $\X_1$ is uniform, $\X_1[s+1, \dots, s + i\cdot t_1]$ remains uniform conditioned on every fixing $\alpha$ of $\X_1[1, \dots, s]$.
By the min-entropy chain rule (\cref{lem:min-entropy-chain-rule}), with probability at least $1 - \gamma$ over fixings $\beta$ of $\X_2[1, \dots, i\cdot t_2]$ and every fixing $\alpha\in \zo^s$:
\begin{align*}
& \minH\left(\X_1[s+1, \dots, s + i\cdot t_1] \mid \X_2[1, \dots, (i-1)\cdot t_2] = \beta, \X_1[1, \dots, s] = \alpha\right)\\
& \ge i\cdot t_1 - ((i-1)\cdot t_2 + \log(1/\gamma))\\
& \ge t_2\\
\end{align*}
By construction, fixing $\X_2[1, \dots, (i-1)\cdot t_2]$ and $\X_1[1, \dots, s]$ fixes $\Y_1, \dots, \Y_{i-1}$.
For every fixing of these blocks, $\minH(\Y_i) \ge \minH(\X_1[s+1, \dots, s + i\cdot t_1])$ and the claim follows.
\end{proof}

\subsection{Deferred proofs of helpful lemmas}\label{subsec:impossibility-deferred-proofs}

The remaining deferred proofs of lemmas follow from the following results:

\begin{lemma}\label{lem:divide blocks cannot condense above fixed rate}
Let $g, \ell, n, g^\pr, \ell^\pr, n^\pr, m\in \N$ be such that $g\le a\cdot g^\pr + \max(b - (\ell^\pr - g^\pr), 0), (a+1)n \le n^\pr$ where $a, b\in \N$ are unique integers such that $\ell = a\cdot \ell^\pr + b$ where $0 \le b < \ell^\pr$.
Let $0 < \eps < 1, \delta > 0$ be such that:
for any function $f:(\zo^{n^\pr})^{\ell'}\to\zo^{m}$, there exists a \uniSHELA[g^\pr, \ell^\pr] (\uniNOSF[g^\pr, \ell^\pr], respectively) $\Y$ so that $\sminH(f(\Y)) \le \frac{g^\pr}{\ell^\pr}\cdot m + \delta$.
Then, for any function $h:(\zo^n)^{\ell}\to\zo^{m}$, there exists a \uniSHELA[g, \ell] (\uniNOSF[g, \ell], respectively) $\X$ such that $\sminH(h(\X))\leq\frac{g^\pr}{\ell^\pr}\cdot m + \delta$.
\end{lemma}

\begin{proof}[Proof of \cref{lem:divide blocks cannot condense above fixed rate}]
    For the sake of contradiction, assume there exists a non-trivial condenser $h:(\zo^n)^{\ell}\to\zo^{m}$ such that for any \uniSHELA[g, \ell] (\uniNOSF[g, \ell], respectively) $\X$, $\sminH(h(\mathbf{X}))\geq \frac{g^\pr}{\ell'}\cdot m + \delta$. We will use $h$ to construct a condenser $f:(\zo^{n^\pr})^{\ell'}\to\zo^{m}$ for \uniSHELA[g^\pr, \ell^\pr] (\uniNOSFs[g^\pr, \ell^\pr], respectively) to get a contradiction.

    Define $f:(\zo^{n^\pr})^{\ell^\pr}\to\zo^m$ as: $f(\mathbf{Y}_1,\dots,\mathbf{Y}_{\ell^\pr}) = h(\mathbf{X}_1,\dots,\mathbf{X}_{\ell=a\ell^\pr + b})$ where the $\mathbf{X}_i$ are constructed by splitting up the $\mathbf{Y}_j$ as evenly as possible.
    Concretely, from each of the $b$ blocks $\Y_1,\dots,\Y_b$, construct $a+1$ blocks of length $n$ each to form $b(a+1)$ blocks of the $\X_i$'s. Furthermore, from each of the $\ell'-b$ remaining blocks $\Y_{b+1},\dots,\Y_\ell^\pr$ construct $a$ blocks of length $n$ each to form $(\ell^\pr-b)a$ blocks of the $\X_i$'s. In total, we constructed $b(a+1) + (\ell^\pr-b)a = a\ell^\pr + b = \ell$ blocks of $\X$ as desired. As $\Y$ contains at least $g^\pr$ good blocks, $\X$ will contain $a\cdot g^\pr$ good blocks if $g^\pr \le \ell^\pr - b$ and at least $a\cdot g^\pr + g^\pr - (b - \ell^\pr)$ otherwise. By assumed constraints on $g$, we infer that $\X$ will contain at least $g$ good blocks as desired.
    We indeed check that this construction preserves one sidedness of the bad blocks --- if $\X$ is a \uniSHELA source, then $\Y$ is also a \uniSHELA source.
    Hence, $\X$ is a \uniSHELA[g, \ell] (\uniNOSF[g, \ell], respectively). Thus, $\sminH(f(\Y) \ge \sminH(h(\X)) \ge \frac{g^\pr}{\ell^\pr}\cdot m + \delta$, a contradiction.
    % and so, $\X$ is a \uniSHELA[g, \ell]. Thus, $\sminH(f(\Y) \ge \sminH(h(\X)) \ge \frac{1}{\ell^\pr}\cdot m + \delta$, a contradiction.
\end{proof}

The deferred proofs of couple of lemmas follow from this result.

\begin{proof}[Proof of \cref{lem:can't condense SHELA above any upper limit 1/l'}]
We apply \cref{lem:divide blocks cannot condense above fixed rate} with $g^\pr = 1$ to infer the claim.
\end{proof}

\begin{proof}[Proof of \cref{lem:can't condense NOSF above any upper limit l'-1/l'}]
We apply \cref{lem:divide blocks cannot condense above fixed rate} with $g^\pr = \ell^\pr-1$ to infer the claim.
\end{proof}

\begin{proof}[Proof of \cref{lem:impossibility scales by arbitrary constant}]
We apply \cref{lem:divide blocks cannot condense above fixed rate} with $g = ag^\pr, \ell = a\ell^\pr$ to infer the claim.
\end{proof}

\dobib

\section{Condensers for oNOSF Sources}

We will prove the following main theorem regarding condensing from \SHELAs in this section:

\begin{theorem}\label{thm:Condensing from low entropy (g;l)-SHELA with g<=l/2}
    For all $g, \ell, r\in \N, \eps>0$ such that $\floor{\frac{\ell-1}{g-1}} = r$ and $r < \frac{\ell-1}{g-1}$, there exists a condenser $\Cond:(\zo^n)^\ell\to\zo^m$ such that for any \SHELA[g,\ell,n,k] $\X$ with $k\geq 2\log(gn/\eps)$, we have that $\sminH(\X)\geq \frac{1}{r}\cdot m - 2(5^{\ell-g}-1)\log\left(\frac{(g-1)k}{8\ell\eps}\right)$ with $m=r\left(\frac{k}{8\ell}-2(5^{\ell-g}-1)\log\left(\frac{(g-1)k}{8\ell}\right)\right)$.
\end{theorem}

This result is tight up to lower order terms as it asymptotically matches the impossibility results of \cref{thm:can't condense SHELA above 1/c}.

We prove this theorem in two steps. First, we show how to transform \SHELAs to \uniSHELAs:

\begin{theorem}\label{thm:existential low min entropy shela to uniform shela}
For any $g, \ell, \eps$, there exists a function $f:(\zo^n)^\ell\to(\zo^m)^{\ell-1}$ with $m=\frac{k}{8\ell}$ such that for any \SHELA[g, \ell, k] $\X$ with $k\geq 2\log(gn/\eps)$ there exists a \uniSHELA[g-1,\ell-1] $\Y$ such that $\abs{f(\X)-\Y}\leq \eps$.
\end{theorem}

Second, we show how to condense from \uniSHELAs.

\begin{theorem}\label{thm:Condensing from uniform (g;l)-SHELA with g<=l/2}
For any $g, \ell, \eps$ such that $\floor{\ell/g}=r$ and $r < \ell/g$, there exists a condenser $\Cond:(\zo^n)^\ell\to\zo^m$ such that for any \uniSHELA[g,\ell] $\X$ we have $\sminH(\Cond(\X))\geq\frac{1}{r}\cdot m - 2(5^{\ell-g}-1)\log(gn/\eps)$ where $m = r(n-2(5^{\ell-g}-1)\log(gn))$.
\end{theorem}

Using these two ingredients, our main theorem follows:

\begin{proof}[Proof of \cref{thm:Condensing from low entropy (g;l)-SHELA with g<=l/2}]
    % We will follow a similar proof as that of  \cref{thm:can condense low min entropy shela with g > l/2+1}.
    Take the transformation function $f$ from \cref{thm:existential low min entropy shela to uniform shela} and let $\mathbf{X}'=f(\X)$ be the resulting source that is $\eps/2$ close to a \uniSHELA[g'=g-1,\ell'=\ell-1,n'=\frac{k}{8\ell}] source.

    By assumption, we have $\floor{\ell'/g'}=r$ and $r < \ell'/g'$. Consequently, we can apply \cref{thm:Condensing from uniform (g;l)-SHELA with g<=l/2} to get a condenser $\Cond':(\zo)^{n'})^{\ell'}\to\zo^m$ where $m=r(n'-2(5^{\ell'-g'}-1)\log(g'n'))$ such that $\minH^{\varepsilon_2}(\Cond(\X'))\geq \frac{1}{r}\cdot m - 2(5^{\ell'-g'}-1)\log(g'n'/\eps_2)$ with $\varepsilon_2=\eps/2$. These expressions simplify to $m=r\left(\frac{k}{8\ell}-2(5^{\ell-g}-1)\log\left(\frac{(g-1)k}{8\ell}\right)\right)$ and $\minH^{\varepsilon_2}(\Cond(\X'))\geq m - 2(5^{\ell-g}-1)\log\left(\frac{(g-1)k}{8\ell\eps_2}\right)$.

    Finally, we put these two steps together to define $\Cond:(\zo^n)^\ell\to\zo^m$ as $\Cond(\X):=\Cond'(f(\X))$ so that $\sminH(\X)\geq m-(5^{\ell-g}-1)\log\left(\frac{(g-1)k}{8\ell\eps}\right)$.
\end{proof}

\subsection{Transforming low entropy oNOSF sources to uniform oNOSF sources}

% Since we know that it is possible to condense from \uniSHELAs[g, \ell] with $g>\ell/2$, one would reasonably ask whether it's possible to condense from \uniSHELAs[g, \ell] where the good blocks are no longer uniform, but rather have only some min-entropy guarantee. We show that this is possible when $g>\ell/2+1$ by

We will prove \cref{thm:existential low min entropy shela to uniform shela} in this subsection. We will use the fact that a random function is a very good  two source extractor. 

\begin{lemma}\label{lem:random-function-two-source-extractor}
Let $n_1, n_2, k_1, k_2, m, \eps$ be such that $k_1 \le n_1, k_2\le n_2, m = k_1 + k_2 - 2\log(1/\eps) - O(1)$, $k_2 \ge \log(n_1 - k_1) + 2\log(1/\eps) + O(1)$, and $k_1 \ge \log(n_2 - k_2) + 2\log(1/\eps) + O(1)$.
Then, a random function $\Ext:\zo^{n_1}\times \zo^{n_2}\to \zo^m$ is a $(k_1, k_2, \eps)$-two source extractor with probability $1 - o(1)$.
\end{lemma}

We defer proof of this to \cref{subsec:proof random function output light two source extractor}.
% We will utilize the well known result that every two-source extractor is also a strong two-source extractor by an argument due to Boaz Barak included in \cite{rao_exposition_2007}. We rephrase it here in the form that we will apply it.
% \begin{lemma}\label{lem:two source extractor is strong}
% Let $n_1, n_2, m, k_1, k_2, k_2', \eps$ be such that $k_1\le n_1, k_2\le k'_2\le n_2$.
% If $2\Ext:\zo^{n_1}\times\zo^{n_2}\to\zo^m$ is a $(k_1,k_2,\varepsilon)$-two-source extractor, then it is a $(k_1,k_2',2^m(\varepsilon+2^{k_2-k_2'}))$ two-source extractor that is strong in the second argument.
% \end{lemma}
% We will focus on strongness in the second argument since that is all that we will need. 
Using this, we will prove our main lemma:

\begin{lemma}\label{lem:low entropy SHELA to uniform SHELA}
    Let $g, \ell, m, n\in \N$ and $k, k_1, k_2, \eps > 0$ be such that $k \ge k_1 + \ell m + \log(1 / \eps), k\ge k_2$.
    Suppose there exists a $(k_1,k_2,\varepsilon)$-two-source extractor $2\Ext:\zo^{(\ell-1)\cdot n}\times\zo^{n}\to\zo^m$. Then we can construct a function $f:(\zo^n)^\ell\to(\zo^m)^{\ell-1}$ such that for any \SHELA[g, \ell, n, k] $\X$, there exists a \uniSHELA[g-1, \ell-1] $\Y$ such that $\abs{f(\X)-\Y}\leq 2(g-1)\eps$.
\end{lemma}

Using this main lemma, \cref{thm:existential low min entropy shela to uniform shela} follows:

\begin{proof}[Proof of \cref{thm:existential low min entropy shela to uniform shela}]
    Applying \cref{lem:random-function-two-source-extractor} with $k_1=k-m\ell-\log(1/\varepsilon)$, $k_2=k$, and $\eps_{\Ext} = \eps / 2(g-1)$, we get that there exists a two-source extractor $2\Ext:\zo^{n_1}\times\zo^{n_2}\to\zo^m$ with these parameters.
    
    % \cref{lem:two source extractor is strong} then gives us that $2\Ext$ is a strong, in the second source, $(k_1,k_2',\varepsilon')$-two-source extractor where $k_2'=2k_2$ and $\varepsilon'=2^m(\varepsilon+2^{k_2-k_2'})=2^{-29k/48}+2^{-k/240}=2^{-\Omega(k)}$.
    Using this two-source extractor as input to \cref{lem:low entropy SHELA to uniform SHELA} gives us the desired result.\footnote{We note that we have not fully optimized our parameters here from \cref{lem:random-function-two-source-extractor}, and it is possible to get $k'_2=(1+\gamma)k_2$ for some $\gamma>0$ at the expense of other constants.}
\end{proof}

One can get an explicit version of this transformation, with polynomial error by using an explicit two-source extractor, such as the one from \cite{chattopadhyay_explicit_2019}.

We now focus on proving \cref{lem:low entropy SHELA to uniform SHELA}. We extend an argument for a somewhere extractor for low entropy \SHELAs from \cite{aggarwal_how_2020}. We do this by using a two source extractor instead of a seeded extractor in their construction.

To achieve this result, we use the notion of \emph{average conditional min-entropy} and use some known results about two-source extractors.

% OK I'm just gonna put a bunch of definitions and lemmas here and we can figure out where they go after.
\begin{definition}
    For any two distributions $\X$ and $\W$, define the \emph{average conditional min-entropy} of $\X$ given $\W$ as 
    \begin{align*}
        \avgcondminH(\X\mid\W)&=-\log\left(\E_{w\sim \W}\left[\max_{x\in\Supp(\X)}\Pr[\X=x\mid\W=w]\right]\right).
    \end{align*}
\end{definition}

We use this notion of average conditional min-entropy to define notions of average-case strongness in two-source extractors:
% Let's first recall from \cref{def:two-source extractor} that if $2\Ext:\zo^{n_1}\times\zo^{n_2}\to\zo^m$ is a $(k_1,k_2,\varepsilon)$-two-source extractor, then it is said to be strong in its first argument if $ 2\Ext(\X_1,\X_2),\X_1\approx_\varepsilon \U_{m}, \X_1$.
% We now define average-case strong two-source extractors:
\begin{definition}
We say that $2\Ext$ is \emph{average-case} strong if 
\[
2\Ext(\X_1,\X_2),\W \approx_\varepsilon \U_m,\W
\]
for every $\X_1$ and $\W$ such that $\avgcondminH(\X_1\mid\W)\geq k_1$ with $\X_2$ independent of $\X_1$ and $\W$.
\end{definition}

One benefit of the average conditional min-entropy in comparison to conditional min-entropy is that the chain rule is simpler:
\begin{lemma}\cite{dodis_fuzzy_2008}\label{lem:avg cond min H chain rule}
    Let $\mathbf{A}$, $\mathbf{B}$, and $\mathbf{C}$ be distributions such that $\Supp(\mathbf{B})\leq 2^\lambda$. Then $\avgcondminH(\mathbf{A}\mid \mathbf{B},\mathbf{C})\geq\avgcondminH(\mathbf{A},\mathbf{B}\mid\mathbf{C})-\lambda\geq\avgcondminH(\mathbf{A}\mid\mathbf{C})-\lambda$.
\end{lemma}

In addition, Lemma 2.3 of \cite{dodis_fuzzy_2008} shows that two-source extractors are average-case-two-source extractors with similar parameters.
\begin{lemma}\cite{dodis_fuzzy_2008}\label{lem:strong two source is average case strong}
    For any $\eta>0$, if $2\Ext$ is a $(k_1,k_2,\varepsilon)$-two-source extractor, then $2\Ext$ is a $(k_1+\log(1/\eta),k_2,\varepsilon+\eta))$-average-case-two-source extractor.
\end{lemma}
% \begin{proof}
%     \Huge if time add an explicit proof here for the appendix
% \end{proof}

% Finally, we introduce a useful fact about average-case strong seeded extractors due to \cite{aggarwal_how_2020} that we slightly generalize to average-case strong two source extractors.
% \begin{lemma}\cite{aggarwal_how_2020}\label{lem:avg case strong two source property}
%     % Let $n_1, n_1, k_1, k_2, m, \eps$ be such that $k_2 \ge k_2'$.
%     Let $2\Ext:\zo^{n_1}\times\zo^{n_2}\to\zo^m$ be a $(k_1,k_2,\varepsilon)$-two-source extractor that is average-case strong in the second argument. If $\W$ is a random variable such that $\avgcondminH(\X_1\mid\W)\geq k_1$ and $\X_2'$ is a $(n_2,k_2')$-source independent of $\X_1$ and $\W$, then 
%     \[
%     2\Ext(\X_1,\X_2'),\X_2',\W\approx_{\varepsilon\cdot 2^{k_2-k_2'}}\U_m,\X_2',\W.
%     \]
% \end{lemma}
% We will prove this lemma later.
We will use it to prove our main theorem in which we provide a general transformation of low min-entropy \SHELAs to \uniSHELAs given a two-source extractor. This transformation is based on a similar transformation in \cite{aggarwal_how_2020}.

\begin{proof}[Proof of \cref{lem:low entropy SHELA to uniform SHELA}]
    For $i\in\{2,\dots,\ell\}$, let $2\Ext_i:(\zo^n)^{i-1}\times\zo^n\to\zo^m$ be defined as $2\Ext_i(x, y) = 2\Ext(x\circ 0^{(\ell-i)n}, y)$. Then, $2\Ext_i$ is a $(k_1, k_2, \eps)$ 
    % strong (in the second argument) 
    two source extractor.
    Using \cref{lem:strong two source is average case strong} with $\eta=\varepsilon$ gives us that $2\Ext_i$ is an average-case $(\overline{k}_1,k_2,\overline{\varepsilon})$-two-source extractor with $\overline{k}_1=k_1+\log(1/\varepsilon)$ and $\overline{\varepsilon}=2\varepsilon$. By assumption, we have that $k\geq k_1+\ell m+\log(1/\varepsilon)=\overline{k}_1+\ell m$, so $k-\ell m\geq\overline{k}_1$. 

    Next, write $\X$ as $\X=\X_1,\X_2,\dots,\X_\ell$ with the $g$ good blocks at indices $G_1,\dots,G_g$.
    Let $G = \{G_2, \dots, G_g\}$.
    For $2\le r\le \ell$, define $G_{\le r} = \{v\in G: v \le r\}$.
    % For notational convenience, we also write $\mathbf{O}_{2:i}=\mathbf{O}_2,\dots,\mathbf{O}_i$ and similarly $\X_{1:i}=\X_1,\dots,\X_i$.
    We define $f(\X)$ as $f(\X) = \mathbf{O} = (\mathbf{O}_2,\dots,\mathbf{O}_{\ell})$ where our $\ell-1$ output blocks are defined as $\mathbf{O}_i:=2\Ext_i(\X_{1:i-1},\X_i)\in\zo^m$.
    We will show that
    \begin{align}
         \mathbf{O}_2,\dots,\mathbf{O}_{\ell} \approx_{2(g-1)\eps} \Y_2, \dots, \Y_l\label{eq:low entropy to uniform shela triangle ineq result}
    \end{align}
    where $\Y = (\Y_2, \dots, \Y_{\ell})$ is a \uniCG[g, \ell] with good blocks at indices $G_2, \dots, G_g$.
    By \cref{prop:uniSHELA equivalent to uniCG}, we infer that $\Y$ is also a \uniSHELA[g, \ell], proving our claim.

    To show \cref{eq:low entropy to uniform shela triangle ineq result}, we will use a hybrid argument.
    For $1\le i\le \ell$, let $\Y^{(i)} = (\mathbf{O}_2, \dots, \mathbf{O}_i, \Y_{i+1}, \dots, \Y_{\ell})$.
    Note that $\Y^{(1)} = \mathbf{O}$ and $\Y^{\ell} = \Y$.
    We will maintain the property that for all $2\le r\le \ell$ and all $j \in G$ with $j \le r$, it holds that $\Y^{(r)}_j$ is uniform conditioned on $\Y^r_{1: (j-1)}$.
    Furthermore, we will show that $\Y^{(r)}\approx_{\abs{G_{\le r}}\eps} \mathbf{O}$.
    We still haven't specified $\Y$, we will do it as we go along in the proof.
    
    % Second, to fulfill the $\uniSHELA$ requirements, we must show that $\mathbf{O}_{G_r}$ for $r\in\{2,\dots,g\}$ is still close to uniform even when conditioned on all previous blocks. That is, we must show
    % \begin{align}
    %     \mathbf{O}_1,\dots,\mathbf{O}_{G_r-1},\mathbf{O}_{G_r}\approx_{2\eps}\mathbf{O}_1,\dots,\mathbf{O}_{G_r-1},\U_m.\label{eq:low entropy to uniform shela triangle ineq goal}
    % \end{align}
    % Notice that by applying the triangle inequality $g-1$ times to \cref{eq:low entropy to uniform shela triangle ineq goal}, we also achieve \cref{eq:low entropy to uniform shela triangle ineq result}. Consequently, we now focus on proving \cref{eq:low entropy to uniform shela triangle ineq goal}.
    
    % We would like to apply \cref{lem:avg case strong two source property} to get \cref{eq:low entropy to uniform shela triangle ineq goal}, so we must show that $\avgcondminH(\X_{1:I_r-1}\mid\mathbf{O}_{2:I_r-1})\geq\overline{k}_1$.
    We now proceed by induction on $r$ where $2\le r\le \ell$.
    If $r\not\in G$, then we let $\Y_r = \mathbf{O}_r$ and observe that the property of uniformity of good blocks conditioned on all previous blocks still holds and the distance property is still maintained.
    If $r\in G$, then we compute:
    \begin{align*}
        \avgcondminH(\X_{1:r-1}\mid\Y^{(r)}_{2:r-1})
        &\geq\avgcondminH(\X_{G_1}\mid\Y^{(r)}_{2:G_r-1})\\
        &\geq\avgcondminH(\X_{G_1}) - \ell m &\text{By \cref{lem:avg cond min H chain rule}}\\
        &\geq k-\ell m. &\text{By assumption}\\
        &\geq (k_1+\ell m+\log(1/\varepsilon))-\ell m &\text{By assumption}\\
        &=k_1+\log(1/\varepsilon)\\
        &=\overline{k}_1,
    \end{align*}
    We observe that $\X_{G_r}$ is independent of $\X_{1:G_r-1}$ and hence, from $\mathbf{O}_{2:G_r-1}$ and $\Y^{(r)}_{2:G_r-1}$.
    So, 
    \[
        \mathbf{O}_r \mid\Y^{(r)}_{2:r-1} = \Ext_{G_r}(\X_{1:r-1}, \X_{r})\mid\Y^{(r)}_{2:r-1} \approx_{2\eps} \U_m
    \]
    Hence, 
    \[
        (\Y^{(r)}_2, \dots, \Y^{(r)}_{r-1}, \mathbf{O}_r) \approx_{2\eps}(\Y^{(r)}_2, \dots, \Y^{(r)}_{r-1}, \U_m)
    \]
    Let $\Y^{(r)}_r = \U_m$.
    Then, by inductive assumption and triangle inequality, we infer that $\mathbf{R}\approx_{\abs{G_{\le r}}\eps} \Y^{(r)}$.
    This completes the inductive proof and \cref{eq:low entropy to uniform shela triangle ineq result} indeed holds.
\end{proof}

% We lastly prove our useful lemma regarding average case strong two-source extractors:

% \begin{proof}[Proof of \cref{lem:avg case strong two source property}]
%     Without loss of generality, assume that $\X_2'$ is a flat $k_2'$ source. Let $S\subseteq\zo^{n_2}$ be such that $S\supseteq\Supp(\X_2')$ and $\abs{S}=2^{k_2}$, and define $\X_2$ to be the flat source on $S$, so $\X_2$ is a $(n_2,k_2)$-source. Because $2\Ext$ is average-case strong in its second argument, we have that 
%     \begin{align*}
%         \abs{2\Ext(\X_1,\X_2),\X_2,\W - \U_m,\X_2,\W}&=\sum_{x_2\in\Supp(\X_2)}2^{-k_2}\abs{2\Ext(\X_1,x_2),\W - \U_m,\W}\\
%         &\leq\varepsilon.
%     \end{align*}
%     Therefore,
%     \begin{align*}
%         \varepsilon&\geq\sum_{x_2\in\Supp(\X_2)}2^{-k_2}\abs{2\Ext(\X_1,x_2),\W - \U_m,\W}\\
%         &\geq\sum_{x_2\in \Supp(\X_2')}2^{-k_2}\abs{2\Ext(\X_1,x_2),\W - \U_m,\W}\\
%         &=2^{k_2'-k_2}\sum_{x_2\in \Supp(\X_2')}2^{-k_2'}\abs{2\Ext(\X_1,x_2),\W - \U_m,\W}\\
%         &=2^{k_2'-k_2}\abs{\Ext(\X_1,\X_2'),\X_2',\W - \U_m,\X_2',\W}.
%     \end{align*}
%     Rearranging the last line gives us the desired inequality.
% \end{proof}

\subsection{Condensing from oNOSF sources using output-light two source extractors}

In this subsection, we will prove \cref{thm:Condensing from uniform (g;l)-SHELA with g<=l/2}.
To obtain the condenser, we will utilize two-source extractors which have an additional property that we call output-light.

Formally, we define output-light two source extractors as follows:
\begin{definition}[Output-light Two Source Extractor]\label{def:output-light}
Let $\Ext: \zo^{n_1} \times \zo^{n_2} \to \zo^m$ be a $(k_1, k_2,\eps)$-two source extractor.
Then, $\Ext$ is $R$-output-light if for every $z\in \zo^m$, it holds that $\abs{\{x\in \zo^{n_1} : \exists y\in \zo^{n_2} (\Ext(x, y) = z)\} } \le R$.
\end{definition}

We will show a random function is a  output-light two source extractor with strong parameters and we will use it with the following parameters:

\begin{lemma}\label{cor:great-output-light-two-source-extractors-exist}
Let $0 < \delta < 1, C \ge 4$ be arbitrary constants. 
Let $n_1$, $k_1$, $n_2$, $k_2$, $m$, $\eps_{\Ext}, \eps$ be such that $n_1$ is arbitrary, $n_2 = C\left(\log(n_1) + \log(1/\eps)\right), k_1 = \delta n_1 - 2n_2, k_2 = 4\left(\log(n_1) + \log(1/\eps)\right), m = k_1 - 2n_2, \eps_{\Ext} = 2^{-k_2/4}$ (note that if $k_2$ is larger than the minimum requirement, then $\eps_{\Ext}$ gets proportionally smaller).
Then, a random function $\Ext: \zo^{n_1}\times\zo^{n_2}\to \zo^m$ is an $R$-output-light $(k_1, k_2, \eps_{\Ext})$ two-source-extractor where $R = 2^{n_1 + n_2 - m + O(1)}$.
\end{lemma}

We defer proof of their existence in \cref{subsec:proof random function output light two source extractor}. Using such an extractor, we will prove the following general condensing result:

\begin{lemma}\label{lem:Condensing from (g;l)-SHELA with g<=l/2 given extractor}
    Let $g, \ell, r, n, \eps$ be such that $r = \floor{\ell/g}$ and $r < \ell/g$.
    Assume that for $c\in\{1,\dots,r\}$, there exists an $R_c$-output-light $(k_{1,c},k_{2,c},\varepsilon_{\Ext_c})$-two-source extractor $2\Ext_c:\zo^{n_{1,c}}\times\zo^{n_{2,c}}\to\zo^{m_c}$ where $n_{1,c} = gn$, $n_{2,c} = \frac{5^{\ell - cg} - 1}{4}\cdot 4\log(gn/\eps)$, $k_{1,c} = n - 2n_{2, c}$, $k_{2,c} = 4\left(\log (gn) + \log(1/\eps)\right)$,  $m_c = n - 2n_{2,c}, \varepsilon_{\Ext_c} = 2^{-k_{2,c}/4}$ and $\log(R_c/\eps)\le n_{1,c} + 2n_{2,c} - m_c$. Then there exists a condenser $\Cond:(\zo^n)^\ell\to\zo^m$ such that for any \uniSHELA[g,\ell] $\X$, we have $\sminH(\Cond(\X))\geq \frac{1}{r}\cdot m - 2n_{2,1}$ here $m = r\cdot m_r$.
\end{lemma}

Using this main lemma, the theorem follows:

\begin{proof}[Proof of \cref{thm:Condensing from uniform (g;l)-SHELA with g<=l/2}]
We plug in the result of \cref{cor:great-output-light-two-source-extractors-exist} to \cref{lem:Condensing from (g;l)-SHELA with g<=l/2 given extractor} to infer our claim.
\end{proof}

Before we prove this main lemma, we prove \cref{thm:Condensing from uniform (g;l)-SHELA with g<=l/2} for the special case when $g > \ell / 2$.

\begin{theorem}\label{thm:can-condense-from-uniform-shela-above-rate-half}
For all $g, \ell, \eps$ such that $g > \ell / 2$, there exists a condenser $\Cond:(\zo^n)^{\ell}\to\zo^m$ such that for any \uniSHELA[g, \ell] $\X$, $\minH^{\eps}(\Cond(\X)) \ge m - (5^{\ell-g}-3)\left(\log(gn)+\log(1/\eps)\right)$ where $m = n - 2(5^{\ell-g}-1)\log(gn)$.
\end{theorem}

As an application of this theorem, we construct a condenser from a low min-entropy \SHELA[g, \ell] with $g > \ell/2 + 1$. We do this by composing our transformation from \cref{thm:existential low min entropy shela to uniform shela} with the condenser from \cref{thm:can-condense-from-uniform-shela-above-rate-half}.

\begin{corollary}\label{thm:can condense low min entropy shela with g > l/2+1}
    For all $g, \ell, \eps$ such that $g>\ell/2+1$, there exists a condenser $\Cond:(\zo^n)^\ell\to\zo^m$ such that for any \SHELA[g, \ell, n, k] $\X$ with $k\geq2\log(n)$, we have $\sminH(\Cond(\X))\geq m-(5^{\ell-g}-3)\log\left(\frac{(g-1)k}{8\ell\eps}\right)$ where $m=\frac{k}{8\ell}-2(5^{\ell-g}-1)\log\left(\frac{(g-1)k}{8\ell}\right)$.
\end{corollary}

\begin{proof}
    We begin by transforming $\X$ into a \uniSHELA[g'=g-1,\ell'=\ell-1,n'=\frac{k}{8\ell}] $\X'$ defined as $\X'=f(\X)$ by taking $f$ from \cref{thm:existential low min entropy shela to uniform shela}. In this step, we accumulate $\varepsilon_1=\eps/2$ error. 
    
    As $g' > \ell'/2$, we apply \cref{thm:can-condense-from-uniform-shela-above-rate-half} to get a condenser $\Cond':(\zo^{n'})^{\ell'}\to\zo^m$ such that $\minH^{\varepsilon_2}(\Cond(\X'))\geq m-(5^{\ell'-g'}-3)\log(g'n'/\eps)$ where $m=n'-2(5^{\ell'-g'}-1)\log(g'n'), \varepsilon_2=\eps/2$.
    Simplifying these expressions then yields $m=\frac{k}{8\ell}-2(5^{\ell-g}-1)\log\left(\frac{(g-1)k}{8\ell}\right)$ and $\minH^{\varepsilon_2}(\Cond(\X'))\geq m-(5^{\ell-g}-3)\log\left(\frac{(g-1)k}{8\ell\eps_2}\right)$.

    We finish by combining these two steps and defining $\Cond:(\zo^n)^\ell\to\zo^m$ as $\Cond(\X):=\Cond'(f(\X))$ so that $\sminH(\Cond(\X))\geq m-(5^{\ell-g}-3)\log\left(\frac{(g-1)k}{8\ell\eps}\right)$.
\end{proof}

\subsubsection{Condensing from \texorpdfstring{\SHELAs[g,\ell]}{(g,l)-oNOSF sources} with \texorpdfstring{$g>\ell/2$}{g>l/2}}

We will prove \cref{thm:can-condense-from-uniform-shela-above-rate-half} that allows us to condense from \uniSHELAs[g, \ell] when $g > \ell/2$. This theorem allows us to condense to almost full entropy.

We will prove this theorem using the following general lemma:

\begin{lemma}\label{lem:shela-condenser-from-output-light-two-source-extractor}
Assume that for some $g, n, \eps$ there exists an $R$-output-light $(k_1, k_2, \eps_{\Ext})$-two-source-extractor $\Ext: \zo^{n_1}\times \zo^{n_2}\to \zo^m$ where $n_1 = gn, n_2 = \frac{5^{\ell-g} - 1}{4}\cdot 4\log(gn/\eps), k_1 = n - 2n_2, k_2 = 4\log(gn/\eps), m = n - 2n_2, \eps_{\Ext} = 2^{-k_2/4}$ (notice that we require that if $k_2$ supplied is larger, then $\eps_{\Ext}$ gets proportionally smaller).
Then, there exists a condenser $\Cond:(\zo^n)^{\ell}\to\zo^m$ such that for any \uniSHELA[g, \ell] $\X$ with $g > \ell/2$, $\minH^{\eps}(\Cond(\X)) = \min\left(m - n_2, n_1 - \log(R / \eps)\right)$.
\end{lemma}

Using this, our theorem directly follows:

\begin{proof}[Proof of \cref{thm:can-condense-from-uniform-shela-above-rate-half}]
We use the output-light two source extractor guaranteed from \cref{cor:great-output-light-two-source-extractors-exist} in \cref{lem:shela-condenser-from-output-light-two-source-extractor} to get our result.
\end{proof}

Towards proving our general lemma, we show that for any flat distribution $\X$ over $n$ bits, if a function $f$ condenses from $\X$, then $f$ also condenses (with a slight loss in parameters) from a distribution $\X^\pr$ which is the same as the distribution $\X$ on most output bits but some output bits are arbitrarily controlled by an adversary. We note that a lemma similar in spirit to this one was shown as Lemma 28 in \cite{ben-aroya_two-source_2019}.

\begin{lemma}\label{lem:control-few-bits-can-still-condense}
Let $\X \sim\zo^n$ be an arbitrary flat distribution and let $\Cond: \zo^n\to\zo^m$ be such that $\sminH(\Cond(\X)) = k$.
Let $G\subset [n]$ with $|G| = n - b$. Let $\X_G\sim \zo^{n-b}$ be the projection of $\X$ onto $G$.
Let $\Xpr\sim \zo^n$ be the distribution where the output bits defined by $G$ equal $\X_G$ and remaining $b$ bits are deterministic functions of the $n-b$ bits defined by $G$ under the restriction that $\supp(\Xpr)\subset \supp(\X)$.
Then, $\minH^{\epspr}(f(\Xpr)) \ge k - b$ where $\epspr = \eps\cdot 2^b$.
\end{lemma}

We will prove this result later. Using this result, we use output-light two-source-extractor to prove our general lemma:

\begin{proof}[Proof of \cref{lem:shela-condenser-from-output-light-two-source-extractor}]
Let the input be $x = (x_1, \dots, x_{\ell})$.
For $g+1\le i\le \ell$, let $y_i$ be the first $5^{\ell-i}\cdot (4\log(gn/\eps))$ bits of $x_i$.
Let $z_1 = x_1\circ \dots \circ x_g, z_2 = y_{g+1}\circ \dots \circ y_{\ell}$.
Then, let $\Cond(x) = \Ext(z_1, z_2)$.
Let $\Y_i$ be the distribution of $y_i$ and let $\ZZ_i$ be the distribution of $z_i$.
We consider cases on the position of the adversary and show that for all such \SHELAs, the output will be condensed:
\begin{casesenum}
\item  At least one source out of $\X_{g+1}, \dots, \X_{\ell}$ is good.\label{case:condensing from g>l/2 uni SHELA - first g sources have at least one good}\\
Let this block be $\X_j$ (so $g+1\le j\le \ell$).
As $g > \ell/2$, at least one source out of $\X_1, \dots, \X_g$ is good and hence, $\minH(\ZZ_1) \ge n$.
Without loss of generality, we assume only these $2$ sources are good in $\X$ and remaining sources are bad.
Let $\A = \Y_{g+1}, \dots, \Y_{j-1}, \B = \Y_{j+1}, \dots, \Y_{\ell}$.
Then, $\ZZ_2 = \A\circ \Y_j\circ \B$.
As $\X_j$ is a good source and $\X$ is a \SHELA, $\X_j$ remains a uniform source conditioned on any fixing of $\A$, so $\Y_j$ does as well.
Also, by the min-entropy chain rule (\cref{lem:min-entropy-chain-rule}),  with probability $1-\eps/2$ over fixings of $\A$, $\minH(\ZZ_1) \ge n - n_2 - \log(2/\eps) \ge k_1$.

Consider $(\ZZ | \A = a)$ where $a$ is such a good fixing of $\A$. 
We will show that for all such good fixings $\minH^{\eps/2}\Ext(\ZZ) \ge m - n_2$.
By assumption, $\minH(\ZZ_1 | \A = a) \ge k_1$ and $\minH(\Y_j | \A = a) = \minH(\Y_j) = 5^{\ell-j}\cdot 4\log (gn/\eps)$.
Moreover, we can without loss of generality assume $\minH(\ZZ_1 | \A = a)$ is a flat source (we can express it as convex combination of such flat sources).
As $\X$ is a \SHELA, $(\ZZ_1 | \A = a)$ and $(\Y_j | \A = a)$ are independent distributions.
Assume for now that $(\B | \A = a)$ were uniform and independent of $\Y_j$ and $\A$.
Then, $(\ZZ_1 | \A = a)$ and $(\ZZ_2 | \A = a) = (a, \Y_j, \B | \A = a)$ will be independent sources with min-entropy at least $k_1$ and $k'_2 = \sum_{i=j}^{\ell} 5^{\ell-j}(4\log (gn/\eps)) = \frac{5^{\ell-j+1}-1}{4}\cdot 4\log(gn/\eps) \ge k_2$, respectively. Hence, $\Ext(\ZZ)$ will be $\eps_{\Ext}$ close to the uniform distribution over $m$ bits where $\eps_{\Ext} = 2^{-k'_2/4}$.
However, in reality, $\B$ might be arbitrarily controlled by an adversary and can depend on $\ZZ_1, \A, \Y_j$. The number of bits controlled by the adversary is $n_b = \sum_{i=j+1}^l 5^{\ell-i}\left(4\log(gn/\eps)\right) = \frac{5^{\ell-j}-1}{4}\cdot 4\log(gn/\eps)$. To overcome this, we apply \cref{lem:control-few-bits-can-still-condense} (using the fact that $(\ZZ_1, \A, \U_{k'_2} ) | \A = a$ is a flat distribution) and infer that $\minH^{\epspr}(\Ext(\ZZ)) \ge m - n_b$ where $\epspr = \eps_{\Ext}\cdot 2^{n_b} \le 2^{-k'_2/4 + n_b}$.
As $n_b \le n_2$, the output min-entropy is at least $m - n_b\ge m - n_2$.
Moreover, $\epspr \le 2^{-k'_2/4 + n_b}$ which is $2^{-\log(gn/\eps)} \le \eps/2$ if $j = \ell$ and which is $2^{-\log(gn/\eps)\cdot (5^{\ell-j}+3)/16}\le 2^{-\log(gn/\eps)/2} \le \eps/2$ if $j \le \ell-1$.

\item All of the sources $\X_{g+1}, \dots, \X_{\ell}$ are bad.\label{case:condensing from g>l/2 uni SHELA - first g sources are all good}\\
This implies sources $\X_1, \dots, \X_g$ are good.
Let $k_{out} = n_1 - \log(R/\eps)$.
Let $N_1 = 2^{n_1}, M = 2^m$, and $K_{out} = 2^{k_{out}}$.
Assume that there exists a \uniSHELA[g, \ell] $\X$ such that $\sminH(\Cond(\X)) < k_{out}$.
By \cref{claim:small smooth entropy implies heavy set}, there exists $H\subset \zo^m$ such that $\abs{H} < K_{out}$ and $\Pr[\Cond(\X)\in H] \ge \eps$.
This implies there exist $h\in H$ and $P\subset \zo^{n_1}$ with $\abs{P} > \frac{\eps N_1}{K_{out}} = R$ such that for all $z_1\in P$, there exists $z_2\in \zo^{n_2}$ so that $\Cond(z_1, z_2) = h$.
However, this contradicts the fact that $\Ext$ is $R$-output-light.
Hence, for all \uniSHELAs[g, \ell]$(g, \ell)$ $\X$, $\sminH(\Cond(\X)) \ge n_1 - \log(R/\eps)$.
\end{casesenum}
\end{proof}

We finally prove our useful lemma that states a condenser for a distribution $\X$ still condenses from a tampered version of $\X$ where some output bits are controlled by an adversary.

\begin{proof}[Proof of \cref{lem:control-few-bits-can-still-condense}]
We first claim that as $\X$ is a flat source, for all $x\in \supp(\Xpr)$, $\Pr[\Xpr = x]\le 2^b\cdot \Pr[\X = x]$.
Indeed, let $S_x = \{z\in \supp(\X): \textrm{$x$ and $z$ equal each other when restricted to bits in $G$}\}$.
Then, $|S_x| \le 2^b$.
Hence, 
\[
\Pr[\Xpr = x] \le \sum_{z\in S_x} \Pr[\X = z] = |S_x|\cdot \Pr[\X = x] \le 2^b\cdot \Pr[\X = x].
\]
We now proceed by contradiction and assume $\minH^{\epspr}(f(\Xpr)) < k - b$.
Let $O = \{y\in \zo^m: \Pr[f(\Xpr) = y] > 2^{k-b}\}$.
As $\minH^{\epspr}(f(\Xpr)) < k-b$, it must be that $\Pr[f(\Xpr)\in O] \ge \epspr + |O|\cdot 2^{b-k}$.
Let $I = \{x\in \supp(\Xpr): f(x)\in O\}$.
We now see that
\[
\Pr[f(\X)\in O] \ge \Pr[\X \in I] = \sum_{x\in I} \Pr[\X = x] \ge \sum_{x\in I} \Pr[\Xpr = x]\cdot 2^{-b} = (\epspr + |O|\cdot 2^{b-k})\cdot 2^{-b} = \eps + |O|\cdot 2^{-k}
\]
where the first inequality follows by our observation.
% By our observation above, for all $x\in I$, it holds that $\Pr[\X = x] \ge \Pr[\Xpr = x]\cdot 2^{-b}$.
% Summing over all $x\in I$, we infer that $\Pr[\X\in I] \ge \epspr\cdot 2^{-b} = \eps$.
For $y\in O$, let $I_y = \{x\in I: f(x) = y\}$.
We see that
\[
\Pr[f(\X) = y] \ge \Pr[\X \in I_y] = \sum_{x\in I_y} \Pr[\X = x] \ge \sum_{x\in I_y} \Pr[\Xpr = x]\cdot 2^{-b} > 2^{-k+b}\cdot 2^{-b} = 2^{-k}
\]
Hence, for all $y\in O$, $\Pr[f(\X) = y] > 2^{-k}$ and $\Pr[f(\X)\in O] \ge \eps + |O|\cdot 2^{-k}$.
These together imply $\sminH(f(\X)) < k$, a contradiction.
% By our observation above, for all $x\in I_y$, it holds that $\Pr[\X = x] \ge \Pr[\Xpr = x]\cdot 2^{-b}$.
% Summing over all $x\in I_y$, 

% By \cref{claim:small smooth entropy implies heavy set}, there exists $H^{\Xpr}_{out}\subset \supp(f(\Xpr))$ such that $|H^{\Xpr}_{out}| < 2^{k-b}$ and $\Pr[f(\Xpr)\in H^{\Xpr}_{out}] \ge \epspr = \eps\cdot 2^b$.
% Let $H^{\Xpr}_{in} = \{x\in \supp(\Xpr): f(x) \in H^{\Xpr}_{out}\}$.
% Let $H^G_{in}$ be the projection of $H^{\Xpr}_{in}$ onto the bits of $G$.
% As the bits outside $G$ in $\Xpr$ are deterministic functions of the bits in $G$, it must be that $|H^G_{in}| = |H^{\Xpr}_{in}|$.
% Let $H^{\X}_{in} = \{x\in \supp(\X): x_G\in H^G_{in}\}$ where $x_G\in \zo^{n-b}$ is the restriction of $x$ to the entries in $G$.
% Then, $|H^{\X}_{in}| \le H^G_{in}\cdot 2^b = H^{\Xpr}_{in}\cdot 2^b$
\end{proof}

\subsubsection{Condensing from uniform oNOSF sources in all regimes}

We finally prove our main lemma of the section - \cref{lem:Condensing from (g;l)-SHELA with g<=l/2 given extractor}. We will use the following simple claim that guarantees projections of high-entropy distributions have high-entropy.

\begin{lemma}\label{lem:removing d bits from source removes d bits of entropy}
    Let $\X$ be an arbitrary $(n,k)$-source and $\pi:\zo^n\to\zo^{n-d}$ be a projection onto $n-d$ bits of $\X$ (i.e., removes $d$ bits of $\X$). Then $\pi(\X)$ is a $(n-d,k-d)$-source.
\end{lemma}
\begin{proof}
    Because $\X$ is an $(n,k)$-source, for any $x\in\Supp(\X)$, we have that $\Pr[\X=x]\leq 2^{-k}$. Furthermore, for any $y\in \zo^{n-d}$, there are at most $2^d$ elements from $\supp(\X)$ that could map to $y$ under $\pi$. Thus, for any $y\in\Supp(\pi(\X))$, we can compute that
    \begin{align*}
        \Pr[\pi(\X)=y]&=\sum_{\substack{x\in\Supp(\X)\\\pi(x)=y}}\Pr[\X=x]\\
        &\leq\sum_{\substack{x\in\Supp(\X)\\\pi(x)=y}}2^{-k}\\
        &\leq 2^d\cdot 2^{-k}=2^{d-k}.
    \end{align*}
    Therefore, $\minH(\pi(\X))\geq k-d$, as required.
\end{proof}

We are finally ready to prove the main lemma. The proof of this main lemma uses a similar strategy as in \cref{lem:shela-condenser-from-output-light-two-source-extractor}.

\begin{proof}[Proof of \cref{lem:Condensing from (g;l)-SHELA with g<=l/2 given extractor}]
    We will proceed inductively on $r\in\N$ with the base case of $r=1$ taken care of by \cref{lem:shela-condenser-from-output-light-two-source-extractor}. For the inductive step, take \cref{lem:Condensing from (g;l)-SHELA with g<=l/2 given extractor} to be true for $r-1$. 

    We will output $r$ output blocks $\mathbf{O}_1, \dots, \mathbf{O}_r$ where each $\mathbf{O}_i\sim \zo^{m_r}$.
    We begin by defining our first output block $\mathbf{O}_1\in\zo^{m}$ by defining $\Y_i$ to be the distribution after $\X_i$ is projected onto its first $5^{\ell-i}\cdot 4\log(gn/\eps)$ bits and setting $\ZZ_1=\X_1,\dots,\X_g$ and $\ZZ_2 = \Y_{g+1},\dots,\Y_\ell$. We let $\mathbf{O}_1$ be the first $m_r$ bits of $2\Ext_1(\ZZ_1,\ZZ_2)$. 

    To define our last $r-1$ output blocks $\mathbf{O}_2,\dots,\mathbf{O}_r$, we pretend for the moment that $\X_1,\dots,\X_g$ are bad blocks of $\X$. Since $\floor{\ell/g}=r$ and $r < \ell/g$, it must be that $\floor{\frac{\ell-g}{g}} = r - 1$ and $r-1 < (\ell-g) / g$. Because $\X_1,\dots,\X_g$ are bad, we see that $\X_{g+1},\dots,\X_\ell$ is a $\uniSHELA[g,\ell-g]$ with $\floor{\frac{\ell-g}{g}}=r-1$ and $r-1 < \frac{\ell-g}{g}$, meaning that we can use existence of $2\Ext_c$ for $c\in\{2,\dots,r\}$ to apply our inductive hypothesis to get the output blocks $\mathbf{O}_2,\dots,\mathbf{O}_r$ with the property that $\sminH(\mathbf{O}_2,\dots,\mathbf{O}_r)\geq m_r - n_{2,2}$. These allow us to define $\Cond(\X)=\mathbf{O}_1,\dots,\mathbf{O}_r$. Of course, we do not necessarily immediately have that this is true, but it will hold in our last case in our case analysis:

    \begin{casesenum}
        \item At least one, but not all, of $\X_1,\dots,\X_g$ is good.
        Because not all of $\X_1,\dots,\X_g$ are good, it must be that at least one of $\X_{g+1},\dots,\X_{\ell}$ is good. Thus, we can use the exact calculations of \ref{case:condensing from g>l/2 uni SHELA - first g sources have at least one good} from \cref{lem:shela-condenser-from-output-light-two-source-extractor} to get that $\sminH(2\Ext_1(\ZZ_1,\ZZ_2)) \geq m_1-n_{2,1}$. Then, because $\mathbf{O}_1$ is just $2\Ext_1(\ZZ_1,\ZZ_2)$ truncated to its first $m_r$ bits, we use \cref{lem:removing d bits from source removes d bits of entropy} to get $\sminH(\mathbf{O}_1)\geq  m_1-n_{2,1}-(m_1-m_r)\ge m_r - 2n_{2,1}$.
        
        \item All of $\X_1,\dots,\X_g$ are good.

        In this case, we get that $\mathbf{O}_1$ is condensed by the $R_1$-output-lightness of $2\Ext_1$. We achieve this by using the exact calculations of \ref{case:condensing from g>l/2 uni SHELA - first g sources are all good} of \cref{lem:shela-condenser-from-output-light-two-source-extractor} to get that $\sminH(2\Ext_1(\ZZ_1,\ZZ_2))\geq n_{1,1}-\log(R_1/\varepsilon) \ge m_1 - 2n_{2, 1}$. We again conclude by using \cref{lem:removing d bits from source removes d bits of entropy} to get that $\sminH(\mathbf{O}_1)\geq m_r - 2n_{2, 1}$.

        \item All of $\X_1,\dots,\X_g$ are bad.

        In this last case, we do not get that $\mathbf{O}_1$ is condensed because $\ZZ_1$ can be arbitrarily bad. Instead, we have that $\X_{g+1},\dots,\X_\ell$ is a $\uniSHELA[g,\ell-g]$ with $\floor{\frac{\ell-g}{g}} = r-1$ and $\frac{\ell-g}{g} < r-1$, so by our inductive hypothesis it must be that $\sminH(\mathbf{O}_2,\dots,\mathbf{O}_r)\geq m_r - 2n_{2,2} \ge m_r - 2n_{2, 1}$. 
    \end{casesenum}

    In all cases, we get that $\sminH(\mathbf{O}_1)\geq m_r - 2n_{2,1}$. Thus, we can conclude that $\sminH(\Cond(\X))=\sminH(\mathbf{O}_1,\dots,\mathbf{O}_r)\geq m_r - n_{2,1} = \frac{1}{r}\cdot m - 2n_{2,1}$ as desired.
\end{proof}

\subsection{Existence of output-light two-source extractors}\label{subsec:proof random function output light two source extractor}

In this subsection, we show a random function is an output-light two-source extractor. Towards showing output lightness, we introduce a related notion, of $R$-invertible functions.

\begin{definition}[$R$-invertible function]\label{def:invertible-function}
A function $f: \zo^n\to \zo^m$ is $R$-invertible if for every $z\in \zo^m$, it holds that $\abs{\{x\in \zo^n : f(x) = z\}} \le R$.
\end{definition}

We record the observation that $R$-invertible functions are also $R$-output light.

\begin{observation}\label{observation:invertible-implies-output-light}
Let $\Ext: \zo^{n_1} \times \zo^{n_2} \to \zo^m$ be a $(k_1, k_2,\eps)$-two source extractor.
If $\Ext$ is $R$-invertible, then $\Ext$ is $R$-output-light.
\end{observation}

We now show that a random function is optimally invertible, hence concluding a random function is also output light.

\begin{lemma}\label{lem:random-function-invertible}
Let $f: \zo^n\rightarrow \zo^m$ be a random function where $m \le n - \log n$. Then, with probability $1-o(1)$, $f$ will be $2^{n - m + c}$-invertible where $c$ is a universal constant.
\end{lemma}

\begin{proof}[Proof of \cref{lem:random-function-invertible}]
Let $R = (1 + \delta) 2^{n - m}$ where $\delta > 0$ is a large constant.
For $z\in \zo^m$, let $E_z$ be the event that $\abs{\{x\in \zo^n: f(x) = z\}} > R$. 
Fix any such $z$.
Let $B_1, \dots, B_{2^n}$ be the events corresponding to whether $f(x) = z$.
Using \cref{claim:chernoff}, we infer that 
\[
\Pr\left[\sum_i B_i > R\right] \le \exp\left(-\frac{\delta^2}{2+\delta}\cdot 2^{n - m}\right)
\]
We union bound over all $z\in \zo^m$ to obtain that the probability that at least one $E_z$ holds is at most 
\[
\exp\left(-\frac{\delta^2}{2+\delta}\cdot 2^{n - m}\right)\cdot 2^m \le 
\exp\left(-\frac{\delta^2}{2+\delta}\cdot m + m\right) \le o(1)
\]
The claim follows.
\end{proof}

We now prove that a random function is a two source extractor with strong parameters:

\begin{proof}[Proof of \cref{lem:random-function-two-source-extractor}]
Let $N_1 = 2^{n_1}, N_2 = 2^{n_2}, K_1 = 2^{k_1}, K_2 = 2^{k_2}, M = 2^{m}$.
It suffices to show that a random function is a two source extractor where the two sources have min-entropies exactly $k_1$, and $k_2$, and are flat.
Using proposition 6.12 in \cite{vadhan_pseudorandomness_2012}, we infer that a random function $\Ext: \zo^{n_1 + n_2}\to \zo^m$ is a $(k_1+k_2, \eps)$ extractor with probability $1 - 2^{-cK_1K_2\eps^2}$ where $c > 0$ is some universal constant.
We union bound over all pairs of sources with min-entropies $k_1$, and $k_2$ out of $n_1$, and $n_2$ bits respectively. The probability that a random function $\Ext$ will not be a $(k_1, k_2, \eps)$-two-source-extractor is:
\begin{align*}
\binom{N_1}{K_1}\binom{N_2}{K_2}2^{-cK_1K_2\eps^2} 
& \le \left(\frac{eN_1}{K_1}\right)^{K_1}\left(\frac{eN_2}{K_2}\right)^{K_2}2^{-cK_1K_2\eps^2}\\
& \le 2^{K_1 \log\left(\frac{eN_1}{K_1}\right) + K_2\log\left(\frac{eN_2}{K_2}\right) - cK_1K_2\eps^2}\\
& = 2^{\log(e)K_1(n_1 - k_1) + \log(e)K_2(n_2 - k_2) - cK_1K_2\eps^2}\\
& \le o(1)
\end{align*}
where the last inequality follows because $k_2 > \log(n_1 - k_1) + 2\log(1/\eps) + O(1)$, and $k_1 > \log(n_2 - k_2) + 2\log(1/\eps) + O(1)$.
Hence, the claim follows.
\end{proof}

Finally, we show  output-light two-source extractors exist, as required for our constructions.

\begin{proof}[Proof of \cref{cor:great-output-light-two-source-extractors-exist}]
Combining \cref{lem:random-function-invertible} and \cref{lem:random-function-two-source-extractor} along with \cref{observation:invertible-implies-output-light}, we infer that   output-light two source extractors with the promised parameters exist.
\end{proof}

 \dobib

\section{Open Questions}

There are several natural questions that are raised by our work.   A few immediate open questions are:
\begin{enumerate}
 \item Explicitly construct output-light two-source extractor. This would immediately imply explicit   condensers for \SHELAs and \uniCGs by \cref{lem:Condensing from (g;l)-SHELA with g<=l/2 given extractor}.
  \item In our condensing possibility results for \uniSHELAs and \uniCGs in \cref{thm:overview-can-condense-from-uniform-shela-above-rate-half} and \cref{thm:overview-Condensing from uniform (g;l)-SHELA with g<=l/2},  and our possibility results for logarithmic min-entropy \SHELAs in \cref{thm:overview-Condensing from low entropy (g;l)-SHELA with g<=l/2}, we require $\ell=o(\log(n))$, that our block size to be much smaller than the total number of blocks. It would be interesting to extend these results to  smaller block sizes, such as the regime achieved for almost CG sources in \cite{doron_almost_2023}.
 \item Is it possible to improve our condenser for \uniCGs in \cref{thm:overview-can-condense-from-uniform-shela-above-rate-half} to have constant entropy gap? 
 \item Can our condensing impossibility result for  CG sources in \cref{thm:can't condense CG sources} be strengthened to close the gap with the results in \cite{doron_almost_2023}?
\end{enumerate} 
\dobib

\section{Acknowledgements}

We want to thank Ran Raz for illuminating discussions.

\dobib

%%% SECTIONS HERE %%%

\printbibliography

\appendix

\section{Explicit Condensers for oNOSF Sources}\label{sec:appendix-explicit condensers}

In this section, we provide explicit constructions of condensers for \uniSHELAs[g, \ell] for \uniSHELAs[2,3] directly in \cref{subsec:possibility_explicit} and for \uniSHELAs[6,9] along with various other settings of parameters via a recursive nesting method in \cref{sec:appendix-recursive condenser compositions}.

Recall that \cite{aggarwal_how_2020} showed that for all $\gamma>0$ there exists an $\ell$ large enough such that it is impossible to extract from \uniSHELAs[\floor{\gamma\ell},\ell] below error $\frac{1-\gamma}{48}$. One may then wonder whether the explicit condensers for \uniSHELAs[g,\ell] that we constructed in the previous couple subsections are for some ``easy'' case of small $g$ and $\ell$, such as uniform $(2,3)$ and $(6,9)$-oNOSF sources in \cref{thm:(2:3) SHELA explicit condense} and \cref{thm:(6;9)oNOSF explicit condenser}.  In \cref{sec:(2;3)-oNOSF extraction impossibility}, we dispel such worries by showing that one cannot extract from rate $2/3$ \uniNOSFs. 

\subsection{An explicit condenser for \texorpdfstring{\uniSHELAs[2,3]}{uniform (2,3)-oNOSF sources}} \label{subsec:possibility_explicit}

In this section, we construct a condenser for \uniSHELAs[2,3]. The following is our main result.
\begin{theorem}
\label{thm:(2:3) SHELA explicit condense}
There exists constant $0 < c_0 < 1$ such that for all $\eps > 2^{-c_0 n}$,
we can explicitly construct a condenser $\Cond:(\zo^n)^3\to\zo^m$, where $m = \frac{n}{16}$ such that for any \uniSHELA[2, 3] $\X$, $\sminH(\Cond(\X))\geq m - O(\log(m / \eps))$.
\end{theorem}

To prove this, we construct an explicit output-light seeded extractor (see \cref{def:output-light}) that works for somewhere-random sources. 
We observe that in the proof of \cref{lem:shela-condenser-from-output-light-two-source-extractor} for $g = 2, \ell = 3$, it suffices to construct an output-light seeded extractor instead of an output-light two-source extractor. And moreover, this output-lights seeded extractor need only extract from somewhere random sources.

We could in fact use use existing seeded extractors that are known to be invertible, such as Trevisan's extractor \cite{cook_explicit_2024}. However,  this requires seed length of $O(\log^2(n))$, which translates into the entropy gap of the condenser. For the specific case of somewhere random sources, we construct a better seeded extractor that has seed length $O(\log(n))$.

\begin{theorem}
\label{lem:(2:3) SHELA somewhere random input}
There exists constant $0 < c_0 < 1$ such that for all $\eps > 2^{-c_0 n}$, there exists a $R$-output-light strong linear seeded $\eps$-extractor $\Ext: \zo^{2n} \times \zo^d\rightarrow \zo^m$ for the class of distributions $\X = (\X_1, \X_2)$, each $\X_i$ being a random variable on $n$ bits and at least one of $\X_1$ or $\X_2$ is guaranteed to be uniform,  with $d = O(\log n / \eps), m = \frac{n}{16}$ and $R = \frac{2^{2n - m}}{\poly(m, 1/\eps)}$.
\end{theorem}

We note that this construction matches the probabilistic bounds (\cref{thm:can-condense-from-uniform-shela-above-rate-half}) as the $m$ bit output is condensed to entropy $m - O(\log(m))$ with $m = O(n)$. We also remark that we have not tried to optimize the constant appearing in the output length of the extractor.

\subsubsection{An explicit output-light seeded extractor for somewhere-random sources}

We prove \cref{lem:(2:3) SHELA somewhere random input} in this section and show

% As discussed in \cref{subsec:proof_overview_explicit}, our construction builds upon a linear-seeded extractor constructed by Li \cite{li2017improved}.  We first introduce some necessary objects for the construction.

\begin{algorithm}
    \caption{$\Ext$ (Output-light Somewhere-extractor)}\label{alg:explicit_output_light_ext}
    \SetKwInput{Input}{Input}
    \SetKwInput{Output}{Output}
    \SetKwProg{Fn}{Function}{}{end}
    \Input{source $X = (X_1, X_2)\in \zo^n\times \zo^n$, seed $S\in \zo^d$}
    \hrulealg
    Let $\Ext_{GUV}: \zo^{7n/4}\times \zo^d\rightarrow \zo^{n/2 - \log(1 / \eps_0)}$ be the GUV extractor from \cref{thm:GUV extractor} instantiated for entropy $3n/4$ and error $\eps_0 = \eps/4$.\\
    \hrulealg
    Let $U = X_1, V = X_2$.\\
    Let $n_1 = \frac{n}{4}, n_2 = \frac{7n}{4}$.\\
    Let $Y = \left(Y_1, Y_2\right)$ where $Y_1 = \left(U_{[1, n_1 / 2]}, V_{[1, n_1 / 2]}\right), Y_2 =  \left(U_{[(n_1/2) + 1, n]}, V_{[(n_1/2)+1, n]}\right)$.\\
    Let $R_2 = \Ext_{GUV}(Y_2, S)$.\\
    Let $R'_{2}$ be a length $n/4$ prefix of $R_2$ with last bit set to $1$.
    Let $R_1 \in \zo^{n / 16} = R'_2\cdot Y_1$ where the operation is done over the finite field $\F_{2^{n/16}}^4$.\\
    Output $R_1$.
\end{algorithm}

\begin{proof}[\textbf{Proof of \cref{lem:(2:3) SHELA somewhere random input}}] We claim that $\Ext$ computed by \cref{alg:explicit_output_light_ext} computes the desired extractor.

Let $\Y = (\Y_1, \Y_2)$ be the distribution of the variable $Y$ above.
Let $\RR_1, \RR_2, \RR'_2$ be the distribution of the variables $R_1, R_2, R'_2$ above.
We will show that $\Y$ is $\eps_0$ close to being a block source.
As either $\X_1$ or $\X_2$ is guaranteed to be uniform, $\minH(\Y_i) \ge \frac{n_i}{2}$.
By the min-entropy chain rule \cref{lem:min-entropy-chain-rule}, with probability $1 - \eps_0$ over fixings of $\Y_1 = \alpha$, it holds that $\minH(\Y_2\restriction {\Y_1 = \alpha}) \ge \frac{n_2}{2} - n_{1} - \log(1/\eps_0) = \frac{3n}{4} - \log(1 / \eps_0)$.
We will add $\eps_0$ to our total error and assume this property about $\Y$ from here on.
By property of $\Ext_{GUV}$, it holds that, $\abs{\RR_2 - \U_{|\RR_2|}} \le \eps_0$. 
We will add $\eps_0$ to our total error and assume $\RR_2$ is uniform from here on.
So, $\RR'_2$ is a distribution over $\zo^{n/4}$ with min entropy $\frac{n}{4} - 1$.
As $\Y_1\sim \zo^{n/4}$ is such that $\minH(\Y_1) \ge \frac{n}{8}$, by \cref{inner product two source}, it holds that $\abs{\RR_1 - \U_{|\RR_1|}} \le 2^{-n/32 + 1}$.
As $\Y$ is a block source, for each fixing $\alpha$ of $\Y_1$, it holds that:
\[
    \abs{\Ext_{GUV}(\Y_2, S) - \U_{|\RR_2|}} \le \eps_0
\]
Hence, it must be that 
\[
    \abs{(\Y_1, \Ext_{GUV}(\Y_2, S)) - (\Y_1, \U_{|\RR_2|})} \le \eps_0
\]
and thus, 
\[
    \abs{\RR_1 - \U_{|\RR_1|}} \le 2\eps_0 + 2^{-n/32 + 1} \le 3\eps_0,
\]
using the fact that $\eps \ge 2^{-c_0 n}$, for some small $c_0 > 0$.
The total error of the extractor on input $\X$ is thus bounded by  $4\eps_0 = \eps$, as desired.

We now prove that this extractor is indeed output-light.
For every fixing of the output $R_1$ of $\RR_1$, $\beta$ of $\Y_2$ and the seed $S$, we can uniquely recover $R'_2$. Given $\frac{3n}{16}$ bits corresponding to first three out of the $4$ intermediate outputs of the inner product, we can use $R_1$ to compute the fourth intermediate outer product and then use $R'_2$ to invert each of the products and recover $R_1$.
Thus for a fixed seed $S$ and output $R_1$, there can be at most $2^{3n/16 + 7n / 4} = 2^{31n / 16}$ such $x\in \zo^{2n}$ so that $\Ext(x, s) = z$.
As there are at most $2^d$ seeds, for a fixed output $R_1\in \zo^{n/16}$, $\abs{\left\{x\in \zo^{2n} : \exists y(\Ext(x, y)) = z\right\}} \le 2^{2n - n / 16 - \log(n/\eps)} = \frac{2^{2n - m}}{\poly(n, 1 / \eps)} = \frac{2^{2n - m}}{\poly(m, 1 / \eps)}$.
\end{proof}

\subsection{Recursive condenser compositions}\label{sec:appendix-recursive condenser compositions}
By composing the explicit condenser from \cref{thm:(2:3) SHELA explicit condense}, we can get explicit condensers for other values of $g$ and $\ell$ as well. We present an explicit computation of parameters for the case of \uniSHELAs[6,9] in \cref{sec:appendix-(6;9) condenser explicit} and sketch of the general recursive composition in \cref{sec:appendix-general recursive composition sketch}.

\subsubsection{An explicit condenser for \texorpdfstring{\uniSHELAs[6,9]}{uniform (6,9)-oNOSF sources}}\label{sec:appendix-(6;9) condenser explicit}
We can take our condenser for \uniSHELAs[2,3] even further to create a condenser for \uniSHELAs[6,9] by nesting it within itself.
\begin{theorem}\label{thm:(6;9)oNOSF explicit condenser}
    There exists a constant $0<c<1$ such that for all $\varepsilon>2^{-cn+2}$, we can explicitly construct a condenser $\Cond:(\zo^n)^{3^2}\to\zo^m$ where $m=\frac{n}{16^2}$ such that for any \uniSHELA[\frac{2}{3}3^2,2^2] $\X$, we have $\sminH(\Cond(\X))\geq m- O(\log(m/\varepsilon)$.
\end{theorem}
\begin{proof}
    To create a condenser for \uniSHELAs[\frac{2}{3}3^2,2^2], we will apply our condenser from \cref{thm:(2:3) SHELA explicit condense} in a nested fashion. Recall that \cref{thm:(2:3) SHELA explicit condense} states that there exists a constant $0<c_1<1$ such that for all $\varepsilon_1>2^{-c_1n}$, we can explicitly construct a condenser $\Cond_1:(\zo^n)^{3}\to\zo^{m_1}$, where $m_1=\frac{n}{16}$ such that for any \uniSHELA[6,9] $\X$, $\sminH(\Cond_1(\X))\geq m_1-O(\log(m_1/\varepsilon_1))$. 

    Let $X$ be a \uniSHELA[6,9]. We will apply $\Cond_1$ on the first, second, and last third of $\X$ to get $\ZZ_1=\Cond_1(\X_1,\X_2,\X_3)$, $\ZZ_2=\Cond_1(\X_4,\X_5,\X_6)$, and $\ZZ_3=\Cond_1(\X_7,\X_8,\X_9)$. To define $\Cond_2:(\zo^n)^9\to\zo^{m_2}$, we again apply $\Cond_1$ to get $\Cond_2(\X)=\Cond_1(\ZZ_1,\ZZ_2,\ZZ_3)$. Now, we analyze the result of this construction.

    We begin by noticing that the output length of $\Cond_2$ is $m_2=\frac{t_1}{16}=\frac{n}{16^2}$ since each of $\ZZ_1$, $\ZZ_2$, and $\ZZ_3$ is on $\zo^{m_{1}}$. Next, because $\X$ is a \uniSHELA[6,9], it has at most $3$ bad blocks. Consequently, if $\Cond_1$ did not work for one of $\ZZ_1$, $\ZZ_2$, or $\ZZ_3$ (i.e., its conditions were not satisfied because 2 or 3 of its given blocks were bad), call this output $\ZZ_k$, then there is at most 1 bad block left over as inputs to $\Cond_1$ for $\ZZ_i$ and $\ZZ_j$ where $i,j\neq k$ and $i\neq j$. Thus, the conditions for $\Cond_1$ are met in the creation of $\ZZ_i$ and $\ZZ_j$ so $\minH^{\varepsilon_1}(\ZZ_i),\minH^{\varepsilon_1}(\ZZ_j)\geq m_1-O(\log(m_1/\varepsilon_1))$.

    By accumulating $2\varepsilon_1$ error, we can consider $\ZZ_i$ and $\ZZ_j$ as having min entropy at least $m_1-O(\log(m_1/\varepsilon_1))$. Applying \cref{lem:CGL15 strong extractor bad seed} allows us to consider $\ZZ_i$ and $\ZZ_j$ as having full min entropy in our application of $\Cond_1(\ZZ)$ by accumulating $2\cdot2^{O(\log(m_1/\varepsilon_1)}=O(\poly(m_1/\varepsilon_1))=\left(\frac{m_1}{\varepsilon_1}\right)^p=:\gamma$ error for some exponent $p\geq1$. Finally, in our application of \cref{thm:(2:3) SHELA explicit condense} in $\Cond_1(\ZZ)$, we set $\varepsilon_2=\left(\frac{m_1}{\varepsilon_1}\right)^{-2p}\gamma+2\varepsilon_1=\left(\frac{m_1}{\varepsilon_1}\right)^{-p}+2\varepsilon_1$. Using that $\varepsilon_1>2^{-c_1n}$, we get that $\left(\frac{m_1}{\varepsilon_1}\right)^{-p}>\left(\frac{n}{16}\right)^{-p}2^{-c_1np}$. If we take $n>16$ then we have $2^{-c_1np}>\left(\frac{n}{16}\right)^{-p}2^{-c_1np}$. Thus, we require that $\varepsilon_2=\left(\frac{m_1}{\varepsilon_1}\right)^{-p}+2\varepsilon_1>2^{-c_1np}+2^{-c_1n+1}>2^{-c_1n+2}$ since $2^{-c_1np}\leq2^{-c_1n+1}$. Setting $\varepsilon>2^{-c_1n+2}$ and $c=c_1$ in the theorem statement gives us our desired result.

\end{proof}
\subsubsection{General recursive composition}\label{sec:appendix-general recursive composition sketch}
At the expense of shorter output length and larger error, we can generalize our explicit recursive composition from \cref{thm:(6;9)oNOSF explicit condenser} to any odd $\ell$. We give a proof sketch of this composition here. We first state a simple corollary from \cref{thm:(6;9)oNOSF explicit condenser}.
\begin{cor}\label{cor:ignore blocks to explicitely condense from (l-1;l)-oNOSF}
    There exists constant $0 < c_0 < 1$ such that for all $\eps > 2^{-c_0 n}$, we can explicitly construct a condenser $\Cond:(\zo^n)^\ell\to\zo^m$, where $m = \frac{n}{16}$ such that for any \uniSHELA[\ell-1,\ell] $\X$ with $\ell\geq3$, we have $\sminH(\Cond(\X))\geq m - O(\log(m / \eps))$.
\end{cor}
\begin{proof}
    We simply apply the condenser from \cref{thm:(6;9)oNOSF explicit condenser} to $\X_1,\X_2,\X_3$ and infer the result because at most one of $\X_1,\X_2,\X_3$ can be bad, so $\X_1,\X_2,\X_3$ is a \uniSHELA[2,3].
\end{proof}
Next, we give a sketch for what happens when we compose two condensers in a nested manner.
\begin{lemma}\label{lem:two condenser composition}
    For $i\in\{1,2\}$, say there exists a condenser $\Cond_i:(\zo^{n_i})^{\ell_i}\to\zo^{m_i}$ for \uniSHELAs[g_i,\ell_i] with $m_i=f_i(n_i)$, entropy gap $\Delta_i=O(\log(m_i/\varepsilon_i))$, and error $\varepsilon_i=2^{-\Omega(n_i)}$. Let $b_i=\ell_i-g_i$. Then there exists a condenser $\Cond:(\zo^n)^\ell\to\zo^m$ for any \uniSHELA[g=\ell-b,\ell] $\X$ where $\ell=\ell_1\cdot\ell_2$, $b=(b_1+1)(b_2+1)-1$, $m=\max(f_1(f_2(n)),f_2(f_1(n)))$, and error $\varepsilon=2^{-\Omega(n)}$ such that $\sminH(\Cond(\X))\geq m-O(\log(m/\varepsilon))$.
\end{lemma}
\begin{proof}
    We will first consider defining $\Cond$ by nesting $\Cond_2$ within $\Cond_1$, although it will turn out that the number of bad blocks that $\Cond$ can handle is independent from the order that we choose to to nest $\Cond_1$ and $\Cond_2$. 

    Because $\X$ has $\ell=\ell_1\cdot\ell_2$ blocks, we can split $\X$ up into $\ell_1$ chunks $\Y_1,\dots,\Y_{\ell_1}$ each with $\ell_2$ blocks from $\X$ in it. Then, we apply $\Cond_2$ to each of these chunks to get $\ZZ_j=\Cond_2(\Y_j)$ for $j\in\{1,\dots,\ell_1\}$. Finally, we define $\Cond(\X):=\Cond_1(\ZZ_1,\dots,\ZZ_{\ell_1})$. We claim that this construction gives us the desired result.

    To compute the number of bad blocks $b$ that $\Cond$ can handle, we will think adversarially as to the fewest number of blocks that are required to break our construction. To make the output of $\Cond_1$ fail, we require that at least $b_1+1$ of $\ZZ_1,\dots,\ZZ_{\ell_1}$ be bad. For a single $\ZZ_j$ to be bad --- that is, for $\Cond_2$ to fail --- we require that at least $b_2+1$ of the blocks of $\X$ used for $\ZZ_j$ be bad. Thus, to make the output of $\Cond$ fail, we require at least $(b_1+1)(b_2+1)$ bad blocks in $\X$. Conversely, this means that $\Cond$ can handle at most $b=(b_1+1)(b_2+1)-1$ bad blocks. In other words, $\Cond$ requires at least $g=\ell-b=\ell-(b_1+1)(b_2+1)+1$ good blocks to succeed. Notice that this equation is symmetric in $b_1$ and $b_2$, demonstrating that the order of composition of $\Cond_1$ and $\Cond_2$ does not matter in computing $b$. 

    Next, we focus on the output length $m$. Each $\ZZ_j$ will be on $\zo^{f_2(n)}$, so $\Cond(\X)$ is on $\zo^{f_1(f_2(n))}$. Since the order of composition of $\Cond_1$ and $\Cond_2$ does not matter for $b$, we can take the optimal order to get $m=\max(f_1(f_2(n)),f_2(f_1(n)))$.

    Finally, to compute $\varepsilon$ and our final min-entropy gap, we defer to the proof of \cref{thm:(6;9)oNOSF explicit condenser} since those computations follow in a similar manner.
\end{proof}
We remark that \cref{thm:(6;9)oNOSF explicit condenser} is an explicitly computed version of \cref{lem:two condenser composition} applied to the explicit \uniSHELA[2,3] condenser from \cref{thm:(2:3) SHELA explicit condense}.

To finish our generalization, we give a sketch for how one could recursively apply \cref{lem:two condenser composition}. For the sake of succinctness, we define a useful prime counting function.
\begin{definition}
    For any $\ell\in\N$, let $\varphi(\ell)$ be the total number of prime factors of $\ell$ counting multiplicity. That is, if $\ell=p_1^{a_1}p_2^{a_2}\cdots p_n^{a_n}$, then $\varphi(\ell)=\sum_{i=1}^na_i$.\footnote{In number theory, this is usually denoted by the prime Omega function $\Omega(\ell)$, but we don't this notation here to avoid confusion with asymptotics.} We use this to define
    \begin{align*}
        \Phi(\ell):=
        \begin{cases}
            \varphi(\ell) & \ell\text{ odd}\\
            \varphi(\ell-1) & \ell\text{ even}
        \end{cases}.
    \end{align*}
\end{definition}
Essentially, $\Phi(\ell)$ returns $\varphi(\ell)$ if $\ell$ is odd and $\varphi(\ell-1)$ if $\ell$ is even. We now state our main theorem.
\begin{theorem}\label{thm:general condenser composition}
    Let $\ell\geq3$.\footnote{Note that we consider $\ell$ as a constant and $\Phi_1(\ell)\leq\ell$, so we can consider $\Phi_1(\ell)$ as a constant as well.} Then there exists an explicit condenser $\Cond:(\zo^n)^\ell\to\zo^m$ for any \uniSHELA[\ell-2^{\Phi(\ell)}+1,\ell] $\X$ such that $\sminH(\Cond(\X))\geq m-O(\log(m/\varepsilon))$ where $m=\frac{n}{16^{\Phi(\ell)}}$ and $\varepsilon=2^{-\Omega(n)}$.
\end{theorem}
\begin{proof}
    Without loss of generality, we take $\ell$ to be odd. If $\ell$ is even, we truncate $\X$ to its first $\ell-1$ blocks. Since $\ell-1$ is odd, meaning that $\Phi(\ell-1)=\varphi(\ell-1)$, we can use our result for the odd case (which we prove below) to get that we can explicitly condense from \uniSHELAs[(\ell-1)-2^{\varphi(\ell-1)}+1,\ell+1]. Thus, since we may be removing a good block when we truncate the last block of $\X$, this means that we can condense from \uniSHELAs[((\ell-1)-2^{\varphi(\ell-1)}+1)+1,\ell] which simplifies to \uniSHELAs[\ell-2^{\Phi(\ell)}+1,\ell], matching our claim.
    
    Factor $\ell$ as $\ell=\ell_1\cdots\ell_{\varphi(\ell)}$ and note that each factor is at least 3 since $\ell$ is odd. By \cref{cor:ignore blocks to explicitely condense from (l-1;l)-oNOSF}, we know that there exists an explicit condenser $\Cond_i$ for \uniSHELAs[\ell_i-1,\ell_i] for $i\in[\varphi(\ell)]$. Let $b_i=\ell_i-1$ and $b'_1=b_1$. Nesting $\Cond_1$ in $\Cond_2$ by \cref{lem:two condenser composition} gives us a new explicit condenser for \uniSHELAs[\ell_1\cdot \ell_2-b_2',\ell_1\cdot \ell_2] that can handle $b_2'=(b_2+1)(b_1'+1)-1$ bad blocks. 

    Repeatedly applying \cref{lem:two condenser composition} gives us explicit condensers for \uniSHELAs[\ell'_i-b_i',\ell_i'] where $\ell_i'=\ell_1\cdot\ell_i$ and $b_i'=(b_i+1)(b_{i-1}'+1)-1$. Taking $i=\varphi(\ell)$ then gives us our desired condenser $\Cond$ for $\ell=\ell'_{\varphi(\ell)}$ blocks that can handle at most $b_{\varphi(\ell)}'=2^{\varphi(\ell)}-1$ bad blocks, meaning $\Cond$ is a condenser for \uniSHELAs[\ell-2^{\varphi(\ell)}+1,\ell], as desired. 

    The output lengths $m=\frac{n}{16^{\varphi(\ell)}}$ of $\Cond$ follows because each application of \cref{lem:two condenser composition} divides the output length by 16 due to our construction in \cref{cor:ignore blocks to explicitely condense from (l-1;l)-oNOSF}. Furthermore, the final error and entropy gap of $\Cond$ again follow similarly from the explicit computations in \cref{thm:(6;9)oNOSF explicit condenser}. 
\end{proof}

As we have done previously in \cref{thm:can condense low min entropy shela with g > l/2+1}, we can prepend our function that transforms low min-entropy \SHELAs to \uniSHELAs from \cref{lem:low entropy SHELA to uniform SHELA} to get a corollary of \cref{thm:general condenser composition} but for low min-entropy \SHELAs. We do note that to use \cref{lem:low entropy SHELA to uniform SHELA} explicitly in this way, we require using a two-source extractor from \cite{chattopadhyay_explicit_2019,li_improved_2016} that has polynomial error which ultimately gives us polynomial instead of exponential error as we have in \cref{thm:general condenser composition}.

\begin{cor}\label{cor:explicit low min entropy oNOSF condenser}
    Let $\ell>3$. Then there exists an explicit condenser $\Cond:(\zo^n)^\ell\to\zo^m$ for any \SHELA[\ell-2^{\Phi(\ell-1)}+1,\ell,n,k] $\X$ with $k\geq \Omega(\log^C(n))$ for some large enough constant $C$,  such that $\sminH(\Cond(\X))\geq m-O(\log(m/\varepsilon))$ where $m=\Omega(\poly(k))$ and $\varepsilon=1/\Omega(\poly(k))$.
\end{cor}
\begin{proof}
    We instantiate \cref{lem:low entropy SHELA to uniform SHELA} with the explicit two-source extractor from \cite{li_improved_2016} which achieves polynomially small error, has output length $\poly(k)$, and can handle min-entropy at least $k\geq \Omega(\log^C(n))$. Prepending this transformation to \cref{thm:general condenser composition} gives us our desired result.
\end{proof}

For a cleaner statement of \cref{thm:general condenser composition}, we can truncate our input to a power of 3 instead.
\begin{cor}
    Let $\ell\geq3$ and take the unique $a,r\in\N$ such that $\ell=3^a+r$ and  $r<2\cdot3^a$. Then there exists an explicit condenser $\Cond:(\zo^n)^\ell\to\zo^m$ for any \uniSHELA[\ell-2^a+1,\ell] $\X$ such that $\sminH(\Cond(\X))\geq m-O(\log(m/\varepsilon))$ where $m=\frac{n}{16^{a}}$ and $\varepsilon=2^{-\Omega(n)}$.
\end{cor}
\begin{proof}
    Truncate $\X$ to its first $3^a$ blocks and use that as input to \cref{thm:general condenser composition} where $\Phi(3^a)=a$. This gives us a condensing possibility result for \uniSHELAs[(\ell-r)+2^a+1,\ell-r]. Since we may remove $r$ good blocks when we truncate $\X$, our result holds for \uniSHELAs[((\ell-r)+2^a+1)+r,\ell], which simplifies to \uniSHELAs[\ell+2^a+1,\ell].
\end{proof}
If we only take powers of 3, then we get:
\begin{cor}
    Let $\ell=3^a$ for some $a\in\N$. Then there exists an explicit condenser $\Cond:(\zo^n)^\ell\to\zo^m$ for any \uniSHELA[3^a-2^a+1,3^a] $\X$ such that $\sminH(\Cond(\X))\geq m-O(\log(m/\varepsilon))$ where $m=\frac{n}{16^{a}}$ and $\varepsilon=2^{-\Omega(n)}$.
\end{cor}
As always, we get similar versions for low min-entropy \SHELAs. Here, we get them as corollaries from \cref{cor:explicit low min entropy oNOSF condenser}.
\begin{cor}
    Let $\ell>3$ and take the unique $a,r\in\N$ such that $\ell=(3^a+1)+r$ and  $r<2\cdot 3^a+1$. Then there exists an explicit condenser $\Cond:(\zo^n)^\ell\to\zo^m$ for any \SHELA[\ell-2^a+1,\ell,n,k] $\X$ with $k\geq\Omega(\log^C(n))$ for some large enough constant $C$, such that $\sminH(\Cond(\X))\geq m-O(\log(m/\varepsilon))$ where $m=\Omega(\poly(k))$ and $\varepsilon=2^{-\Omega(\poly(k))}$.
\end{cor}
\begin{proof}
    Truncate $\X$ to its first $3^a+1$ blocks and use that as input to \cref{cor:explicit low min entropy oNOSF condenser} where $\Phi((3^a+1)-1)=a$. This gives us a condensing possibility result for low min-entropy \SHELAs[(\ell-r)+2^a+1,\ell-r]. Since we may remove $r$ good blocks when we truncate $\X$, our result holds for low min-entropy \SHELAs[((\ell-r)+2^a+1)+r,\ell], which simplifies to low min-entropy  \SHELAs[\ell+2^a+1,\ell].
\end{proof}
We can also restrict $r=0$ to get an analogous result.
\begin{cor}
    Let $\ell=3^a+1$ for some $a\in\N$. Then there exists an explicit condenser $\Cond:(\zo^n)^\ell\to\zo^m$ for any \SHELA[3^a-2^a+2,3^a,n,k] $\X$ with $k\geq\Omega(\log^C(n))$ for some large enough constant $C$, such that $\sminH(\Cond(\X))\geq m-O(\log(m/\varepsilon))$ where $m=\Omega(\poly(k))$ and $\varepsilon=2^{-\Omega(\poly(k))}$.
\end{cor}

\section{Extraction impossibility for rate \texorpdfstring{$2/3$ \SHELAs}{2/3 oNOSF sources}}\label{sec:(2;3)-oNOSF extraction impossibility}

We end our appendix by showing that one cannot extract from rate $2/3$ \uniSHELAs. Importantly, we note that this result is distinct from a similar result in \cite{koppartyn23multimerger} where the authors showed that extracting from rate $2/3$ \uniNOSFs is impossible. Since \uniSHELAs are a strict subset of the class of \uniNOSFs, their impossibility result does not transfer to \uniSHELAs and we must prove our own. To do so, we first claim the case of \uniSHELAs[2,3].

\begin{theorem}
\label{thm:can't extract (2;3)-SHELA}
For any function $f:(\zo^n)^3\to\zo$ there exists a \uniSHELA[2,3] $X$ such that $\abs{f(\X)-\U_1}\geq 0.08$.
\end{theorem}
Then our desired result follows as a corollary.
\begin{cor}
    For any function $f:(\zo^n)^\ell\to\zo$ where $\ell$ is divisible by 3, there exists a \uniSHELA[\frac{2}{3}\cdot \ell,\ell] $X$ such that $\abs{f(\X)-\U_1}\geq 0.08$.
\end{cor}
\begin{proof}
    For the sake of contradiction, say there exists such an $\ell$ and function $f$ such that $\abs{f(\X)-\U_1}< 0.08$ for any \uniSHELA[\frac{2}{3}\cdot \ell,\ell,n] $X$. Then if we let $\X$ be a \uniSHELA[2,3,\frac{n\ell}{3}] but consider it as a \uniSHELA[\frac{2}{3}\cdot \ell,\ell,n] by splitting up each block into $\ell/3$ sub-blocks to get $3\cdot\frac{\ell}{3}=\ell$ total blocks, we get that $f$ is a extractor for \uniSHELAs[2,3,\frac{n\ell}{3}], a contradiction to \cref{thm:can't extract (2;3)-SHELA}. 
\end{proof}
We now prove the main theorem.
\begin{proof}[Proof of \cref{thm:can't extract (2;3)-SHELA}]
    To show that extraction is impossible, we will attempt to fix the output of $f$ with constant probability over its inputs. We begin by classifying the points in the first two coordinates of $f$ as follows.
    \begin{align*}
        S_0&=\{(x_1,x_2)\in[N]^2\mid \forall x_3\in[N],\ f(x_1,x_2,x_3)=0\}\\
        S_1&=\{(x_1,x_2)\in[N]^2\mid \forall x_3\in[N],\ f(x_1,x_2,x_3)=1\}\\
        S_{0,1}&=\{(x_1,x_2)\in[N]^2\mid \exists x_3,x_3'\in[N],\ f(x_1,x_2,x_3)=0\text{ and }f(x_1,x_2,x_3')=1\}.
    \end{align*}
    Note that we can write $S_{0,1}=[N]^2\setminus(S_0\cup S_1)$. In order, these are the sets of points in $\X_1$ and $\X_2$ that fix the output of $f$ to 0, to 1, and that do not fix the output of $f$. We now take constants $0.5\leq c_0,c_1\leq 1$ and look at two cases that allow us to fix the output of $f$ by putting an adversary in the third coordinate, $\X_3$.
    \begin{casesenum}
        \item We have $\abs{S_0}+\abs{S_{0,1}}\geq c_0N^2$. Here, we know that for $(x_1,x_2)\in S_0\cup S_{0,1}$ there exists some $x_3$ such that $f(x_1,x_2,x_3)=0$. Define $a(x_1,x_2)$ be this $x_3$ for $(x_1,x_2)\in  S_0\cup S_{0,1}$ and 0 otherwise. Consequently, if we let $\X_1$ and $\X_2$ be random and define our uniform $(2,3)$-SHELA source as $\X=\X_1,\X_2,a(\X_1,\X_2)$, then we have that $\Pr[f(\X)=0]\geq c_0$. It follows that $\abs{f(\X)-\U_1}\geq c_0-\frac{1}{2}$.

        \item We have $\abs{S_1}+\abs{S_{0,1}}\geq c_0N^2$. This case follows similarly since for $(x_1,x_2)\in S_1\cup S_{0,1}$ there exists some $x_3$ such that $f(x_1,x_2,x_3)=1$. Therefore, we can define an adversary $a(x_1,x_2)$ such that when $\X=\X_1,\X_2,a(\X_1,\X_2)$ with $\X_1$ and $\X_2$ uniform we have $\abs{f(\X)-\U_1}\geq c_0-\frac{1}{2}$.

        \item We are in neither of the previous two cases. Thus, because $\abs{S_0}+\abs{S_1}+\abs{S_{0,1}}=N^2$ we have that $(1-c_0)N^2<\abs{S_0},\abs{S_1}<c_0N^2$ and $(2c_0-1)N^2<\abs{S_{0,1}}<c_0N^2$. To proceed, we will set up two sub-cases in which we either make $\X_1$ our bad block or $\X_2$ our bad block. 

        Consider the bipartite graph $H=(U,V)$ with $\abs{U}=N$ left vertices representing the values of $\X_1$ and $\abs{V}=N$ vertices representing the values of $\X_2$. We place an edge $(u,v)$ with label $t$ if $(u,v)\in S_t$ and do not place an edge otherwise. Consequently, the number of edges $E$ in $H$ is at least $E=\abs{S_0}+\abs{S_1}\geq2(1-c_0)N^2$. For any $u\in U$, define its normalized degree (counting edges with either label) as $d_u=\deg(u)/N$. We then see that $\E_{u\sim U}[d_u]= E/\abs{U}\geq2(1-c_0)$. To split into our two sub-cases, we will consider the set of heavy vertices $U_H=\{u\in U\mid d_u > c_1\}$ in $U$. By \cref{claim:reverse-markov}, we get that $\Pr_{u\in U}[d_u> c_1]\geq\frac{\E_u[d_u]-c_1}{1-c_1}\geq\frac{2(1-c_0)-c_1}{1-c_1}=:c_2$, meaning that $\abs{U_H}\geq c_2N$.
        \begin{casesenum}
            \item For all $u\in U_H$ we have $u\in S_0\cap S_1$ (i.e., $u$ has at least one edge labeled with a 0 and another with a 1). this means that for any $u\in U_H$ there exists an $x_2\in[N]$ such that for all $x_3\in[N]$ we have that $f(u,x_2,x_3)=0$. Let $a(x_1)$ be defined as outputting this $x_2$ that fixes $f$ to 0 for $x_1\in U_H$ and to be 0 otherwise. Defining $\X=\X_1,a(\X_1),\X_3$ with $\X_1$ and $\X_3$ uniform gives us a uniform $(2,3)$-SHELA source for which $\Pr[f(\X)=0]\geq\abs{U_H}/N\geq c_2$, so $\abs{f(\X)-U_1}\geq c_2-\frac{1}{2}$.

            \item There exists a $u\in U_H$ such that $u\notin S_0\cap S_1$. Without loss of generality, say $u\in S_0$, so all of the edges of $u$ are labeled 0, meaning that for all $x_2\in\mathcal{N}(u)$ and any $x_3\in[N]$ we have that $f(u,x_2,x_3)=0$. Because $u\in U_H$, we have that $d_u > c_1$, so defining $\X=u,\X_2,\X_3$ with $\X_2$ and $\X_3$ uniform gives us that $\Pr[f(\X)=0]\geq c_1$. Therefore, $\abs{f(\X)-U_1}\geq c_1-\frac{1}{2}$.
        \end{casesenum}
    \end{casesenum}

    Combining all of our cases and recalling that $c_2=\frac{2(1-c_0)-c_1}{1-c_1}$, we have that we can construct a \uniSHELA[2,3] $\X$ such that $\abs{f(\X)-U_1}\geq\varepsilon$ where $\varepsilon=\min(c_0,c_1,c_2)-\frac{1}{2}$. Setting $c_0=0.58$ and $c_1=0.6$ gives us $\varepsilon=0.58-0.5=0.08$. 
\end{proof}

% \dobib

\end{document}